\documentclass[preprint,3p,12pt]{elsarticle}

\usepackage{mathrsfs}
\usepackage{amsmath}
\usepackage{stmaryrd}
\usepackage{bbding}
\usepackage{dcolumn}
\usepackage{graphicx}
\usepackage{amsfonts}
\usepackage{amssymb}
\usepackage{psfrag}
\usepackage{wrapfig}
\usepackage{subfigure}
\usepackage{makeidx}
\usepackage{bm}
\usepackage{epsf}
\usepackage{epsfig}
\usepackage{setspace}
\usepackage{graphicx}
\usepackage{epstopdf}
\usepackage{psfrag}
\usepackage{subfigure}
\usepackage{color}
\usepackage{comment}
\usepackage{algorithm}
\usepackage{algorithmic}
\usepackage{booktabs}

\usepackage[amsmath,thmmarks]{ntheorem}

\newcommand \td {\mathrm{~d}}
\begin{document}

\title{An implicit gas-kinetic scheme for internal and external flows}

\author[HKUST1]{Yue Zhang}
\ead{yzhangnl@connect.ust.hk}	
\author[xijiao]{Xing Ji} 
\ead{jixing@xjtu.edu.cn}
\author[HKUST1,HKUST2,HKUST3]{Kun Xu\corref{cor1}}
\ead{makxu@ust.hk}

\address[HKUST1]{Department of Mathematics, Hong Kong University of Science and Technology, Clear Water Bay, Kowloon, Hong Kong}
\address[HKUST2]{Department of Mechanical and Aerospace Engineering, Hong Kong University of Science and Technology, Clear Water Bay, Kowloon, Hong Kong}
\address[HKUST3]{Shenzhen Research Institute, Hong Kong University of Science and Technology, Shenzhen, China}
\cortext[cor1]{Corresponding author}
\address[xijiao]{Shaanxi Key Laboratory of Environment and Control for Flight Vehicle, State Key Laboratory for Strength and Vibration of Mechanical Structures, School of Aerospace Engineering, Xi'an Jiaotong University, Xi'an, China}

\begin{abstract}
The gas-kinetic scheme(GKS) is a promising computational fluid dynamics (CFD) method for solving the Navier-Stokes equations. It is based on the analytical solution of the BGK equation, which enables accurate and robust simulations. While GKS has demonstrated excellent properties (e.g., unified treatment of inviscid and viscous fluxes, inherent adaptive dissipation control), its application to classical engineering problems, such as aerodynamic flows and fluid machinery, remains underdeveloped compared to conventional CFD methods. This study bridges this gap by advancing GKS capabilities for real-world engineering challenges.
First, the GKS is extended to a rotating coordinate frame, enabling efficient simulations of internal flows in turbomachinery. Second, the computational inefficiency of explicit GKS is addressed through an implicit time discretization using the generalized minimal residual method. The Jacobian matrices for inviscid/viscous fluxes are approximated using the first-order kinetic flux vector splitting scheme and the thin shear layer approximation to enhance robustness and computational efficiency further. Third, the shear-stress transport turbulence model is coupled to expand GKS's applicability to industrial turbulent flows.
Numerical tests, including internal compressor rotor flow and external flow over a 3-D wingbody, validate the proposed method's accuracy and efficiency. Via our implicit scheme, the force coefficients of the 3-D wing-body flow with about five million mesh elements can converge after 500 steps. This work represents a practical advancement of GKS, demonstrating its potential to compete with established CFD solvers in high-Reynolds-number external and internal turbulent flows.
\end{abstract}

\begin{keyword}
   gas-kinetic scheme, rotating coordinate frame, generalized minimal residual, turbulence model, aerodynamic, turbomachinery
\end{keyword}

\maketitle

\section{Introduction}

The advancement of the aviation industry has led to an increasing demand for enhanced aerodynamic performance, improved propulsion efficiency, and heightened safety in aircraft design. Computational fluid dynamics (CFD) has emerged as a fundamental technology in this field, enabling engineers to simulate fluid flow characteristics numerically. By providing high-resolution data in complex flow scenarios, CFD significantly reduces the cost and duration associated with traditional wind tunnel testing. It addresses challenges such as external aerodynamics in aircraft shape design, wing-engine interactions, and internal flow issues in compressor and turbine components. This work aims to develop the gas-kinetic scheme (GKS) method specifically for aviation applications.

The GKS is a numerical approach rooted in kinetic theory, designed to solve the Euler and Navier–Stokes equations \cite{xuGKS2001}. Unlike conventional solvers that rely on Riemann problems to compute inviscid flux and the central difference method for viscous fluxes, GKS operates at the mesoscopic scale, utilizing kinetic theory to capture the underlying physical processes with greater accuracy. By integrating both normal and tangential gradients of flow variables around cell interfaces during the time evolution of the gas distribution function, GKS effectively incorporates multi-dimensional effects \cite{xu2005multidimensional}. The time-accurate surface distribution function allows for direct updates of flow variables without additional equations, facilitating the implementation of compact reconstruction methods. 
A third-order compact scheme was first developed by combining cell-averaged flow variables with pointwise variables at cell interfaces \cite{pan2015compact}. HWENO-type methods and two-step multi-resolution WENO reconstructions have been developed within the GKS framework to achieve high-order spatial accuracy \cite{zhuNewFiniteVolume2018, jiMultiresolution2021}. Furthermore, the compact GKS (CGKS) has been successfully implemented on both structured and unstructured meshes for two- and three-dimensional problems \cite{ji4order2018, jiThreedimensionalCompactHighorder2020, zhaoCompactHigherorderGaskinetic2019, zhaoAcousticShockWave2020, zhaoDirectModelingComputational2021, jiGC2021, zhaoCompactHighorderGaskinetic2022a}, which shows the potential of GKS in high-order accuracy and computational efficiency.

Accurate and efficient predictions of steady-state flow are of vital importance in design and optimization problems within aviation applications. However, explicit time integration methods can be computationally expensive when applied to steady problems, particularly when dealing with small-scale meshes for resolving boundary layers. To improve computational efficiency, implicit schemes are often employed for temporal discretization. Two widely used implicit methods are the lower-upper symmetric Gauss-Seidel (LU-SGS) and generalized minimal residual (GMRES) methods. The LU-SGS method, introduced by Yoon \cite{yoon1988lower}, decomposes the coefficient matrix into lower triangular, diagonal, and upper triangular components, solving the system through forward and backward sweeps. While LU-SGS is stable and straightforward to implement, it often exhibits slow convergence. In contrast, the GMRES method \cite{brown1990hybrid,saad1986GMRES}, a Krylov subspace technique, minimizes the residual norm within a Krylov subspace, offering enhanced robustness and superior performance for ill-conditioned systems, albeit with higher memory requirements. Various preconditioners have been applied to GMRES to improve convergence further \cite{crivellini2011implicit,luo1994implicit,xiaoquan2019robust,zhou2013evaluation}. Significant research has focused on enhancing the efficiency of GKS solvers, including the development of LU-SGS-based \cite{cao2019implicit,liu2024compact} and GMRES-based \cite{tan2017time,yang2024implicit,yang2025implicit} methods. In this paper, the GMRES method is employed to enhance convergence. As demonstrated in \cite{xiaoquan2019robust,jia2023robust}, using an accurate Jacobian matrix can significantly improve the performance of the GMRES method. Therefore, this study utilizes a more precise Jacobian matrix than previous research for implicit GKS.

In addressing challenges in aviation applications, it is important to consider typical rotating components such as compressors and turbines in engines. A non-inertial rotating coordinate system is well-suited for analyzing the intricate flow dynamics associated with these components. In this framework, the effects of centrifugal and Coriolis forces become significant and must be incorporated into the governing kinetic equations \cite{zhouThreeDimensionalGasKineticBGK2019}. Our previous work developed the GKS for rotating mesh problems using this approach \cite{zhang2023slidingmesh}. The corresponding macroscopic governing equations are systematically derived from the BGK equation via the Chapman-Enskog expansion, facilitating a rigorous transition from kinetic to continuum descriptions. This study focuses on steady-state problems, ignoring the acceleration term in the gas distribution function to simplify flux calculations.

The structure of this paper is organized as follows: In Section 2, the gas-kinetic scheme in a rotating coordinate frame is derived. Section 3 presents the reconstruction methods for non-equilibrium and equilibrium distribution functions. Section 4 discusses the implicit time discretization using the GMRES method in detail. Section 5 introduces the shear-stress transport (SST) turbulence model and its discretization. In Section 6, the numerical results are presented, and finally, Section 7 concludes the paper.

\section{Gas-Kinetic Finite Volume Scheme}

\subsection{BGK equation in rotating framework}
In this section, the gas-kinetic BGK equation in a rotating frame is derived to deal with the rotation problem. The system returns to the inertial frame when the rotation speed equals zero. The gas-kinetic BGK equation in a rotating frame is
\begin{equation*}
  \frac{\partial f}{ \partial t}+ \mathbf{w}\cdot \nabla_x f + \mathbf{a}_w\cdot \nabla_w f = \frac{g-f}{\tau},
\end{equation*}
where $f=f(\mathbf{x},t,\mathbf{w},\xi)$ is the gas distribution function, $g$ is the corresponding equilibrium state, and $\tau$ is
the collision time. $\mathbf{w}=(w_1,w_2,w_3)$ is the particle velocity in the rotating frame. And the acceleration in the rotating frame is
\begin{equation*}
  \mathbf{a}_w =\frac{\td\mathbf{w}}{\td t} = -\mathbf{\Omega}\times(\mathbf{\Omega}\times \mathbf{r})-2(\mathbf{\Omega}\times \mathbf{w}),
\end{equation*}
where $\mathbf{\Omega}$ is the angular velocity of the rotating frame, and $\mathbf{r}$ is a position vector from the origin of rotation to the particle's position. $-\mathbf{\Omega}\times(\mathbf{\Omega}\times \mathbf{r})$ is centrifugal force and $-2(\mathbf{\Omega}\times \mathbf{w})$ is Coriolis force.    Denote $\mathbf{v}=(v_1,v_2,v_3)$ as the particle velocity in the absolute inertial reference frame; the relationship among velocities is
\begin{equation*}
  \mathbf{v} =\mathbf{U}+\mathbf{w},
\end{equation*}
where $\mathbf{U}=\mathbf{\Omega}\times\mathbf{r}$ is convection velocity due to the frame rotation.
According to $\mathbf{a}_v= \frac{\td\mathbf{v}}{\td t} = \frac{\td\mathbf{w}}{\td t}+ (\mathbf{\Omega}\times \mathbf{w}) $, the acceleration term in the rotating frame can be expressed as $-(\mathbf{\Omega}\times \mathbf{v}) \cdot\nabla_v f $. Then the BGK equation becomes
\begin{equation}\label{BGK}
  \frac{\partial f}{ \partial t}+ \mathbf{w}\cdot \nabla_x f - (\mathbf{\Omega}\times \mathbf{v})\cdot\nabla_v f = \frac{g-f}{\tau},
\end{equation}
where $f$ can be defined by absolute velocity $\mathbf{v}$, such as $f=f(\mathbf{x},t,\mathbf{v},\xi)$.
The collision term in the above equation describes the evolution process from a non-equilibrium state to an equilibrium one, with the satisfaction of the compatibility condition
\begin{equation*}
  \int \frac{g-f}{\tau} \Psi \td\mathbf{v}\td\Xi = 0,
\end{equation*}
where $\Psi=(1,v_1,v_2,v_3,\frac{1}{2}(v_1^2+v_2^2+v_3^2+\xi^2))^T$ and $\td\Xi=\td\xi_1\cdots\td\xi_K$ ($K$ is the number of internal degree of freedom, i.e. $K=2$ for three-dimensional diatomic gas). Based on the Chapman-Enskog Expansion \cite{zhang2023slidingmesh}, the Euler and N-S equations in the rotating frame can be obtained. The N-S equations in a rotating frame are
\begin{equation*}
  \begin{aligned}
    \frac{\partial \rho }{\partial t} &+\nabla\cdot\rho(\mathbf{V}-\mathbf{U})=0,\\
    \frac{\partial \rho \mathbf{V}}{\partial t} &+\nabla\cdot[\rho (\mathbf{V}-\mathbf{U}) \mathbf{V}+pI -\overline{\overline{\sigma}}]=-\mathbf{\Omega}\times \rho\mathbf{V},\\
    \frac{\partial \rho E }{\partial t} &+\nabla\cdot[\rho H(\mathbf{V}-\mathbf{U})+p\mathbf{V} -\kappa \nabla T -\overline{\overline{\sigma}}\cdot \mathbf{V}]=0,\\
  \end{aligned}
\end{equation*}
where $\rho,\mathbf{V}, p, T, E, H$ and $\overline{\overline{\sigma}}$ are the density, absolute velocity, pressure, temperature, energy, enthalpy, and viscosity stress of fluid. With the gradient of temperature $\nabla T$ and viscosity stress $\overline{\overline{\sigma}}$ equal to zero, the N-S equations become the Euler Equations.

\subsection{Finite volume discretization}
The finite volume method is used to discretize the system. The whole domain $\Omega$ is discretized into small cells $\Omega_i$
  \begin{equation*}
    \Omega =\bigcup \Omega_i,\ \Omega_i \bigcap \Omega_j=\phi (i \neq j).
  \end{equation*}
 For each cell $\Omega_i$, the boundary faces can be expressed as
  \begin{equation*}
    \partial \Omega_i=\bigcup_{p=1}^{N_f}\Gamma_{ip}.
  \end{equation*}
 Taking moments of the BGK equation (\ref{BGK}) and integrating over the cell $\Omega_i$, the semi-discretized form of the finite volume scheme can be written as 
\begin{equation}
 \label{fvm_rigid}
 |\Omega_i|\frac{\td \mathbf{W_i}}{\td t} = -\sum_{p=1}^{N_f}\int_{\Gamma_{ip}} (\mathbf{F}(\mathbf{W}) - \mathbf{W}\mathbf{U})\cdot \mathbf{n}_p \td S + \mathbf{S}(\mathbf{W})
 := \mathcal{L}_F\left(\mathbf{W}\right) + \mathbf{S}(\mathbf{W}).
    \end{equation}
where $\mathbf{W}_i$ is the cell average conservative value, $|\Omega_i|$ is the volume of cell $\Omega_i$, $\mathbf{F}(\mathbf{W})$ is the flux passing through the fixed element surface, $\mathbf{W}\mathbf{U}$ presents the flux due to the mesh motion, $\mathbf{n}_p=(n_1,n_2,n_3)^T$ is the normal direction of cell surface and $\mathbf{S}=|\Omega_i|[0,-\mathbf{\Omega}\times \rho\mathbf{V},0]^T$ is the source term due to frame rotation.

The integration of flux can be approximated (the index $i$ is omitted )
\begin{equation*}
  \int_{\Gamma_{p}} (\mathbf{F}(\mathbf{W}) - \mathbf{W}\mathbf{U}) \cdot \mathbf{n}_p dS  \approx |S_p| (\mathbf{F}(\mathbf{W})-\mathbf{W}\mathbf{U})_{\mathbf{x}_{p}}\cdot \mathbf{n}_{p},
\end{equation*}
where $|S_p|$ is the cell surface area, and $\mathbf{x}_{p}$ is the coordinate of the cell surface center. The flux through the surface is expressed as
\begin{equation*}
  \mathbb{F}(\mathbf{W},\mathbf{x})= (\mathbf{F}(\mathbf{W}) - \mathbf{W}\mathbf{U})_{\mathbf{x}_{p}}\cdot \mathbf{n}_{p} = \int f(\mathbf{x},t,\mathbf{v},\xi) \mathbf{w}\cdot \mathbf{n}_p \Psi \td\mathbf{v}\td\Xi.
\end{equation*}
There is a conflict that the particle velocity across the cell surface is the relative velocity, but the momentum and energy that particles carry is about the absolute velocity. To solve this problem, the transformation from the flux with relative velocity to the flux with absolute velocity is used here.
The components of $\mathbb{F}$ are determined through the following transformation relations:
\begin{equation}\label{fluxTransform}
  \begin{cases}
    \mathbb{F}_\rho=\mathbb{F}_\rho^\prime ,\\
    \mathbb{F}_{\rho V_1}=\mathbb{F}_{\rho W_1}^\prime + U_1 \mathbb{F}_\rho^\prime ,\\
    \mathbb{F}_{\rho V_2}=\mathbb{F}_{\rho W_2}^\prime + U_2 \mathbb{F}_\rho^\prime ,\\
    \mathbb{F}_{\rho V_3}=\mathbb{F}_{\rho W_3}^\prime + U_3 \mathbb{F}_\rho^\prime ,\\
    \mathbb{F}_{\rho E}=\mathbb{F}_{\rho E}^\prime+U_1\mathbb{F}_{\rho W_1}^\prime +U_2\mathbb{F}_{\rho W_2}^\prime +U_3\mathbb{F}_{\rho W_3}^\prime + \frac{1}{2}(U_1^2+U_2^2+U_3^2) \mathbb{F}_\rho^\prime . \\
  \end{cases}.
\end{equation}
The flux $\mathbb{F}^\prime$ is defined by the relative velocity, and can be evaluated through the following integral formulation:
  \begin{equation}\label{fluxIntegralW}
    \mathbb{F}^\prime(\mathbf{W},\tilde{\mathbf{x}} ) =\mathbf{T}^{-1}\int f(\tilde{\mathbf{x}},t,\tilde{\mathbf{w}},\xi) \tilde{w}_1 \tilde{\mathbf{ \Psi}}^\prime \td\tilde{\mathbf{w}}\td\Xi,
  \end{equation}
where $\tilde{\mathbf{x}}=(0,0,0)$ denotes the origin of the local coordinate system with x-direction aligned with $\mathbf{n}$, and $\widetilde{\mathbf{ \Psi}}^\prime =(1,\tilde{w}_1,\tilde{w}_2,\tilde{w}_3,\frac{1}{2}(\tilde{w}_1^2+\tilde{w}_2^2+\tilde{w}_3^2+\xi^2))^T$, and $\mathbf{T} = \text{diag}(1,\mathbf{T}^{\prime},1)$ denotes the rotation matrix, with
\begin{equation*}
  \mathbf{T}^{\prime}=\left(\begin{array}{ccccc}
 n_{1} & n_{2} & n_{3}\\
 -n_{2} & n_{1}+\frac{n_{3}^{2}}{1+n_{1}} & -\frac{n_{2} n_{3}}{1+n_{1}} \\
 -n_{3} & -\frac{n_{2} n_{3}}{1+n_{1}} & 1-\frac{n_{3}^{2}}{1+n_{1}}
  \end{array}\right), \quad n_{1} \neq-1,
  \end{equation*}
and $\mathbf{T}^\prime$ reduces to $\text{diag}(-1,-1,1)$ when $n_1=-1$.
The microscopic velocities in the local coordinate system are obtained through the transformation $\tilde{\mathbf{w}} = \mathbf{T}^\prime \mathbf{w}$.
 
\subsection{Gas evolution model}
 In order to construct the numerical fluxes at $\mathbf{x} = (0,0,0)^T$, the integral solution of the BGK equation Eq.(\ref{BGK}) is
  \begin{equation}\label{solbgk}
 f(\mathbf{x},t, \mathbf{w},\xi)=\frac{1}{\tau}\int_0^t g(\mathbf{x}^{\prime},t^{\prime},\mathbf{w},\xi)e^{-(t-t^{\prime})/\tau}\td t^{\prime}+ e^{t/\tau}f_0(\mathbf{x}_{0},\mathbf{w}),
  \end{equation}
with the trajectory
  \begin{equation*}
    \mathbf{x}=\mathbf{x}^{\prime}+\mathbf{w}(t-t^{\prime}).
  \end{equation*}
In Eq.(\ref{solbgk}), $f_0$ is the initial gas distribution function, and $g$ is the corresponding equilibrium state. $\mathbf{x}_0$ are the initial position by tracing back particles $\mathbf{x}$ at time $t$ back to $t=0$.

Before constructing the initial distribution function $f_0$ and equilibrium state $g$, first denote
  \begin{equation*}
    \mathbf{a}=(a_1,a_2,a_3) = \nabla_x g/g, A = g_t/g.
  \end{equation*}
In the following derivation, quadratic terms of time will be ignored directly.
With the consideration of possible discontinuity at an interface, the initial distribution is constructed as
  \begin{equation}\label{ini_gas}
 f_0(\mathbf{x}_{0},\mathbf{w}_0)=f_0^l(\mathbf{x}_{0},\mathbf{w}) (1 -\mathbb{H}(x_1)) +f_0^r(\mathbf{x}_{0},\mathbf{w})\mathbb{H}(x_1),
  \end{equation}
 where $\mathbb{H}$ is the Heaviside function. $f_0^{l}$ and $f_0^{r}$ are the initial gas distribution functions on the left and right sides of the interface, which are determined by corresponding initial macroscopic variables and their spatial derivatives.
 With the second-order accuracy, $f_0^k(\mathbf{x},\mathbf{w})$ is constructed by Taylor expansion around $(\mathbf{x},\mathbf{w})$
  \begin{equation}\label{taylor_ex}
 f_0^k(\mathbf{x}_{0},\mathbf{w})=f_G^k (\mathbf{x},\mathbf{w}) - \mathbf{w}t\cdot \nabla_xf_G^k
  \end{equation}
 for $k=l,r$. Based on the Chapman-Enskog expansion, $f_G^k$ is given by
  \begin{equation}\label{ce_ex}
 f_G^k= g^k[1-\tau(A^k +  \mathbf{a}^k \cdot \mathbf{w})],
  \end{equation}
where $g^k$ is the equilibrium distribution function defined by the macroscopic variables $\mathbf{W}^k$ at both sides of a cell interface, $\mathbf{a}^k$ are defined by the spatial derivatives of $g^k$
\begin{equation*}
  \begin{aligned}
a_i^k&= (\frac{\partial g^k}{\partial \rho^k }\frac{\partial \rho^k}{\partial x_i}+ \frac{\partial g^k}{\partial W_1^k }\frac{\partial W_1^k}{\partial x_i} + \frac{\partial g^k}{\partial W_2^k }\frac{\partial W_2^k}{\partial x_i} + \frac{\partial g^k}{\partial W_3^k }\frac{\partial W_3^k}{\partial x_i}+ \frac{\partial g^k}{\partial \lambda^k}\frac{\partial \lambda^k}{\partial x_i})/g^k
\\&= a_{i1}^k +a_{i2}^k w_1 + a_{i3}^k w_2 + a_{i4}^k w_3 +a_{i5}^k \frac{1}{2}(w_1^2+w_2^2+w_3^2\xi^2),
  \end{aligned}
\end{equation*}
and
  \begin{equation*}
A^k=A_{1}^k +A_{2}^k w_1 + A_{3}^k w_2 + A_{4}^k w_3 +A_{5}^k \frac{1}{2}(w_1^2+w_2^2+w_3^2\xi^2)
\end{equation*}
are determined  by the compatibility condition
  \begin{equation}\label{compatibility}
\int(f^k_G-g^k) \Psi\td\mathbf{w}\td\Xi=0.
\end{equation}
 Substituting Eq.(\ref{taylor_ex}) and (\ref{ce_ex}) into (\ref{ini_gas}), the initial gas distribution has following form
  \begin{equation}\label{inidis}
    \begin{aligned}
 f_{0}= \begin{cases}g^{l}\left[1-\left(\mathbf{a}^{l} \cdot \mathbf{w}\right) t-\tau\left(A^{l}+\mathbf{a}^{l} \cdot \mathbf{w} \right)\right], & x_1<0 , \\
 g^{r}\left[1-\left(\mathbf{a}^{r} \cdot \mathbf{w}\right) t-\tau\left(A^{r}+\mathbf{a}^{r} \cdot \mathbf{w}\right)\right], & x_1 \geq 0 ,\end{cases}
    \end{aligned}
    \end{equation}
Then, the equilibrium distribution is defined by the Taylor expansion
    \begin{equation}\label{equdis}
      \begin{aligned}
 g(\mathbf{x^{\prime}},t^{\prime},\mathbf{w})
 =&\overline{g}(\mathbf{x},0,\mathbf{w}) + \nabla_x \overline{g}\cdot (\mathbf{x^{\prime}} -\mathbf{x}) + \overline{g}_t t^{\prime} \\
 =&\overline{g}(\mathbf{x},0,\mathbf{w}) - \nabla_x \overline{g}\cdot \mathbf{w}(t-t^{\prime})   + \overline{g}_t t^{\prime} \\
 =&\overline{g}(\mathbf{x},0,\mathbf{w}) \left\{ 1 - \overline{\mathbf{a}}\cdot \mathbf{w}(t-t^{\prime})+A t^{\prime} \right\} , \\
       \end{aligned}
    \end{equation}
where $\overline{g}$ and $\overline{\mathbf{a}}$ are determined from the reconstruction of macroscopic flow variables presented in section \ref{equ_res}, and $A$ is obtained by the compatibility condition (\ref{compatibility}).

In a smooth flow region, the collision time for viscous flow is determined by $\tau=\mu/p$, where $\mu$ is the dynamic viscosity coefficient and $p$ is the pressure at the cell interface, and for inviscid flow, the collision time should be set as zero. In order to properly capture the unresolved shock structure, additional numerical dissipation is needed. The physical collision time $\tau$ in the exponential function part can be replaced by a numerical collision time 
\begin{equation*}
  \tau_n=\frac{\mu}{p} + 5.0\frac{|p_l-p_r|}{p_l+p_r}\Delta t ,
\end{equation*}
where $p_l$ and $p_r$ denote the pressure on the left and right sides of the cell interface.

By substituting  Eq. (\ref{inidis}) and Eq. (\ref{equdis}) into Eq. (\ref{solbgk}) with $\tau$ and $\tau_n$, the second-order time-dependent gas distribution function at a cell interface becomes
  \begin{equation*}
    \begin{aligned}
 f\left(\mathbf{x}, t, \mathbf{w}, \xi\right)
 &=(1-e^{-t / \tau_n})\bar{g}+e^{-t / \tau_n}\left[\mathbb{H}\left(w_1\right) g^{l}+\left(1-\mathbb{H}\left(w_1\right)\right) g^{r}\right] +t\bar{A} \bar{g}\\
 &-\tau\left(1-e^{-t / \tau_n}\right)\bar{g}\left(\overline{\mathbf{a}} \cdot \mathbf{w} +\bar{A} \right)\\
 &-\tau e^{-t / \tau_n} \mathbb{H}\left(w_1\right)g^{l}\left(\mathbf{a}^{l}\cdot  \mathbf{w} + A^{l}\right)\\
 &-\tau e^{-t / \tau_n} \left(1-\mathbb{H}\left(w_1\right)\right)g^{r} \left(\mathbf{a}^{r} \cdot \mathbf{w} + A^{r}\right)\\
 &+t e^{-t / \tau_n}\left[ \left(\overline{\mathbf{a}} \cdot \mathbf{w}\right)\bar{g}- \mathbb{H}\left(w_1\right)\left(\mathbf{a}^{l} \cdot \mathbf{w}\right) g^{l}-\left(1-\mathbb{H}\left(w_1\right)\right)\left(\mathbf{a}^{r} \cdot \mathbf{w}\right) g^{r}\right].\\
  \end{aligned}
    \end{equation*}
The fluxes in Eq.(\ref{fluxIntegralW}) can be obtained by taking the moments of the above distribution function. The calculation of moments can be found in \cite{xuGKS2001}.

In order to account for the Prandtl number not equal to 1, additional correction heat flux should be added to the energy flux:
\begin{equation*}
 (\frac{1}{\text{Pr}}-1)q,
\end{equation*}
where $q = F_{v4}-W_1F_{v1}-W_2F_{v2}-W_3F_{v3}$ can be obtained through the viscous flux $\mathbf{F}_{vis}=[0,F_{v1},F_{v2},F_{v3},F_{v4}]$. The viscous flux can be obtained by
\begin{equation*}
  \mathbf{F}_{vis}=-\tau \int w_1\bar{g}\left[ \left(\overline{\mathbf{a}} \cdot \mathbf{w}\right) +\bar{A} \right]\td{\mathbf{w}}\td\Xi.
\end{equation*}

\section{Spatial Reconstruction}
In our work, the Green-Gauss reconstruction is used to reconstruct the macroscopic flow variables, which is more robust than the least square method when dealing with a large aspect ratio mesh. Furthermore, the discontinuity feedback factor is introduced to enhance the shock-capturing capabilities.
\subsection{Green-Gauss reconstruction}
The target cell is denoted as $\Omega_0$, with the cell center represented by $\mathbf{x}_0=(x_0,y_0,z_0)$. The neighboring cells are denoted as $\Omega_m$ for $m = 1, 2, \cdots, N_f$.
The cell-averaged values of cell $\Omega_m$ are denoted as $\overline{Q}_m$.
The classical Green-Gauss reconstruction with only cell-averaged values is adopted to provide the linear polynomial $p^1=q_0+q_1(x-x_0)+q_2(y-y_0)+q_3(z-z_0)$ for the sub-stencil.
\begin{equation*}
 p^1=\overline{Q}_0+\frac{1}{|\Omega_0|}(\mathbf{x}-\mathbf{x}_0)\cdot \sum_{m=1}^{N_f}\frac{\overline{Q}_m+\overline{Q}_0}{2}S_m\mathbf{n} _m,
\end{equation*}
where $S_m$ is the area of the cell's surface and $\mathbf{n}_m$ is the surface's normal vector.
\subsection{Discontinuity feedback factor}

A discontinuity feedback factor (DF) is introduced to modify the reconstruction process to enhance shock-capturing capabilities adaptively. This approach analyzes the reconstructed interface values to identify regions where discontinuities will likely propagate in subsequent timesteps, allowing for more accurate shock tracking. It was first proposed in \cite{jiGC2021}, and several improvements have been made in \cite{zhang2023slidingmesh}. Denote $\alpha_i \in[0,1]$ as discontinuity feedback factor at targeted cell $\Omega_i$
\begin{equation*}
  \alpha_i = \prod_{p=1}^{n_t}\prod_{k=0}^{M_p}\alpha_{p,k},
\end{equation*}
where $\alpha_{p,k}$ is the DF obtained at the $k$th Gaussian point of the interface $p$ around cell $\Omega_i$, which can be calculated by
\begin{equation*}
  \begin{aligned}
 &\alpha_{p,k}=\frac{1}{1+D^2}, \\
 &D=\frac{|p^l-p^r|}{p^l} +\frac{|p^l-p^r|}{p^r}+(\text{Ma}^{l}_n-\text{Ma}^{r}_n)^2+(\text{Ma}^{l}_t-\text{Ma}^{r}_t)^2,
  \end{aligned}
\end{equation*}
where $p$ is pressure, $\text{Ma}_n$ and $\text{Ma}_t$ are the Mach numbers defined by normal and tangential velocity, and superscript $l,r$ denote the left and right values of the Gaussian points.

Then, the Green-Gauss reconstruction is modified as
\begin{equation*}
 p^1=\overline{Q}_0+\alpha_0 \frac{1}{|\Omega_0|}(\mathbf{x}-\mathbf{x}_0)\cdot \sum_{m=1}^{N_f}\frac{\overline{Q}_m+\overline{Q}_0}{2}S_m\mathbf{n} _m.
\end{equation*}

\subsection{Reconstruction of equilibrium states}\label{equ_res}
After reconstructing the non-equilibrium state, a kinetic weighted average method can be used to get the equilibrium states 
and their derivatives \cite{jiHWENOReconstructionBased2020},
\begin{equation*}
  \int \bar{g}\Psi\td\mathbf{v}\td\Xi =\mathbf{W}_0 = \int_{\tilde{w}_1>0} g^{l} \Psi\td\mathbf{v}\td\Xi +\int_{\tilde{w}_1<0} g^{r}\Psi\td\mathbf{v}\td\Xi,
\end{equation*}
\begin{equation*}
  \int \bar{a}_i \bar{g}\Psi\td\mathbf{v}\td\Xi = \frac{\partial \mathbf{W}_{0} }{\partial \tilde{x}_i}= \int_{\tilde{w}_1>0}a_i^l g^{l} \Psi\td\mathbf{v}\td\Xi +\int_{\tilde{w}_1<0} a_i^r g^{r}  \Psi\td\mathbf{v}\td\Xi , i=1,2,3.
\end{equation*}
For the normal derivatives, the above solution is further modified according to the idea in linear diffusive generalized Riemann problem (dGRP) \cite{dgrp}
\begin{equation}\label{dgrp}
  \int \bar{a}_1\bar{g}\Psi\td\mathbf{v}\td\Xi = \frac{\partial \mathbf{W}_{0} }{\partial \tilde{x}_1}=  \int_{\tilde{w}_1>0} a_1^l g^{l}  \Psi\td\mathbf{v}\td\Xi +\int_{\tilde{w}_1<0} a_1^r g^{r}  \Psi\td\mathbf{v}\td\Xi + \frac{\mathbf{W}^r-\mathbf{W}^l}{(\mathbf{x}_{rc}-\mathbf{x}_{lc}) \cdot \mathbf{n}},
\end{equation}
where $\mathbf{x}_{rc}$ and $\mathbf{x}_{lc}$ are the coordinates of the left and right cell centroid, and $\mathbf{n}$ is the normal vector of the interface. By adding a penalty term in Eq. (\ref{dgrp}), the whole scheme is essentially free from the odd-even decoupling phenomenon \cite{BLAZEK201573}.

\section{Solution updates and temporal discretization}

\subsection{Implicit scheme for the fluid governing equations}
According to the semi-discretization Eq. (\ref{fvm_rigid}), the back Euler method can be write as
\begin{equation*}
 |\Omega_i|\frac{\mathbf{W}_i^{n+1}-\mathbf{W}_i^n}{\Delta t} = \mathcal{L}_F(\mathbf{W}_i^{n+1}) + \mathbf{S}(\mathbf{W}_i^{n+1})
\end{equation*}
With the approximation of
\begin{equation*}
  \begin{aligned}
  \mathcal{L}_F(\mathbf{W}^{n+1})-\mathcal{L}_F(\mathbf{W}^{n})\approx( \frac{\partial \mathcal{L}_F}{\partial \mathbf{W}})^{n}(\mathbf{W}^{n+1}-\mathbf{W}^{n}),\\
  \mathbf{S}(\mathbf{W}^{n+1})-\mathbf{S}(\mathbf{W}^{n})\approx( \frac{\partial \mathbf{S}}{\partial \mathbf{W}})^{n}(\mathbf{W}^{n+1}-\mathbf{W}^{n}).
  \end{aligned}
\end{equation*}
The scheme can be
\begin{equation}\label{implicit_scheme}
\left(\frac{|\Omega_i|}{\Delta t} - \left( \frac{\partial \mathcal{L}_F}{\partial \mathbf{W}}\right)^{n}-\left( \frac{\partial \mathbf{S}}{\partial \mathbf{W}}\right)^{n}\right)(\mathbf{W}_i^{n+1}-\mathbf{W}_i^n) = \mathcal{L}_F(\mathbf{W}_i^n) + \mathbf{S}(\mathbf{W}_i^n).
\end{equation}
The spatial residual $\mathcal{L}_F(\mathbf{W}_i^n)$ is the time-averaged value of the flux residual over the global time step of CFL number 0.5. Moreover, the GKS flux through flux translation ( Eq. (\ref{fluxTransform}) and Eq. (\ref{fluxIntegralW}) ) is used to calculate the flux residual.
To obtain the $\partial \mathcal{L}_F/\partial \mathbf{W}$, the flux Jacobian matrix is calculated by the first-order KFVS scheme for inviscid flux, and thin shear layer (TSL) approximation for viscous flux. The derivation of the Jacobian matrix of the face flux is shown in \ref{flux_jacobian}. And the source term Jacobian matrix is given by
  \begin{equation*}
    \left(\frac{\partial \mathbf{S}}{\partial \mathbf{W}}\right)_i = |\Omega_i|\left[\begin{matrix}
      0 & 0 & 0 & 0 & 0\\
      0 & 0 & \Omega_z & -\Omega_y & 0\\
      0 & -\Omega_z & 0 & \Omega_x & 0\\
      0 & \Omega_y & -\Omega_x & 0 & 0 \\
      0 & 0 & 0 & 0 & 0
    \end{matrix}\right].
  \end{equation*}

\subsection{Treatment of boundary conditions}
The ghost cells are used to realize boundary conditions. The ghost cell values $\mathbf{W}_g$ and inner cell values $\mathbf{W}_i$ has the relationship
\begin{equation*}
  \mathbf{W}_g=f(\mathbf{W}_i).
\end{equation*}
Take a differential on both sides
\begin{equation*}
  \Delta \mathbf{W}_g=\frac{\partial f}{\partial \mathbf{W}_i}\Delta \mathbf{W}_i = \frac{\partial \mathbf{W}_g}{\partial \mathbf{W}_i}\Delta \mathbf{W}_i,
\end{equation*}
e.g.
\begin{equation}\label{boundary_condition}
  \Delta \mathbf{W}_g-\frac{\partial \mathbf{W}_g}{\partial \mathbf{W}_i}\Delta \mathbf{W}_i =0,
\end{equation}
The $\frac{\partial \mathbf{W}_g}{\partial \mathbf{W}_i}$ is obtained by the chain rule. In this case, the implicit treatment for the boundary condition is constructed.

\subsection{GMRES method}
The complete linear system for steady-state flow calculations is described by Eq. (\ref{implicit_scheme}) and Eq. (\ref{boundary_condition}).
The generalized minimal residual (GMRES) approach can be employed to solve this system, which belongs to the family of Krylov subspace techniques. The detailed implementation steps of our GMRES solver are outlined in Algorithm \ref{alg:GMRES}.
\begin{algorithm}
  \renewcommand{\algorithmicrequire}{\textbf{Input: $A,b,M_L,M_R,x_0,k_{max}$}}
  \renewcommand{\algorithmicensure}{\textbf{Output: $x$}}
  \caption{Implementation of GMRES method}\label{alg:GMRES}
  \algorithmicrequire \\
  \algorithmicensure
  \begin{algorithmic}[1]
    \STATE $X_0=0$
    \FOR{$i=1,n$}
    \STATE $r_0=b-A\cdot X_0,r_1=M_L^{-1}r_0,\beta =||r_1||,q_1=r_1/\beta $
    \FOR{$m=1,k_{max}$}
     \STATE $w=M_L^{-1}\cdot A\cdot M_R^{-1}\cdot q_m$
     \STATE commumicate $w$ to other processors
     \STATE $ h_{l,m}=(w,q_l),l=1,2,\cdots,m$
      \STATE $\widetilde{q}_{m+1}=w-\sum_{l=1}^m h_{l,m}q_l$
     \STATE $ h_{m+1,m}=||\widetilde{q}_{m+1}||,q_{m+1}=\frac{\widetilde{q}_{m+1}}{h_{m+1,m}}$
    \ENDFOR
    \STATE solve $y=\min_y||\beta e_1-\tilde{H}y||_2,(e_1=[1,0,0,\cdots,0]^T)$
    \STATE $x^\prime = \sum q_ky_k, x = x_0+M_R^{-1}x^\prime$
    \STATE $X_0=x$
    \ENDFOR
    \STATE return $x$
  \end{algorithmic}
  \end{algorithm}

The implementation incorporates preconditioning techniques to enhance convergence. Specifically, a Lower-Upper Symmetric Gauss-Seidel (LUSGS) scheme is utilized as the left preconditioner, while an identity matrix is maintained as the right preconditioner. For a given matrix $A$, a decomposition into $A = L + D + U$ is performed, where $L$, $D$, and $U$ represent the lower triangular, diagonal, and upper triangular components, respectively. The left preconditioner $M_L$ is defined as:
\begin{equation*}
 M_L=(L+D)\cdot D^{-1}\cdot(D+U)
\end{equation*}
To solve the linear system $M_L\cdot X=B$, it can be expressed as: 
\begin{equation*}
 Y=D^{-1}(B-L\cdot Y),X=Y-D^{-1}U\cdot X
\end{equation*}

The parallelization of the code is based on the Message Passing Interface (MPI). The physical domain is partitioned using the METIS library to achieve load balancing across processors. Recursive bisection is employed through METIS\_PartGraphRecursive() for partitioning with eight or fewer subdomains. The multilevel k-way partitioning algorithm METIS\_PartGraphKway() is utilized for larger numbers of partitions. This partitioning strategy minimizes communication overhead while maintaining balanced computational loads across processors. The recursive bisection method is preferred for smaller partition counts as it produces more geometrically compact subdomains, while k-way partitioning is more effective for larger numbers of partitions.
In the context of MPI-based parallelization, the GMRES algorithm can be naturally extended to parallel computing. However, the LUSGS preconditioning method cannot achieve parallel consistency. Therefore, off-diagonal elements in the parallel buffer are ignored, and the LUSGS method is applied independently within each processor core, requiring only one communication per iteration. Although this approach does not maintain serial-parallel consistency, experimental results indicate that the method's convergence remains unaffected. Additionally, to enhance the efficiency of LUSGS, Reverse Cuthill-McKee (RCM) reordering is employed.

\section{Turbulence Models}
\subsection{Shear-stress transport (SST) turbulence model}
The Shear-Stress Transport (SST) turbulence model, developed by Menter and first presented in \cite{menter1994two}, marks a notable improvement in turbulence modeling. This model refines the original Menter model by modifying the formulation of turbulent eddy viscosity and adjusting one constant ($\sigma_{k1}$). The governing equations, expressed in conservation form, can be represented as a coupled system:
\begin{equation*}
  \begin{aligned}
\frac{\partial (\rho k)}{\partial t} + \frac{\partial (\rho u_j k)}{\partial x_j}
 &= \tau_{ij} \frac{\partial u_i}{\partial x_j} - \beta^* \rho \omega k  + \frac{\partial}{\partial x_j}
  \left[\left(\mu + \sigma_k \mu_t \right)\frac{\partial k}{\partial x_j}\right] \\
  \frac{\partial (\rho \omega)}{\partial t} + \frac{\partial (\rho u_j \omega)}{\partial x_j}
 &= \frac{\gamma}{\nu_t} \tau_{ij} \frac{\partial u_i}{\partial x_j} -
  \beta \rho \omega^2 + \frac{\partial}{\partial x_j}
  \left[ \left( \mu + \sigma_{\omega} \mu_t \right)
  \frac{\partial \omega}{\partial x_j} \right] +
  2(1-F_1) \frac{\rho \sigma_{\omega 2}}{\omega} \frac{\partial k}{\partial x_j}
  \frac{\partial \omega}{\partial x_j}.
  \end{aligned}
\end{equation*}
The viscous stress tensor $\tau_{ij}$ is given by:
\begin{equation*}
\tau_{ij} = \mu_t \left(2S_{ij} - \frac{2}{3} \frac{\partial u_k}{\partial x_k} \delta_{ij} \right)
 - \frac{2}{3} \rho k \delta_{ij}, \quad S_{ij} = \frac{1}{2} \left( \frac{\partial u_i}{\partial x_j} + \frac{\partial u_j}{\partial x_i} \right).
\end{equation*}
The turbulent eddy viscosity is calculated as:
\begin{equation*}
\mu_t = \frac{\rho a_1 k}{\max(a_1 \omega, \Omega F_2)}.
\end{equation*}
Constants are blended from inner (1) and outer (2) regions using:
\begin{equation*}
\phi = F_1 \phi_1 + (1-F_1) \phi_2,
\end{equation*}
where $\phi_1$ and $\phi_2$ are the constants for the inner and outer regions, respectively. Additional functions are defined as follows:
\begin{equation*}
  \begin{aligned}
F_1 &= \tanh(\text{arg}_1^4), \\
\text{arg}_1 &= \min\left[\max\left(\frac{\sqrt{k}}{\beta^*\omega d},
   \frac{500 \nu}{d^2 \omega}\right), \frac{4 \rho \sigma_{\omega 2} k}{\text{CD}_{k \omega} d^2}\right], \\
\text{CD}_{k \omega} &= \max\left(2 \rho \sigma_{\omega 2} \frac{1}{\omega}
\frac{\partial k}{\partial x_j} \frac{\partial \omega}{\partial x_j}, 10^{-20}\right), \\
F_2 &= \tanh(\text{arg}_2^2), \\
\text{arg}_2 &= \max\left(2\frac{\sqrt{k}}{\beta^* \omega d}, \frac{500 \nu}{d^2 \omega}\right),
\end{aligned}
\end{equation*}
where $\rho$ represents the density, $\nu_t = \mu_t/\rho$ is the turbulent kinematic viscosity, $\mu$ is the molecular dynamic viscosity, $d$ is the distance from the field point to the nearest wall, and $\Omega = \sqrt{2 W_{ij} W_{ij}}$ denotes the vorticity magnitude, defined as
\begin{equation*}
W_{ij} = \frac{1}{2}\left(\frac{\partial u_i}{\partial x_j} -
   \frac{\partial u_j}{\partial x_i}\right).
\end{equation*}

By \cite{menter1993zonal}, a production limiter is introduced by modifying the production term in the $k$-equation; thus, the modified SST model is given by
\begin{equation*}
 \begin{aligned}
\frac{\partial (\rho k)}{\partial t} + \frac{\partial (\rho u_j k)}{\partial x_j}
 &= \underbrace{  \tau_{ij} \frac{\partial u_i}{\partial x_j}, 10\beta^* \rho \omega k)}_{\tilde{P}_k} - \underbrace{\beta^* \rho \omega k}_{D_k} + \frac{\partial}{\partial x_j}
 \left[\left(\mu + \sigma_k \mu_t \right)\frac{\partial k}{\partial x_j}\right], \\
 \frac{\partial (\rho \omega)}{\partial t} + \frac{\partial (\rho u_j \omega)}{\partial x_j}
 &= \underbrace{\frac{\gamma}{\nu_t} \tilde{P}_k}_{\tilde{P}_{\omega}} -
 \underbrace{\beta \rho \omega^2}_{D_{\omega}} + \frac{\partial}{\partial x_j}
 \left[ \left( \mu + \sigma_{\omega} \mu_t \right)
 \frac{\partial \omega}{\partial x_j} \right] +
 \underbrace{2(1-F_1) \frac{\rho \sigma_{\omega 2}}{\omega} \frac{\partial k}{\partial x_j}
 \frac{\partial \omega}{\partial x_j}}_{P_{cross}},
 \end{aligned}
\end{equation*}
where $\tilde{P}k$ and $D_k$ denote the production and destruction terms in the $k$-equation, and $\tilde{P}\omega$, $D_\omega$, and $P_{\text{cross}}$ denote the production, destruction, and cross production terms in the $\omega$ equation.

The model also requires specific boundary conditions. For farfield or inlet conditions, the following constraints apply:
\begin{equation*}
\frac{U_{\infty}}{L} < \omega < 10\frac{U_{\infty}}{L},\quad \frac{10^{-5}U_{\infty}^2}{Re_L} < k < \frac{0.1U_{\infty}^2}{Re_L}.
\end{equation*}
Wall boundary conditions are defined as:
\begin{equation*}
k_{\text{wall}} = 0,\quad \omega_{\text{wall}} = 10\frac{6\nu}{\beta_1(\Delta d_1)^2}.
\end{equation*}

The complete set of model constants is as follows:
\begin{equation*}
 \begin{aligned}
\gamma_1 &= \frac{\beta_1}{\beta^*} - \frac{\sigma_{\omega 1} \kappa^2}{\sqrt{\beta^*}}, \\
\gamma_2 &= \frac{\beta_2}{\beta^*} - \frac{\sigma_{\omega 2} \kappa^2}{\sqrt{\beta^*}}, \\
\sigma_{k 1} &= 0.85, \quad
\sigma_{\omega 1} = 0.5, \quad
\beta_1 = 0.075, \\
\sigma_{k 2} &= 1.0, \quad
\sigma_{\omega 2} = 0.856, \quad
\beta_2 = 0.0828, \\
\beta^* &= 0.09, \quad
\kappa = 0.41, \quad
a_1 = 0.31.
\end{aligned}
\end{equation*}

After determining the turbulent viscosity $\mu_t$, the gas-kinetic scheme is integrated with the SST model by adjusting the collision time $\tau$ as follows:
\begin{equation*}
 \tau = \frac{\mu}{p}+\frac{\mu_t}{p},
\end{equation*}
where $p$ denotes the pressure. The heat flux correction term is expressed as:
\begin{equation*}
\left[\left(\frac{1}{0.72}-1\right)+\left(\frac{1}{0.9}-1\right)\frac{\mu_t}{\mu}\right]q,
\end{equation*}
with $q$ representing the heat flux, where $0.72$ and $0.9$ are the Prandtl numbers for air and turbulence, respectively.

\subsection{Discrete form of the turbulence model}
The advection, diffusion, and source terms of the turbulence model are treated separately in this work. A simple first-order upwind scheme is utilized for inviscid flux, with the velocity at the cell interface obtained from the kinetic equilibrium state at that interface. For viscous flux, a second-order central scheme is applied.

The $k$-equation and $\omega$-equation are addressed separately in the two-equation SST turbulence model. The Jacobian of the inviscid term and viscous term can be derived directly by ignoring the derivative of the conservative variables. For the source term, implicit treatment is applied to the negative part to ensure the positive definiteness of the turbulence variables, while explicit treatment is used for the positive part. The Jacobian of the source term is approximated by
\begin{equation*}
\begin{aligned}
 \frac{\partial}{\partial \rho k}(\tilde{P}_k-D_k)\approx -\frac{2D_k}{\rho k}+\frac{\min(0,\tilde{P}_k)}{\rho k}, \\
 \frac{\partial}{\partial \rho \omega}(\tilde{P}_{\omega}-D_{\omega}+P_{\text{cross}})\approx -\frac{2D_{\omega}+|P_{\text{cross}}|}{\rho \omega}.
\end{aligned}
\end{equation*}

\section{Test case}
The time step $\Delta t_i$ is the time step defined in each cell
\begin{equation*}
  \Delta t_i=C_{\text{CFL}}\frac{h_i}{|\mathbf{V}-\mathbf{U}|_i+a_i+2(\nu_i+\nu_t)/h_i},
\end{equation*}
where $C_{\text{CFL}}$ is the CFL number, and $|\mathbf{V}-\mathbf{U}|_i$, $a_i$, and $\nu_i=(\mu/\rho)_i$ represent the magnitude of related velocities, sound speed and kinematic viscosity coefficient of cell $i$, respectively.

All simulations were conducted on the SUGON computation platform using dual AMD EPYC 7285 processors, each equipped with 32 cores at a base frequency of 2.5 GHz. The system architecture leverages the Zen microarchitecture, and the computation network is supported by a 100Gb InfiniBand (IB) connection. The code is compiled using GCC version 7.3.1 with the -O3 optimization flag and is linked to Open MPI version 4.1.5.

\subsection{The turbulent flow past the Onera M6 wing}
This test case is the three-dimensional turbulent transonic flow at a Mach number of $\text{Ma}=0.8395$ over an Onera M6 wing. The Reynolds number based on the mean aerodynamic chord ($\text{L}=0.64607m$) of the wing,  is $\text{Re}_{\text{L}}=11.72\times10^6$, and the angle of attack ($\text{AoA}$) is $3.06^\circ$. This case serves as a benchmark CFD validation problem for external flows due to its complex flow features, including local supersonic regions, shock waves, and turbulent boundary layer separation, while maintaining a relatively simple geometry.

\begin{figure}[htb!]
    \centering
    \subfigure[The surface mesh\label{fig:m6mesh}]{\includegraphics[width=0.4\textwidth]{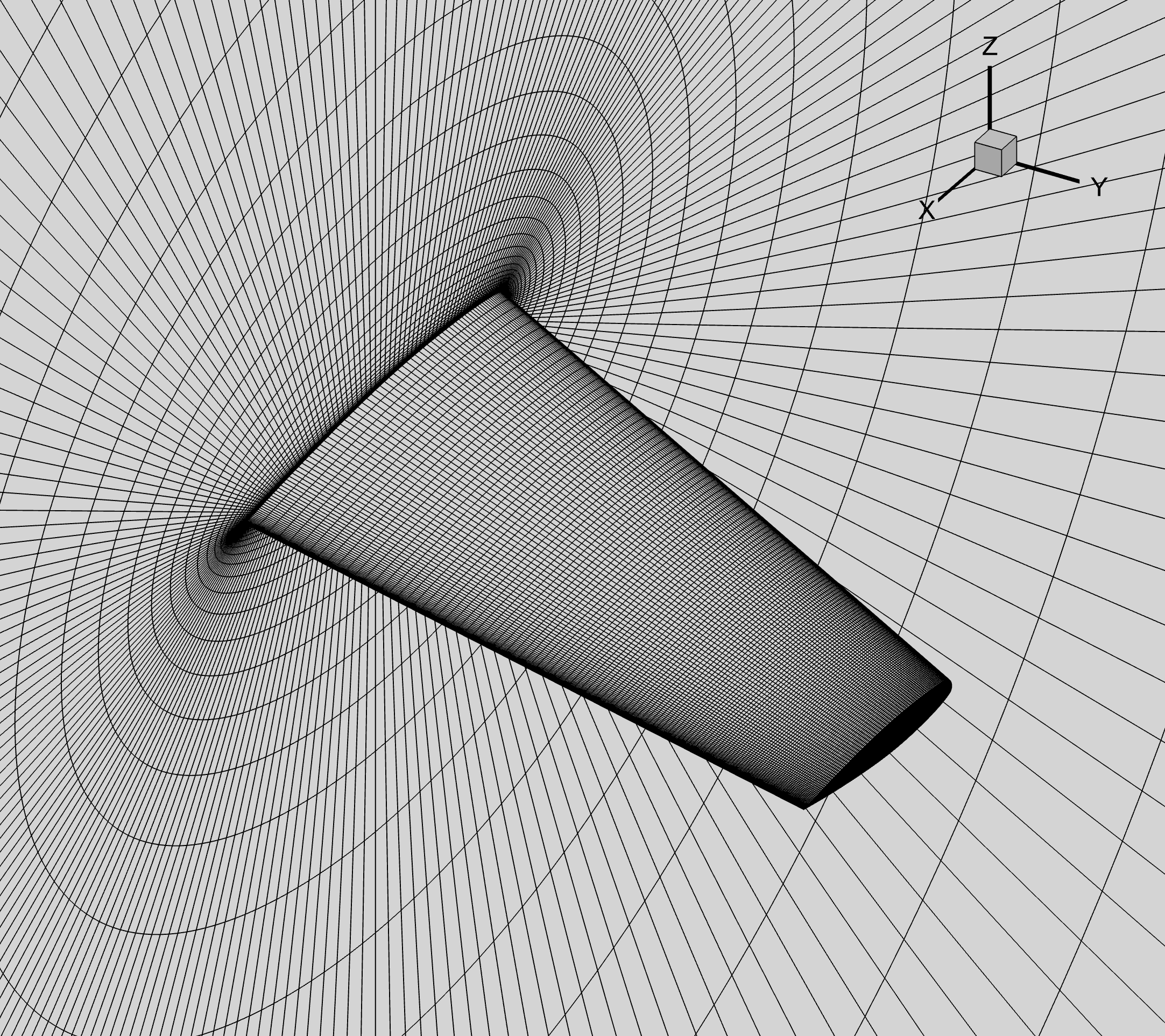}}
    \subfigure[The surface $y^+$ distribution\label{fig:m6surfaceyplus}]{\includegraphics[width=0.4\textwidth]{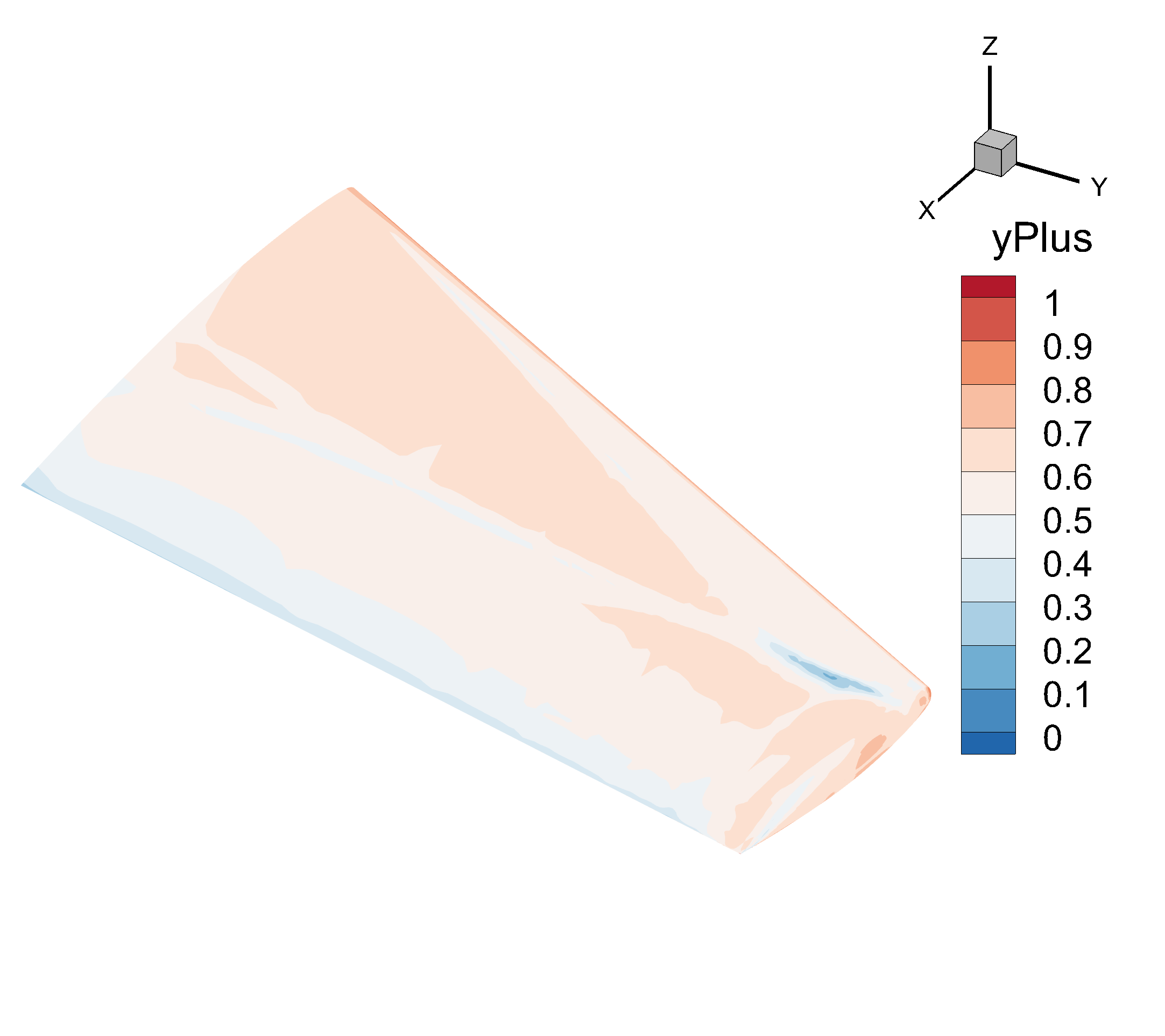}}
    \caption{The surface mesh of the M6 wing with 2,608,200 cells}
\end{figure}

\begin{figure}[htb!]
    \centering
    \subfigure[The convergence history of the M6 wing test case\label{fig:M6Res}]{\includegraphics[width=6cm]{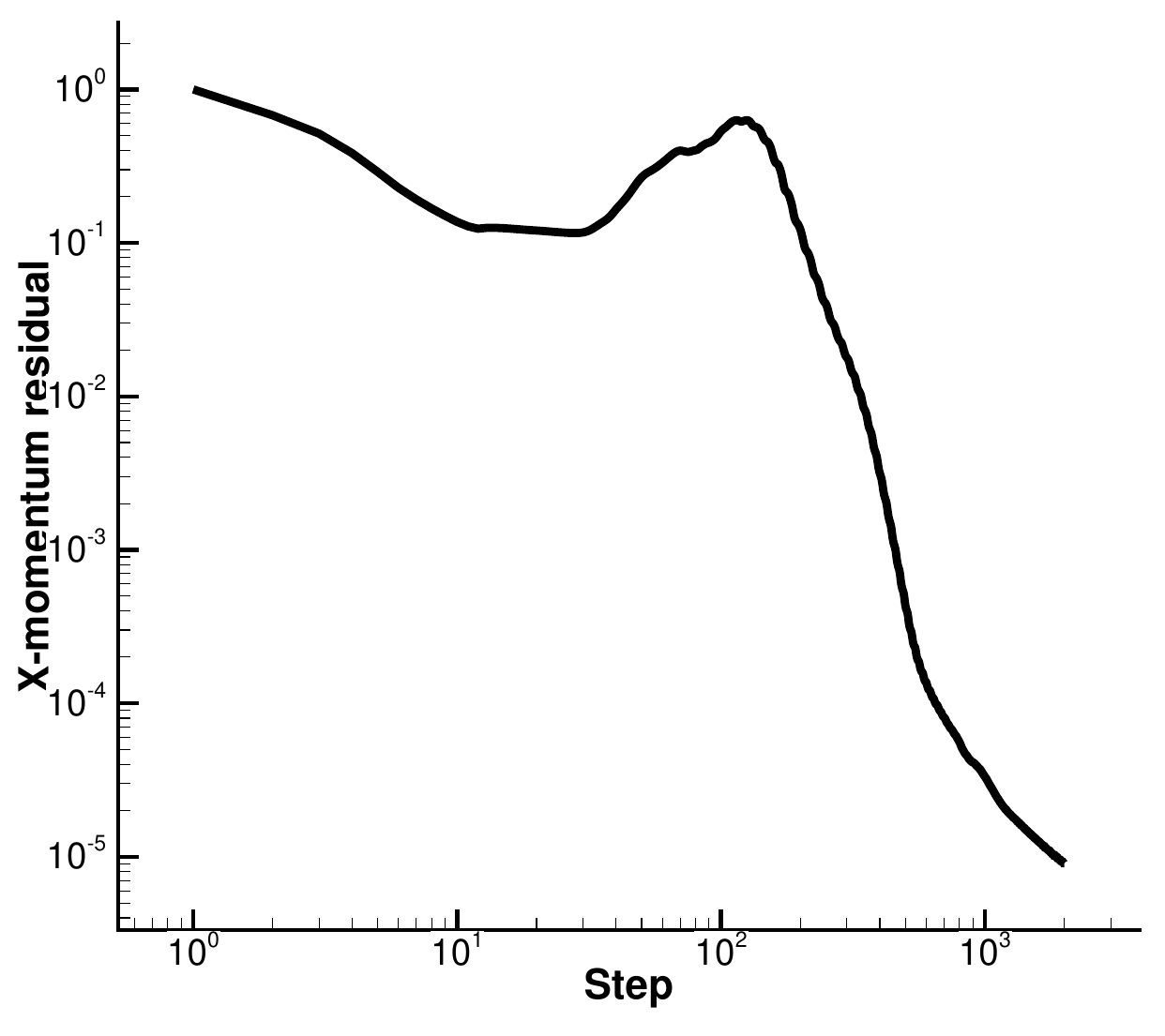}}
    \subfigure[The coefficient of pressure contour of flow Ma=0.8395, Re=11.72E+06, 3.06 degrees attack angle past the M6 wing\label{fig:M6Cp}]{\includegraphics[width=6cm]{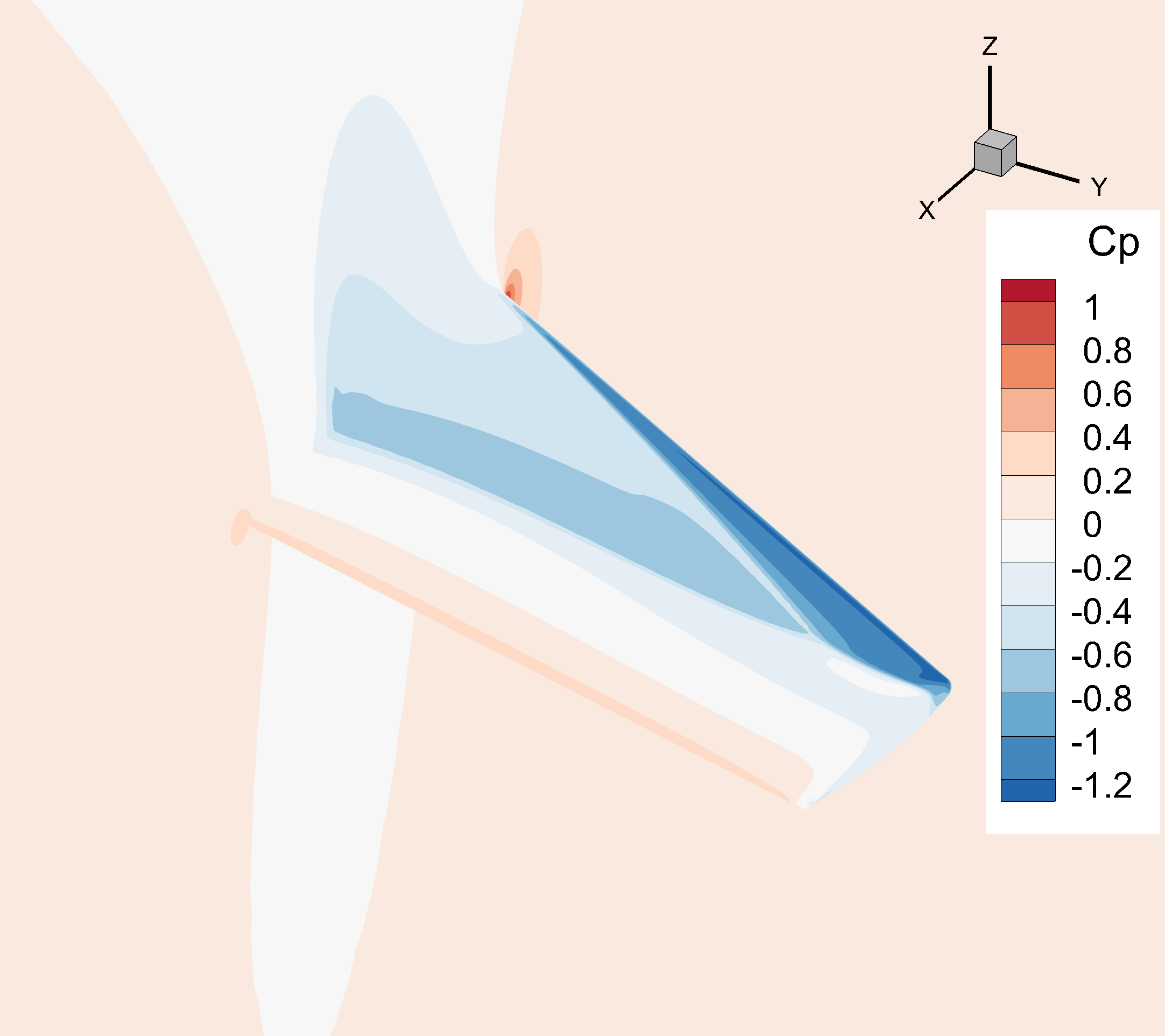}}
    \caption{Results of the M6 wing test case }
\end{figure}

\begin{figure}[!ht]
    \centering
     \subfigure[y=0.2b]{\includegraphics[width=5cm]{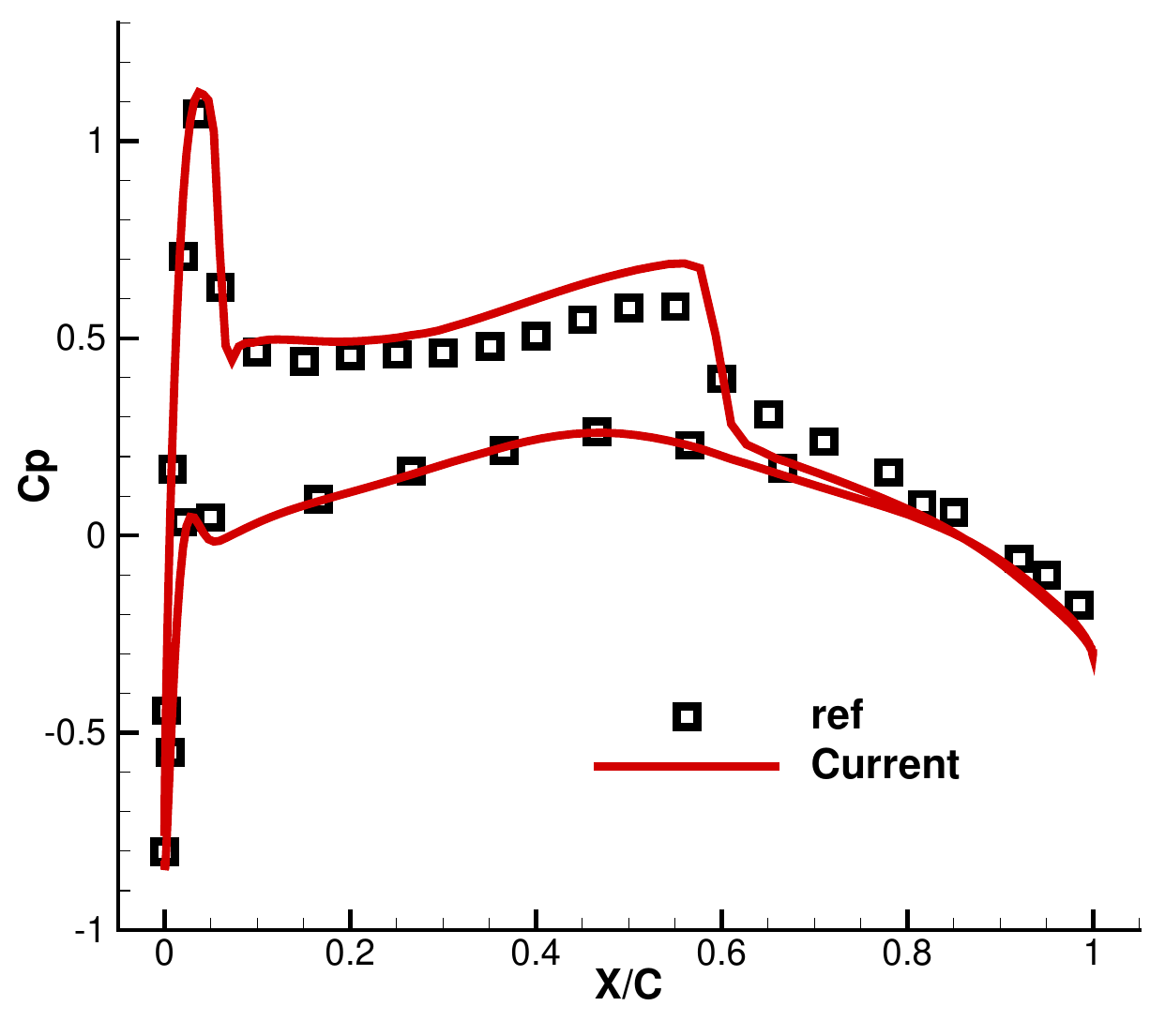}}
     \subfigure[y=0.44b]{\includegraphics[width=5cm]{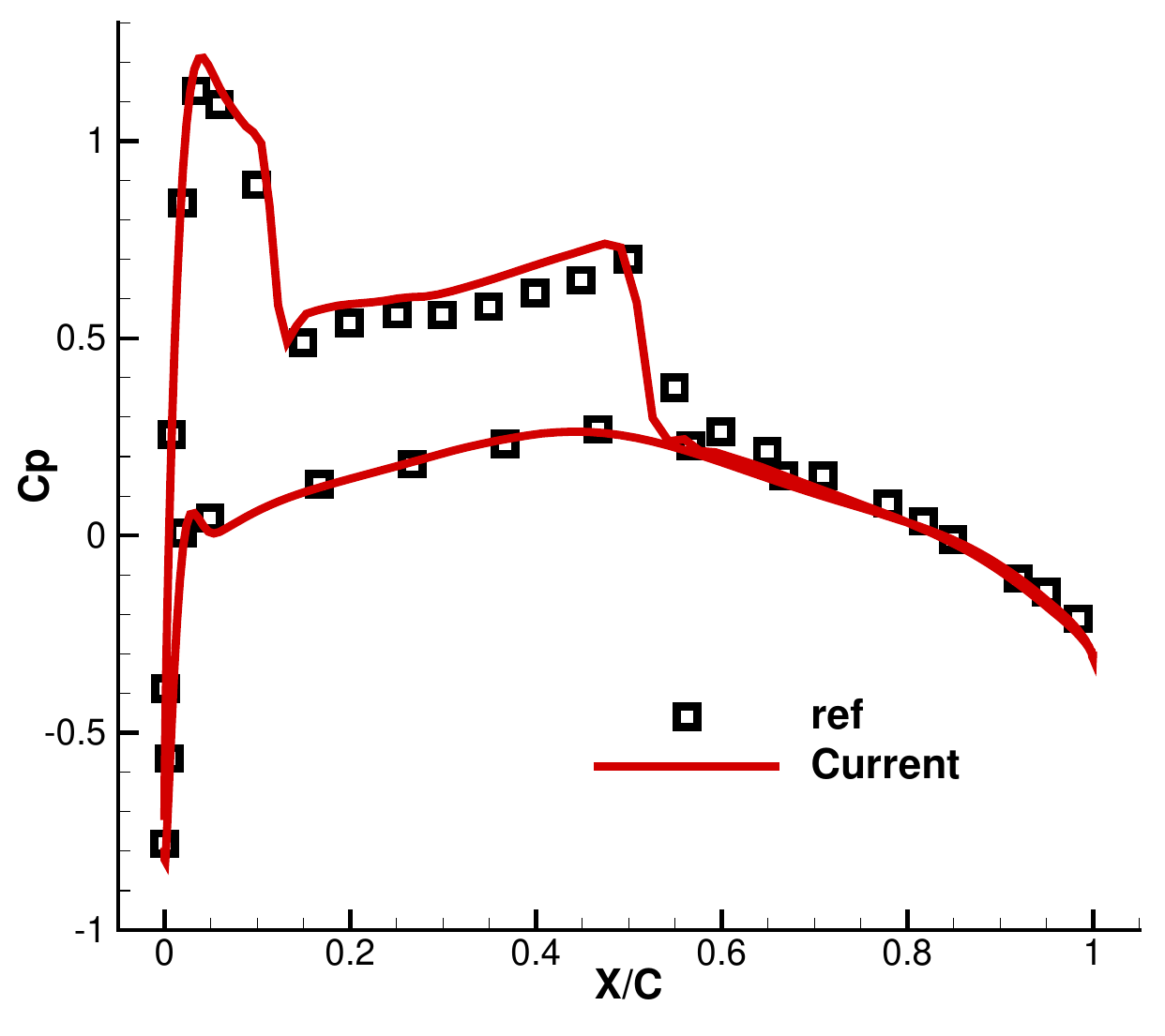}}
     \subfigure[y=0.65b]{\includegraphics[width=5cm]{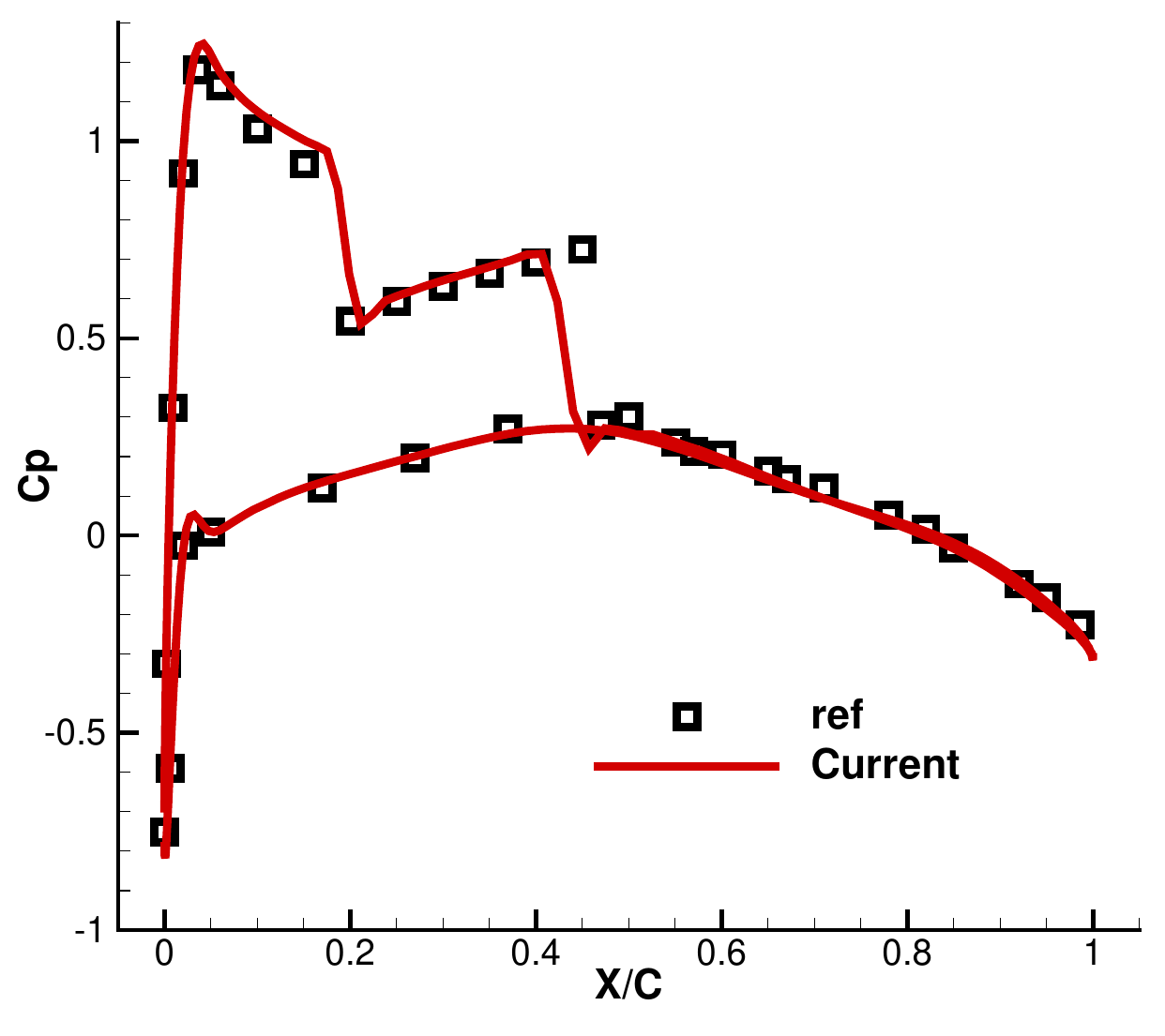}}
     \subfigure[y=0.8b]{\includegraphics[width=5cm]{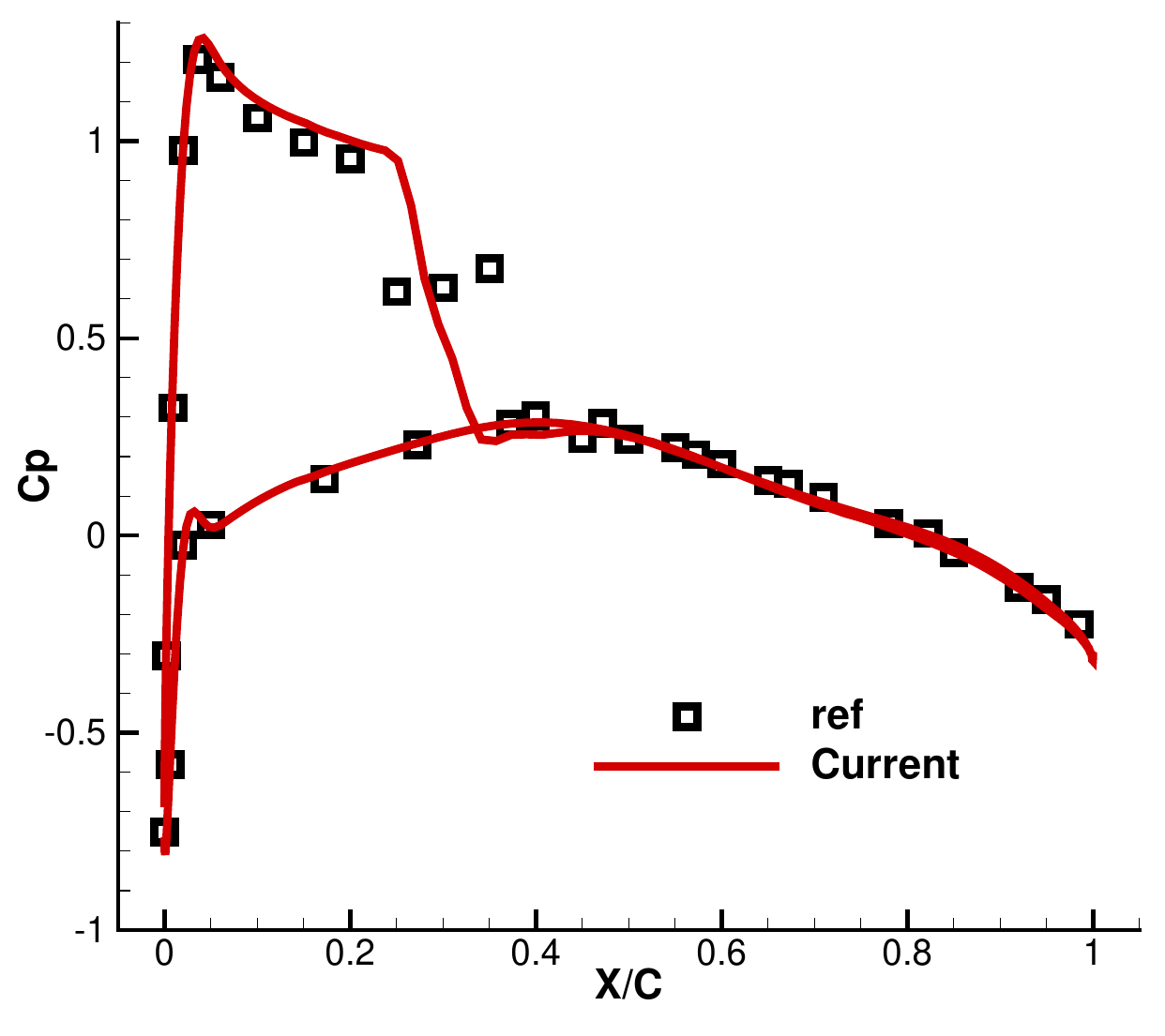}}
     \subfigure[y=0.9b]{\includegraphics[width=5cm]{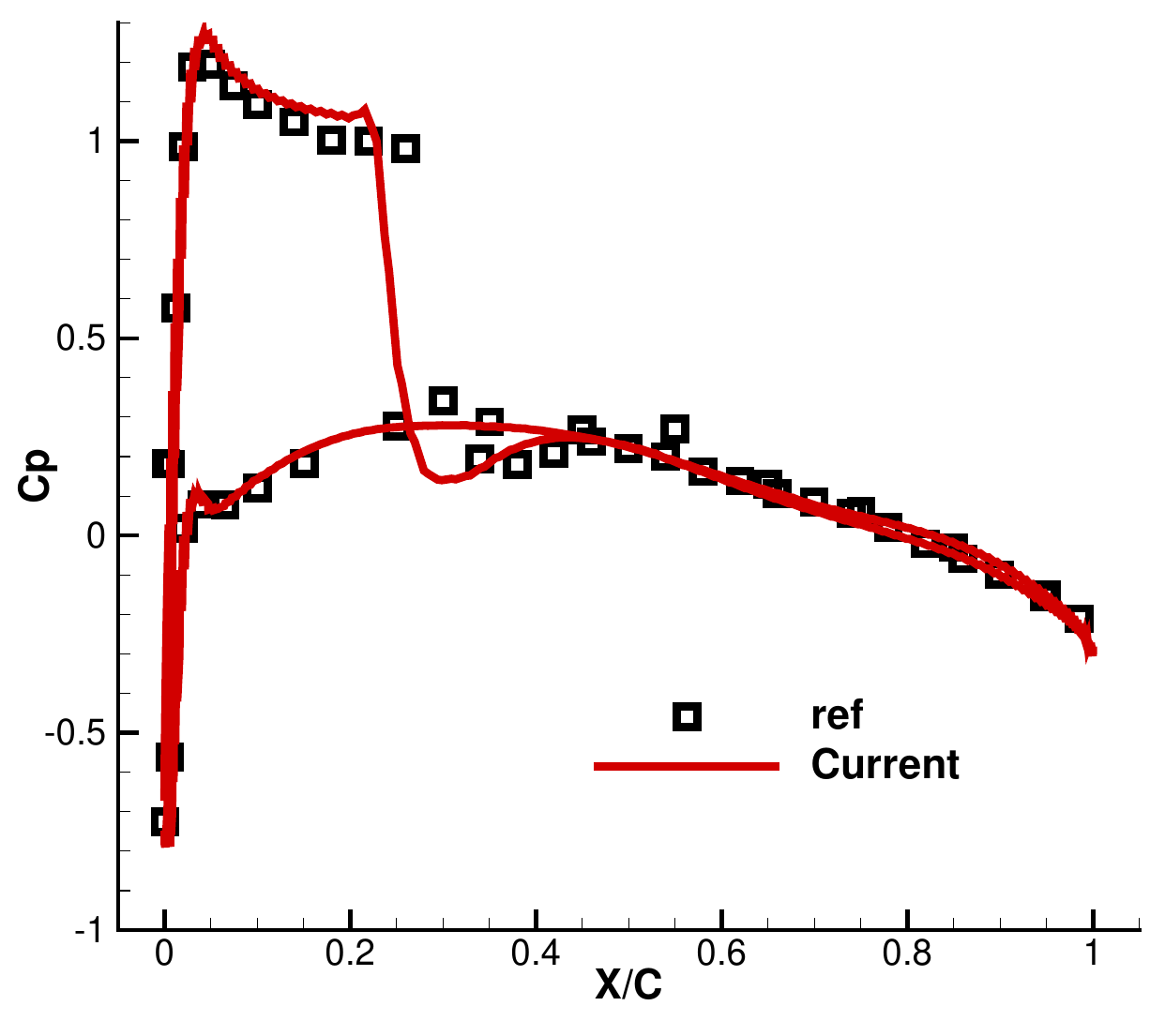}}
     \subfigure[y=0.95b]{\includegraphics[width=5cm]{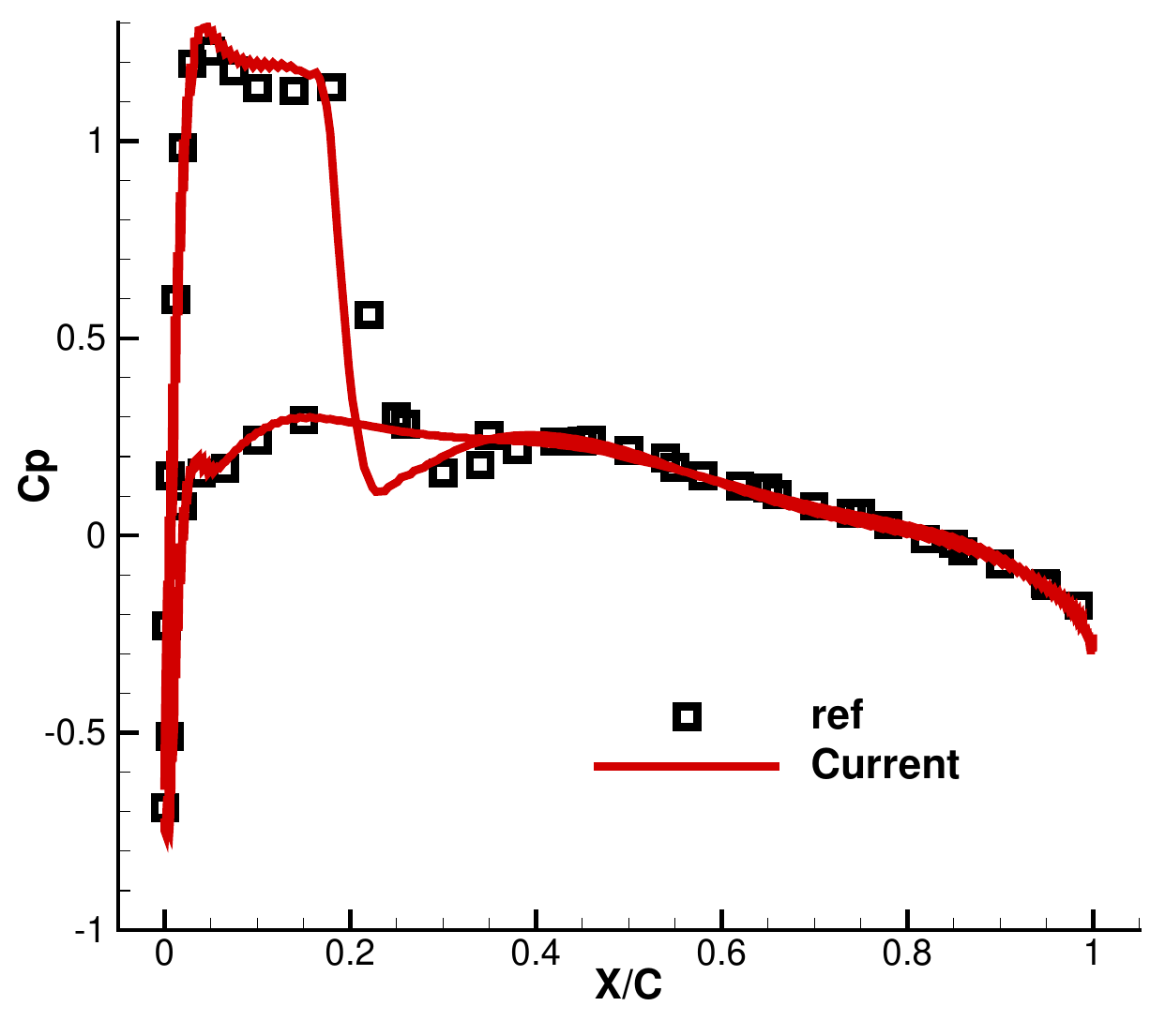}}
    \caption{The coefficient of pressure distribution of the M6 wing at different spanwise positions}
    \label{fig:M6CPdistribution}
\end{figure}

The website provides the computational mesh \cite{Slater2008_ONERAM6}. The total mesh number used in this test case is $2,608,200$. The first layer of cells is located at a distance of $2\times10^{-6}$ from the wall. The surface mesh is shown in Fig. \ref{fig:m6mesh}. The surface $y^+$ distribution defined by the distance of the cell center and face is shown in Fig. \ref{fig:m6surfaceyplus}, which shows the first layer of cells is located at the wall with a distance of $y^+$ less than 1.
The parallel computing efficiency of the GMRES solver with 100 iterations in the M6 wing test case is shown in Table \ref{M6Eff}. The results show that the GMRES solver can remain 90\% parallel efficiency up to 256 cores.
  \begin{table}[htb!]
 \centering
 \caption{Wall clock time and parallel computing efficiency of the GMRES solver with 100 iterations in the M6 wing test case}\label{M6Eff}
  \begin{tabular}{cccc}
    \toprule
 Cores & Wall time (s)& Actual speedup & Parallel efficiency \\
  \midrule
  64 & 732 & 1.00 & 100.00\%\\  
  128 & 380 & 1.93 & 96.32\%\\
  256 & 203 & 3.61 & 90.15\%\\
  512 & 132 & 5.55 & 69.32\%\\
  \bottomrule
\end{tabular}
\end{table}
The convergence history of the M6 wing test case is shown in Fig. \ref{fig:M6Res}. The CFL number is set as an exponential function increasing from 1 to 100 during the first 100 steps, and remains at 100 afterwards. The convergence history shows the residual of the M6 wing test case is reduced to 10$^{-4}$ after 600 steps, and reduces slowly afterwards. Finally, 2000 steps are run for this case. The coefficient of pressure distribution of the M6 wing is shown in Fig. \ref{fig:M6Cp}.
The coefficient of pressure distribution at different spanwise positions of the M6 wing is shown in Fig. \ref{fig:M6CPdistribution}, which shows the pressure distribution is similar to the experimental data \cite{schmitt1979pressure} and can accurately capture the position of the shock wave.

\subsection{NASA Common Research Model (CRM) Wing-Body buffet study}
\begin{figure}[htb!]
  \centering
  \subfigure[The surface mesh\label{fig:CRMMesh}]{\includegraphics[width=0.4\textwidth]{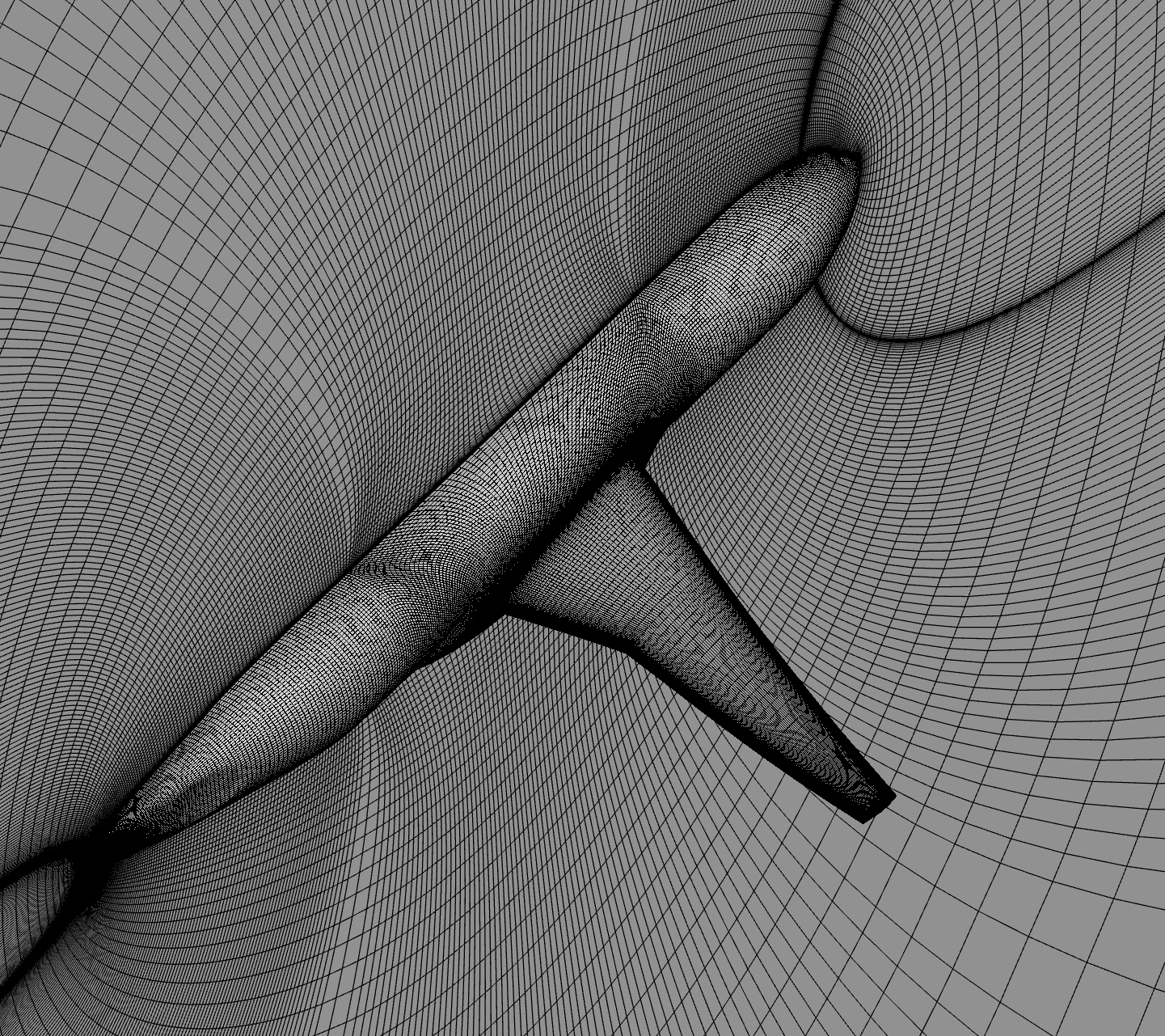}}
  \subfigure[The surface $y^+$ distribution\label{fig:CRMyplus}]{\includegraphics[width=0.4\textwidth]{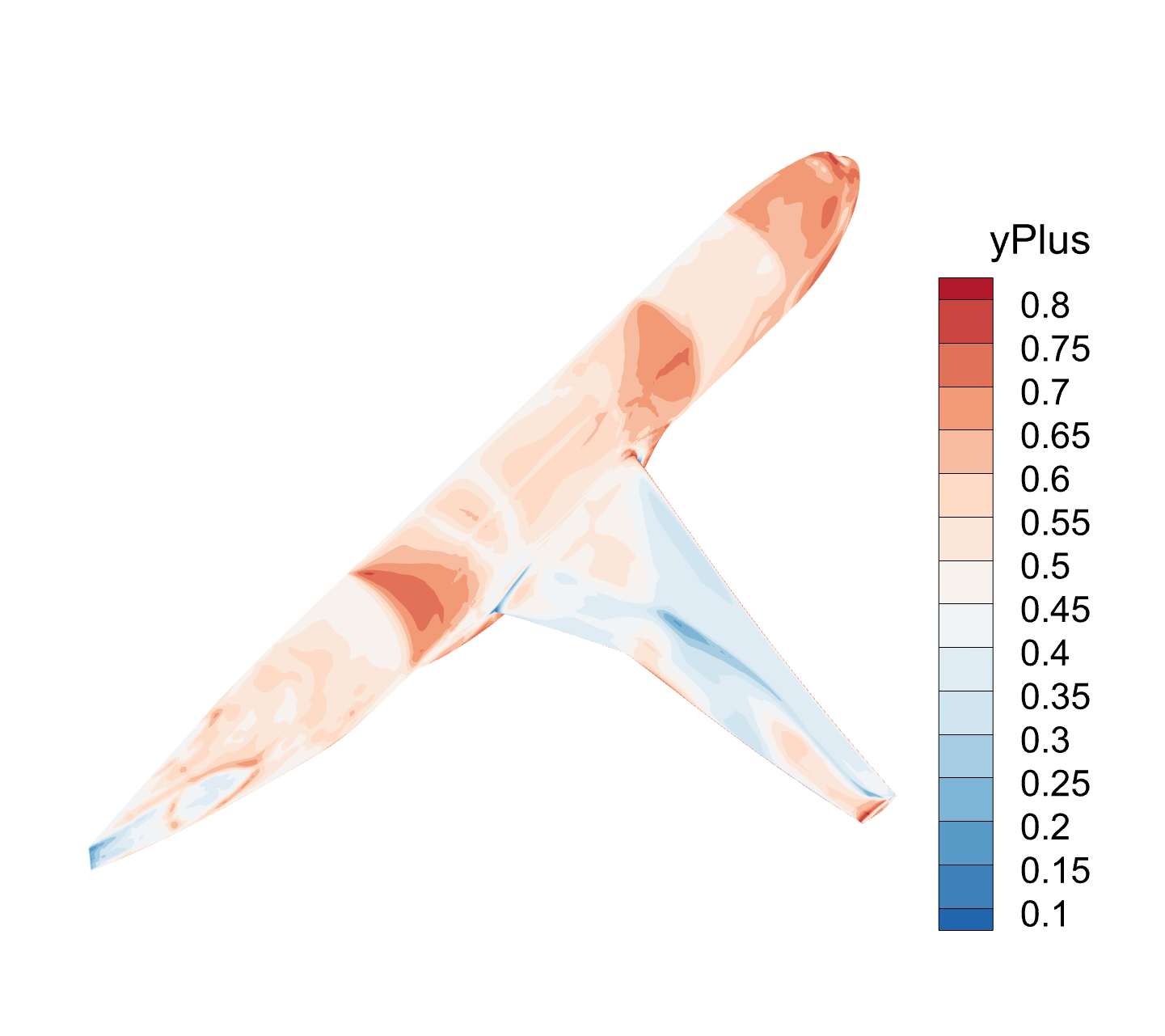}}
  \caption{The computational mesh of the CRM Wing-Body configuration}
\end{figure}

The buffet study presented in this paper corresponds to test case 2 of the 5th AIAA Drag Prediction Workshop. This case is a genuine guide to test the simulation capacity of the CFD solver in transonic flow with complex geometry. The geometry is still the Common Research Model (CRM) Wing-Body configuration. The incoming flow is a subsonic flow with a Reynolds number of 5.0E+06 and a Mach number of 0.85. Moreover, the reference temperature is 310.92K, which is set as the incoming total temperature. Seven different angles of attack are considered, which are 2.5$^\circ$, 2.75$^\circ$, 3$^\circ$, 3.25$^\circ$, 3.5$^\circ$, 3.75$^\circ$, and 4$^\circ$. The medium point-matched grid (L3, $5,111,808$ hexahedral elements) provided by the DPW Committee is used in this test, and the surface grid is shown in Fig. \ref{fig:CRMMesh}. The first layer of cells is located at the wall with a distance of $10^{-6}$ from the wall, while the $y^+$ contour is shown in Fig. \ref{fig:CRMyplus}.

The parallel computing efficiency of the GMRES solver with 100 iterations in the CRM Wing-Body Buffet Study is shown in Table \ref{CRMEff}. The results show that the GMRES solver can remain about 80\% parallel efficiency up to 512 cores.
  \begin{table}[htb!]
 \centering
 \caption{Wall clock time and parallel computing efficiency of the GMRES solver with 100 iterations in the CRM Wing-Body Buffet Study}\label{CRMEff}
  \begin{tabular}{cccc}
    \toprule
 Cores & Wall time (s)& Actual speedup & Parallel efficiency \\
  \midrule
  64 & 1507 & 1.00 & 100.00\%\\  
  128 & 784 & 1.92 & 96.11\%\\
  256 & 412 & 3.66 & 91.44\%\\
  512 & 244 & 6.18 & 77.20\%\\
  1024 & 181 & 8.33 & 52.04\%\\
  \bottomrule
\end{tabular}
\end{table}

\begin{figure}[htb!]
  \centering
  \subfigure[The lift coefficient vs. angle of attack]{\includegraphics[width=0.4\textwidth]{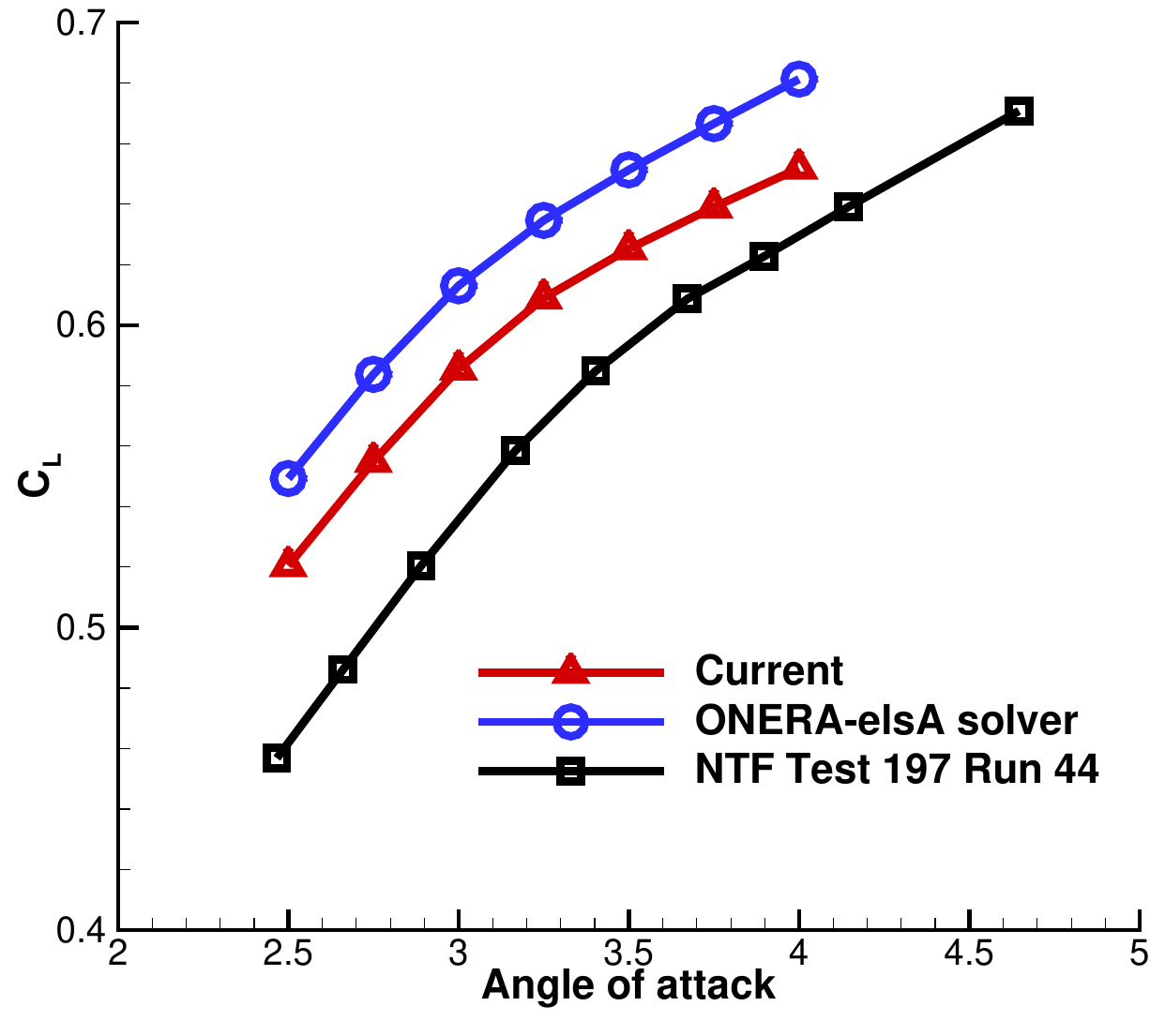}}
  \subfigure[The drag coefficient vs. lift coefficient]{\includegraphics[width=0.4\textwidth]{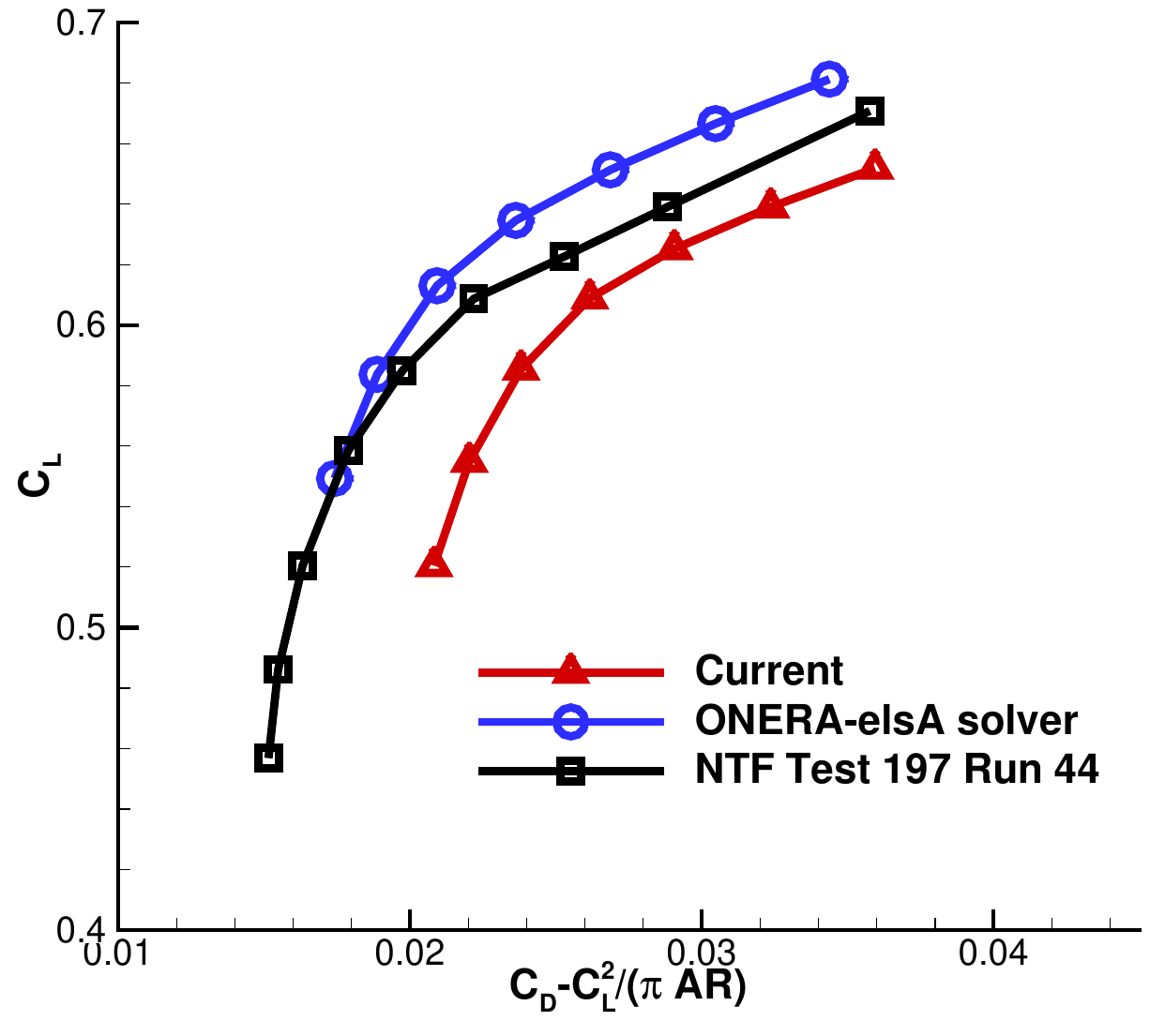}}
  \caption{The characteristic of the CRM Wing-Body Buffet Study of Reynolds number 5.0E+06 and Mach number 0.85}
  \label{fig:CRMChar}
\end{figure}

The CFL number is set as an exponential function increasing from 1 to 100 during the first 100 steps, and remains at 100 afterwards. The characteristic of the CRM Wing-Body Buffet Study is shown in Fig. \ref{fig:CRMChar}. The results are compared with the  NTF test data \cite{levy2014summary} and the ONERA-elsA solver \cite{hue2014fifth} results. The results show that the lift coefficient of our solver is closer to the NTF test data than the ONERA-elsA solver. The drag characteristics are plotted in terms of the idealized profile drag defined as $(C_D-C_L^2)/(\pi AR)$, where $AR=9$ is the aspect ratio, $C_D$ is the drag coefficient, and $C_L$ is the lift coefficient. The results show that the drag coefficient of our solver is lower than the NTF test data, and the ONERA-elsA solver is higher than the NTF test data.

\begin{figure}[htb!]
  \centering
  \subfigure[The convergence history of x-momentum residual]{\includegraphics[width=0.3\textwidth]{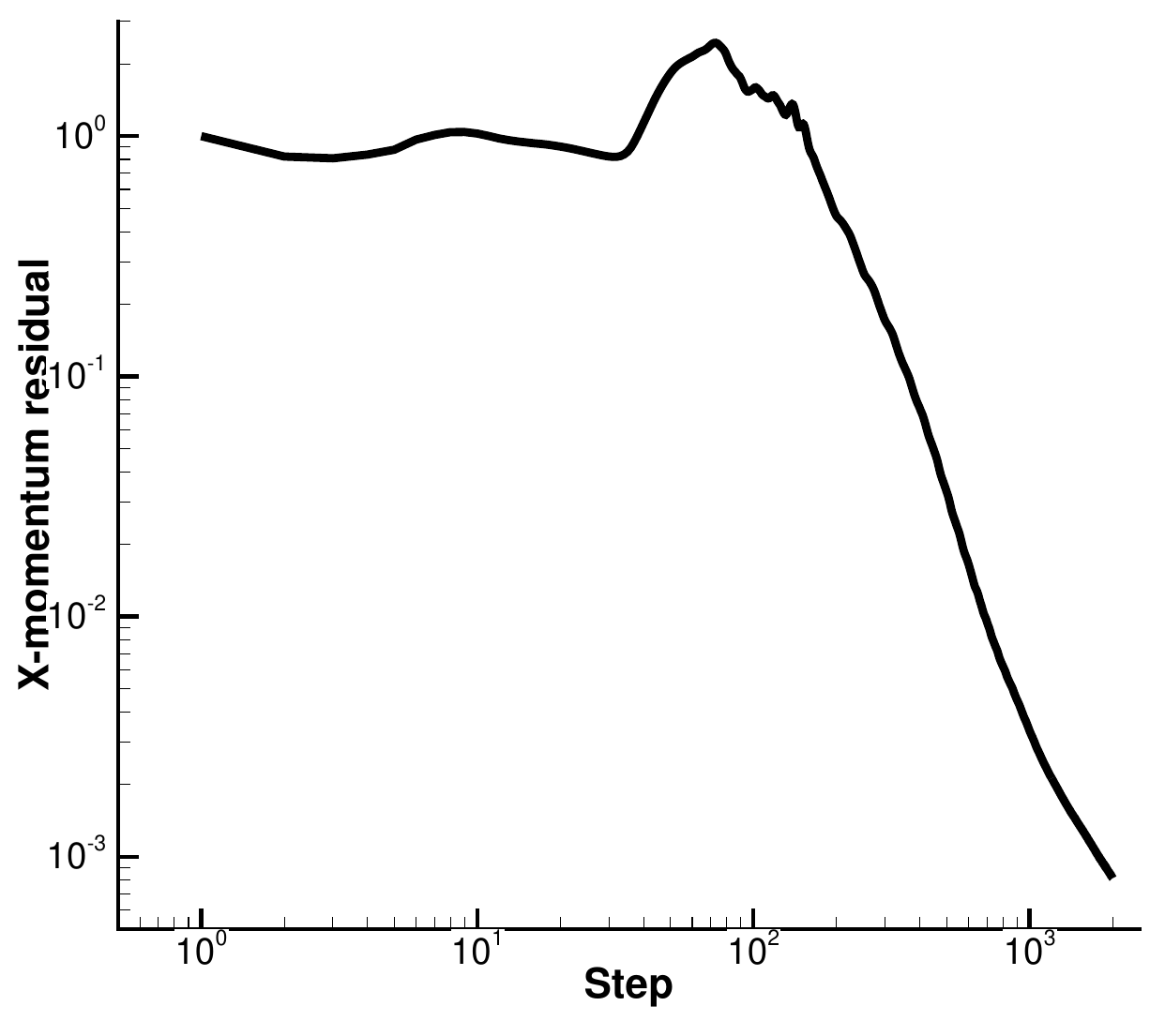}}
  \subfigure[The convergence history of lift coefficient]{\includegraphics[width=0.3\textwidth]{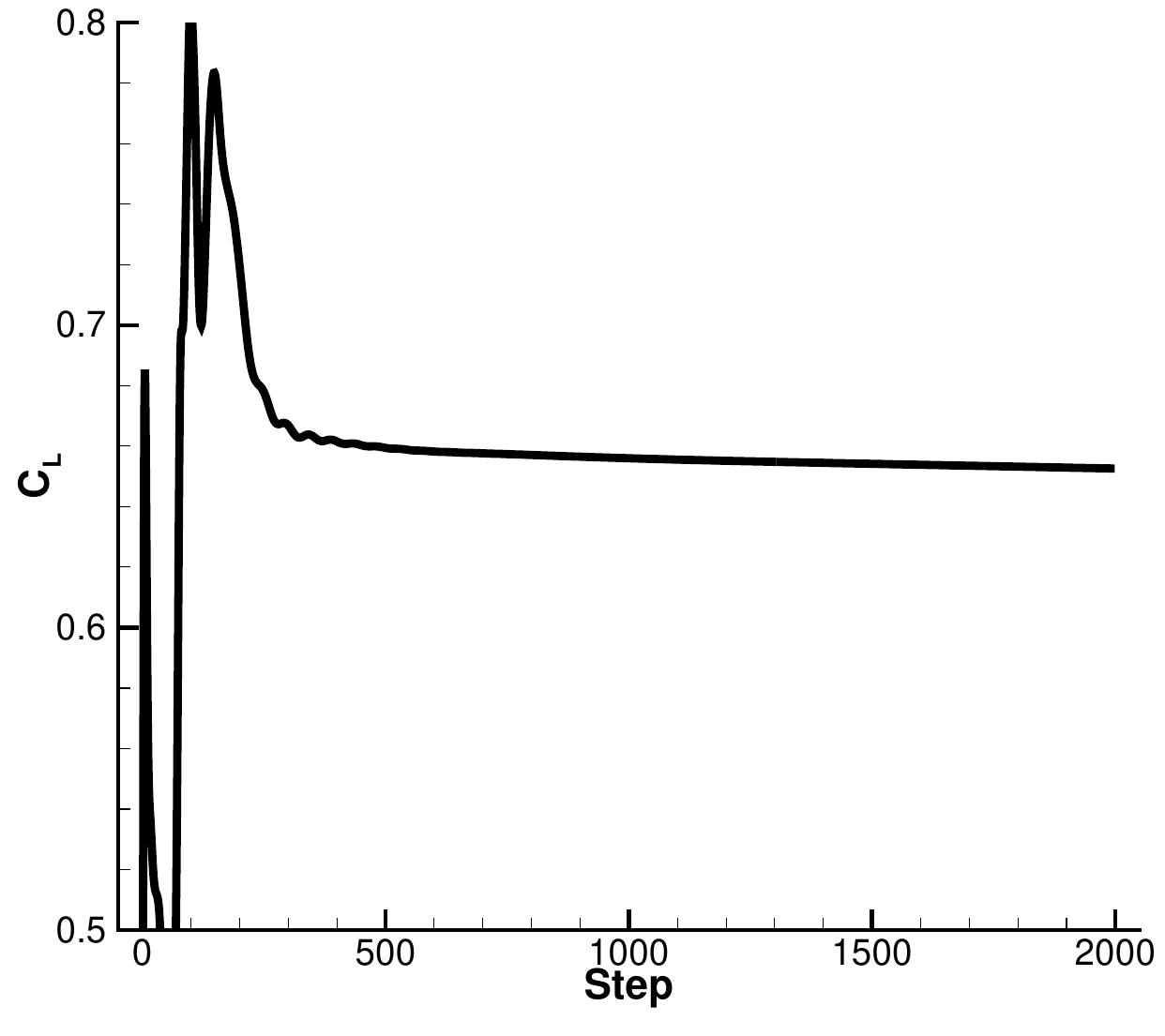}}
  \subfigure[The convergence history of drag coefficient]{\includegraphics[width=0.3\textwidth]{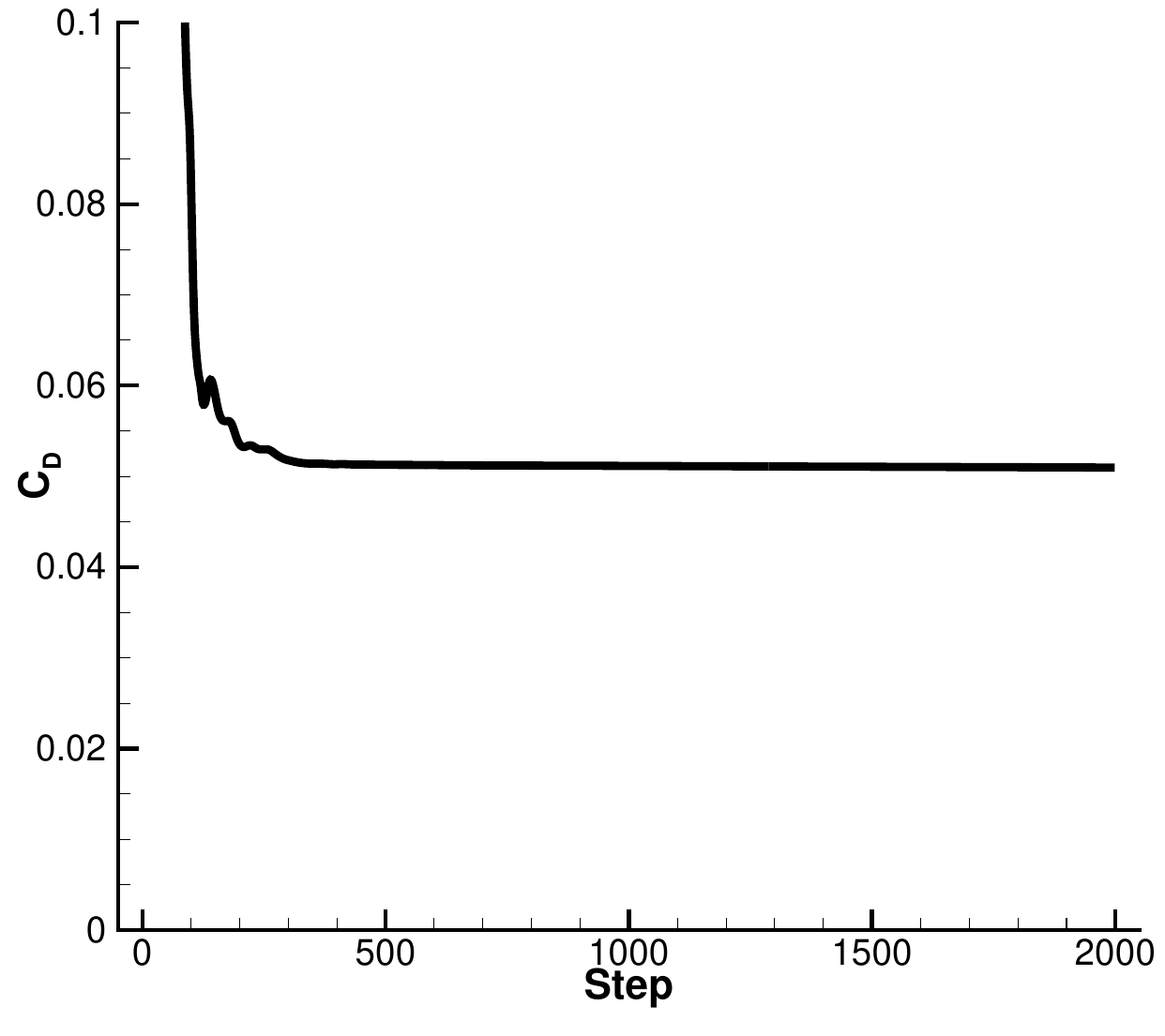}}
  \caption{The convergence history of the CRM Wing-Body Buffet Study at $\alpha=4^\circ$ angle of attack}
  \label{fig:CRMRes}
\end{figure}

The convergence history of the CRM Wing-Body Buffet Study at $\alpha=4^\circ$ angle of attack is shown in Fig. \ref{fig:CRMRes}. The results show that the lift and drag coefficients converge after 500 steps. A total of 2000 steps are run to reach the convergence.

\begin{figure}[htb!]
  \centering
  \subfigure[Angle of attack $\alpha=2.75^\circ$]{\includegraphics[width=0.3\textwidth]{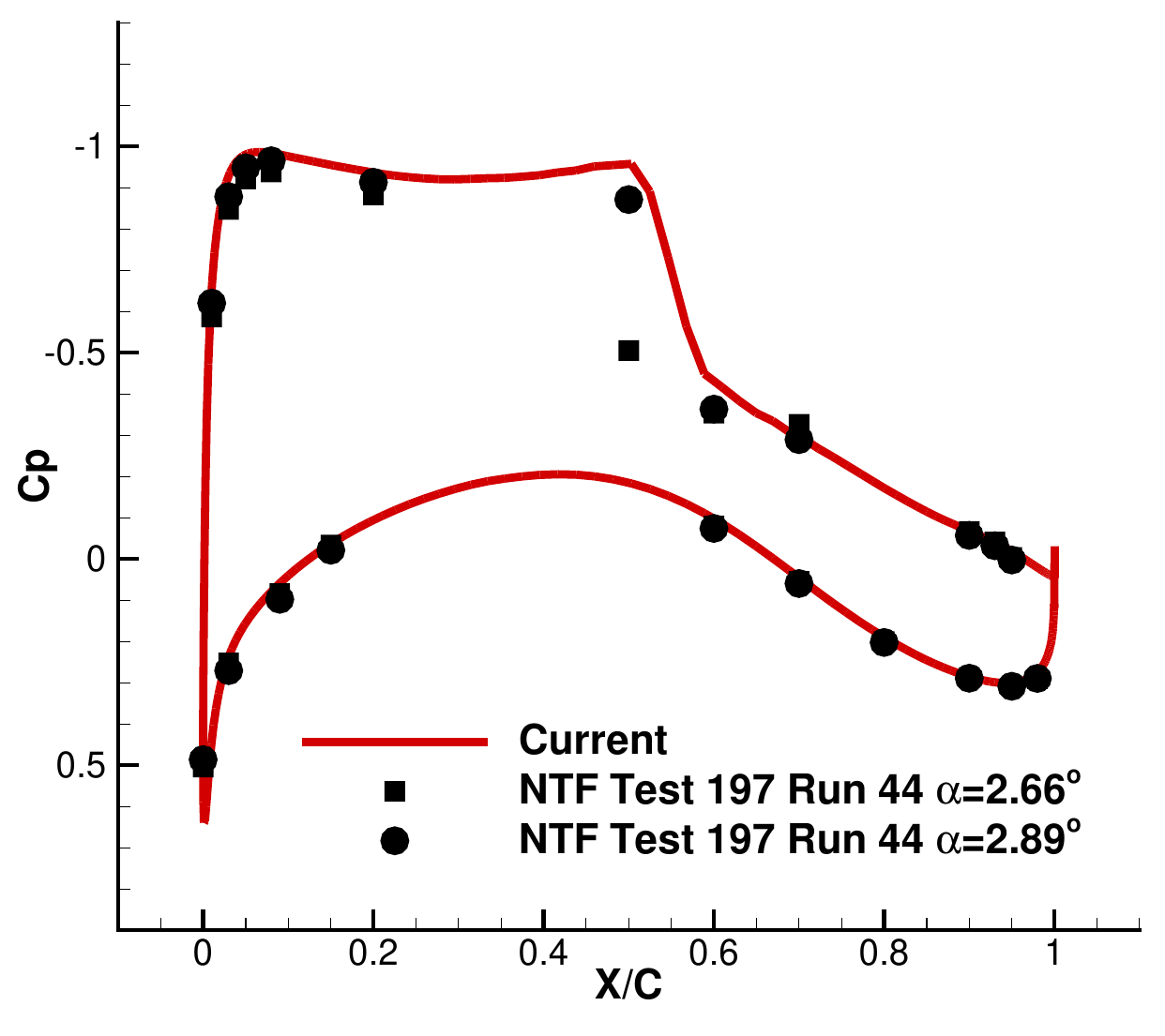}}
  \subfigure[Angle of attack $\alpha=3^\circ$]{\includegraphics[width=0.3\textwidth]{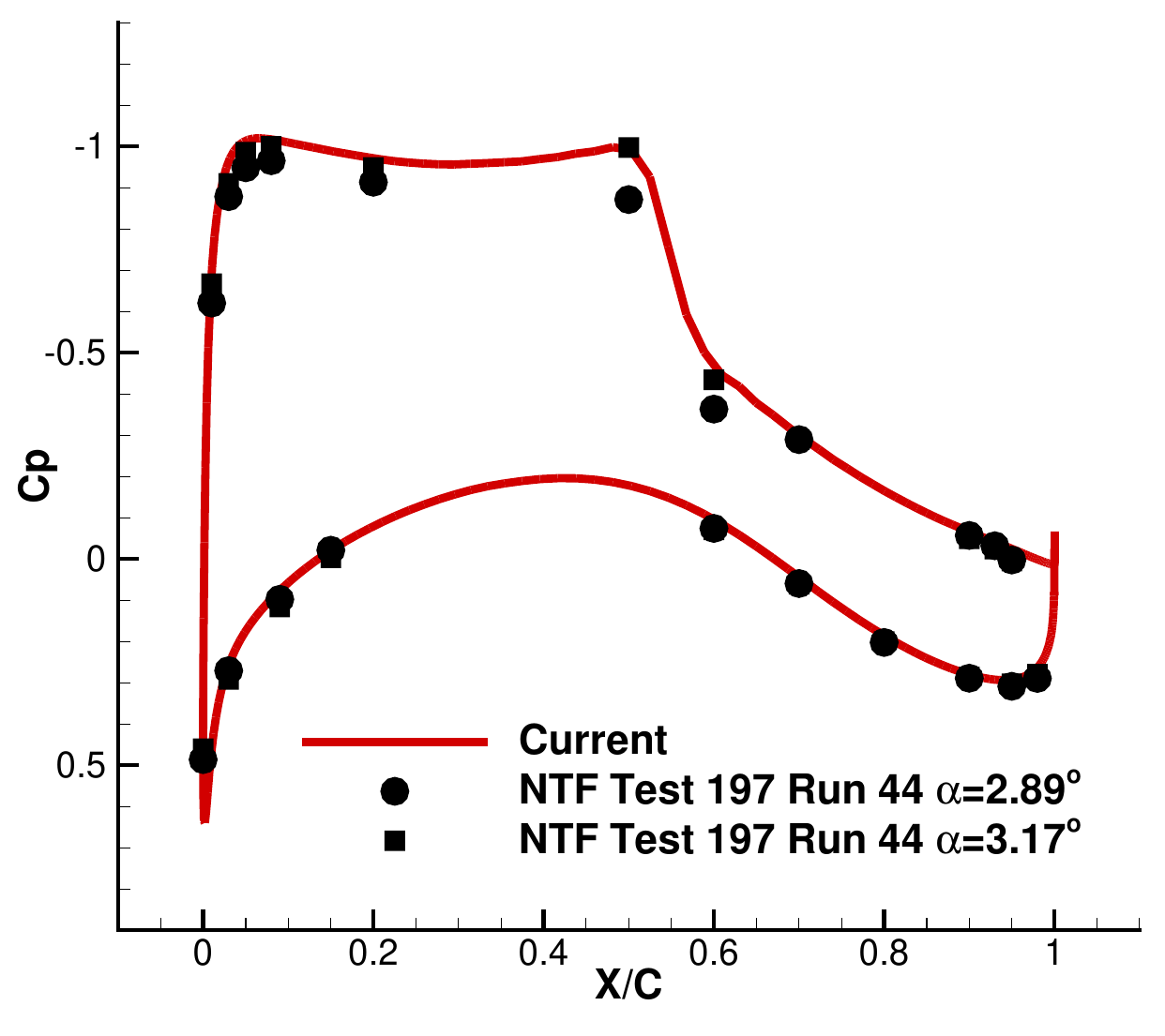}}
  \subfigure[Angle of attack $\alpha=3.25^\circ$]{\includegraphics[width=0.3\textwidth]{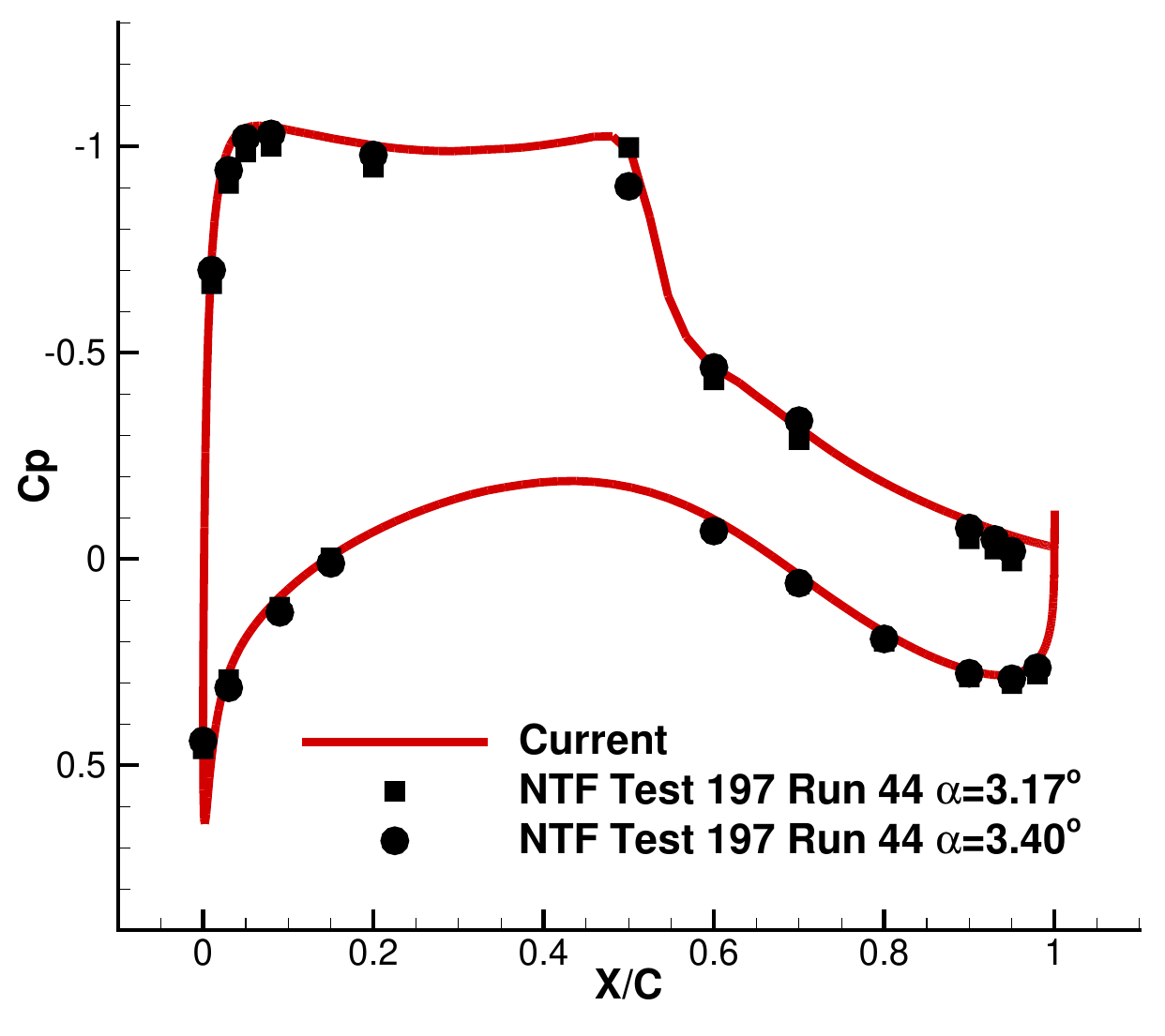}}
  \subfigure[Angle of attack $\alpha=3.5^\circ$]{\includegraphics[width=0.3\textwidth]{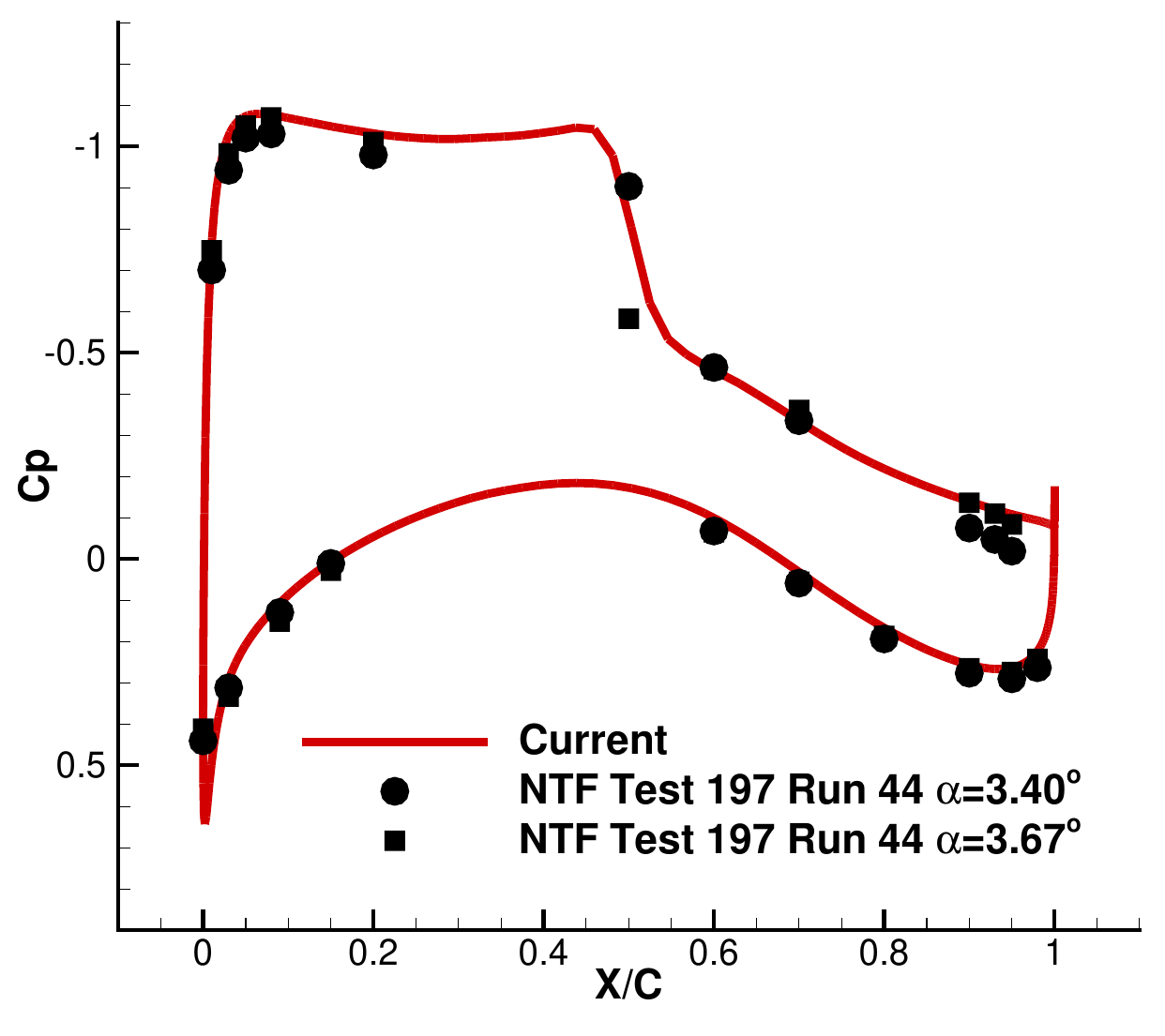}}
  \subfigure[Angle of attack $\alpha=3.75^\circ$]{\includegraphics[width=0.3\textwidth]{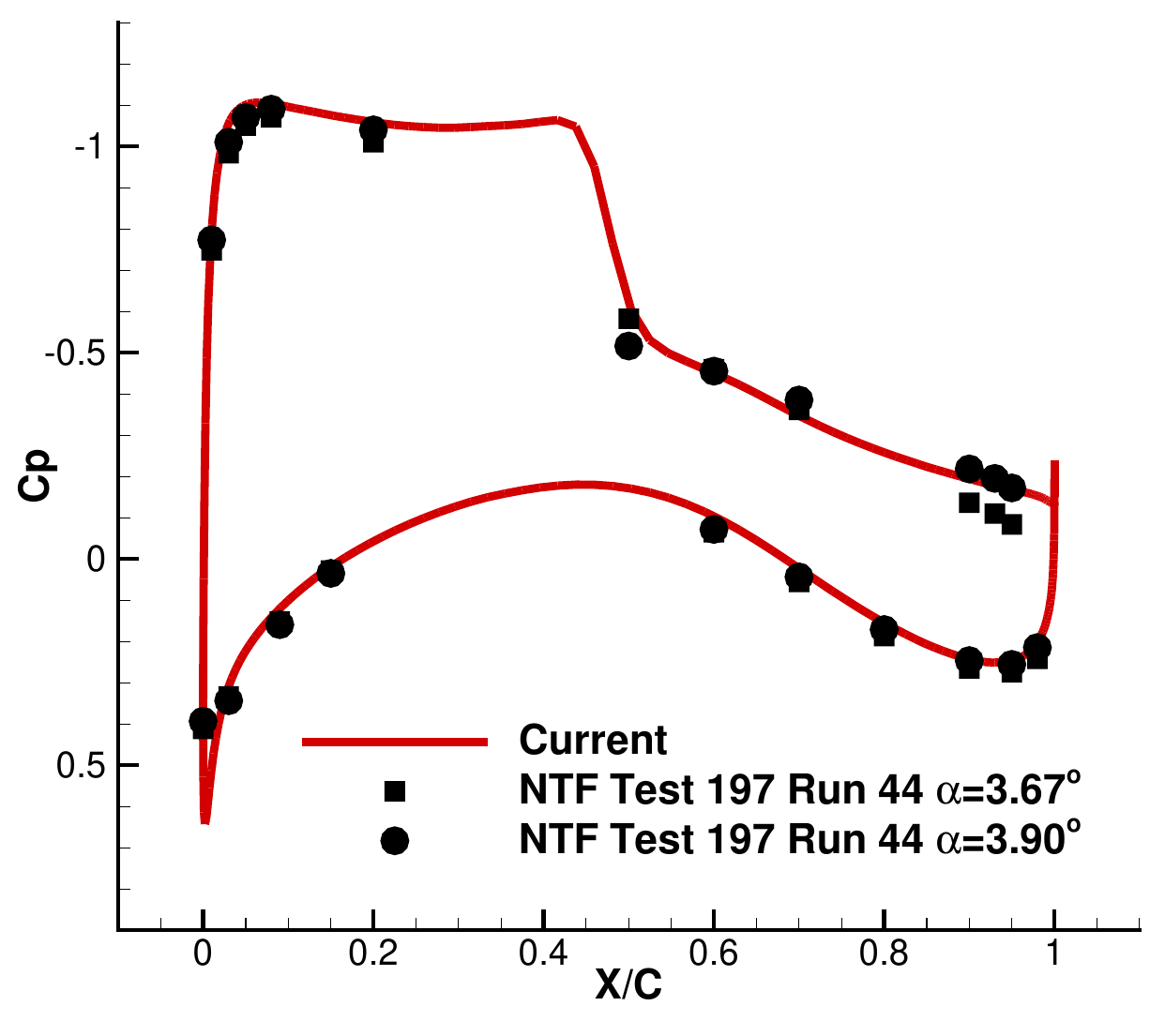}}
  \subfigure[Angle of attack $\alpha=4^\circ$]{\includegraphics[width=0.3\textwidth]{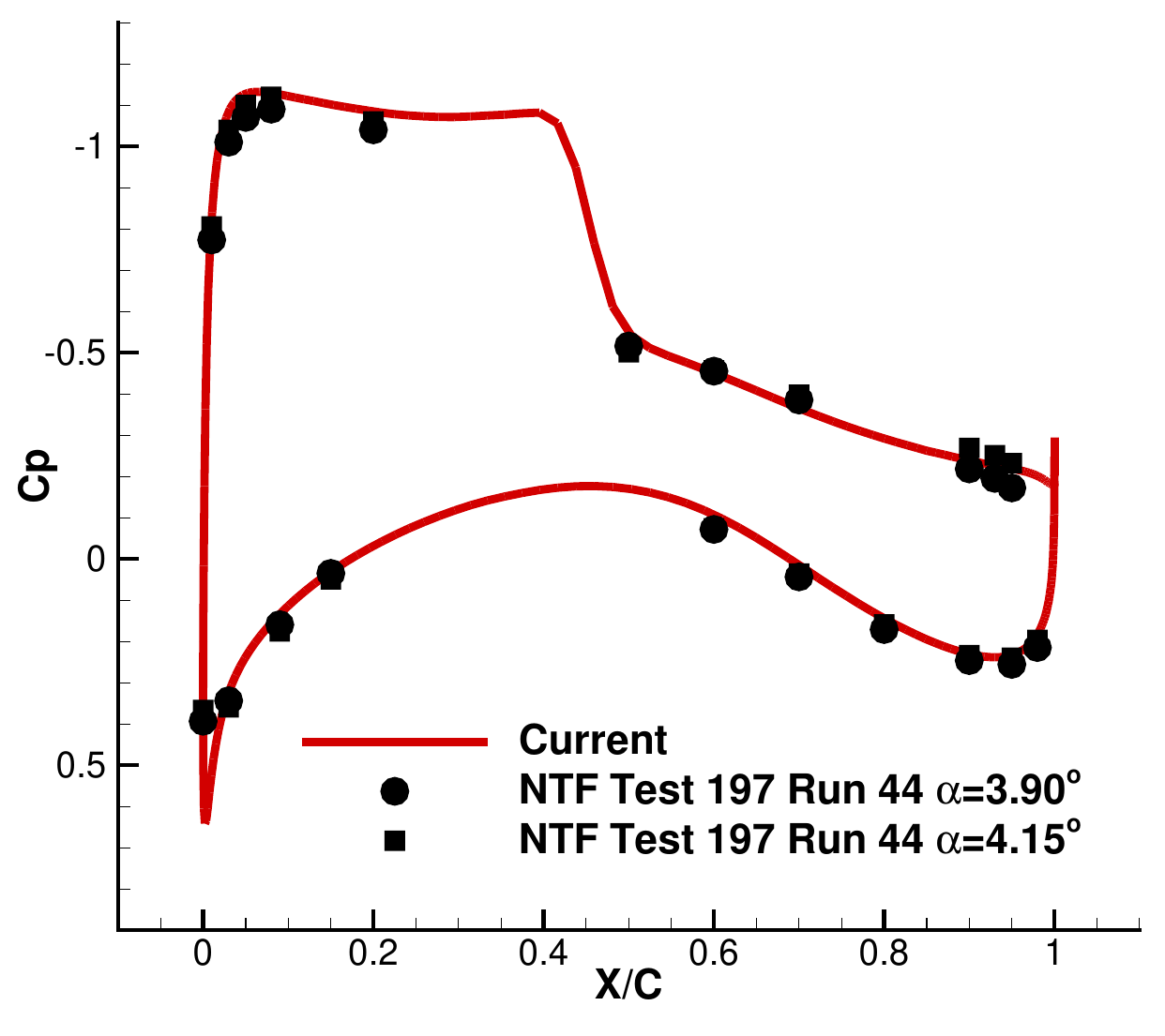}}
  \caption{The wing pressure coefficient distribution at station 10 ($\eta=0.5024$) at different angle of attack}
  \label{fig:CRMCP}
\end{figure}

The pressure coefficient distribution at station 10 ($\eta=0.5024$) at different angles of attack is shown in Fig. \ref{fig:CRMCP}. The results show that the pressure coefficient distribution is similar to the experimental data \cite{levy2014summary} and can capture the position of the shock wave.

\begin{figure}[htb!]
  \centering
  \subfigure[The station 4 ($\eta=0.1306$)]{\includegraphics[width=0.4\textwidth]{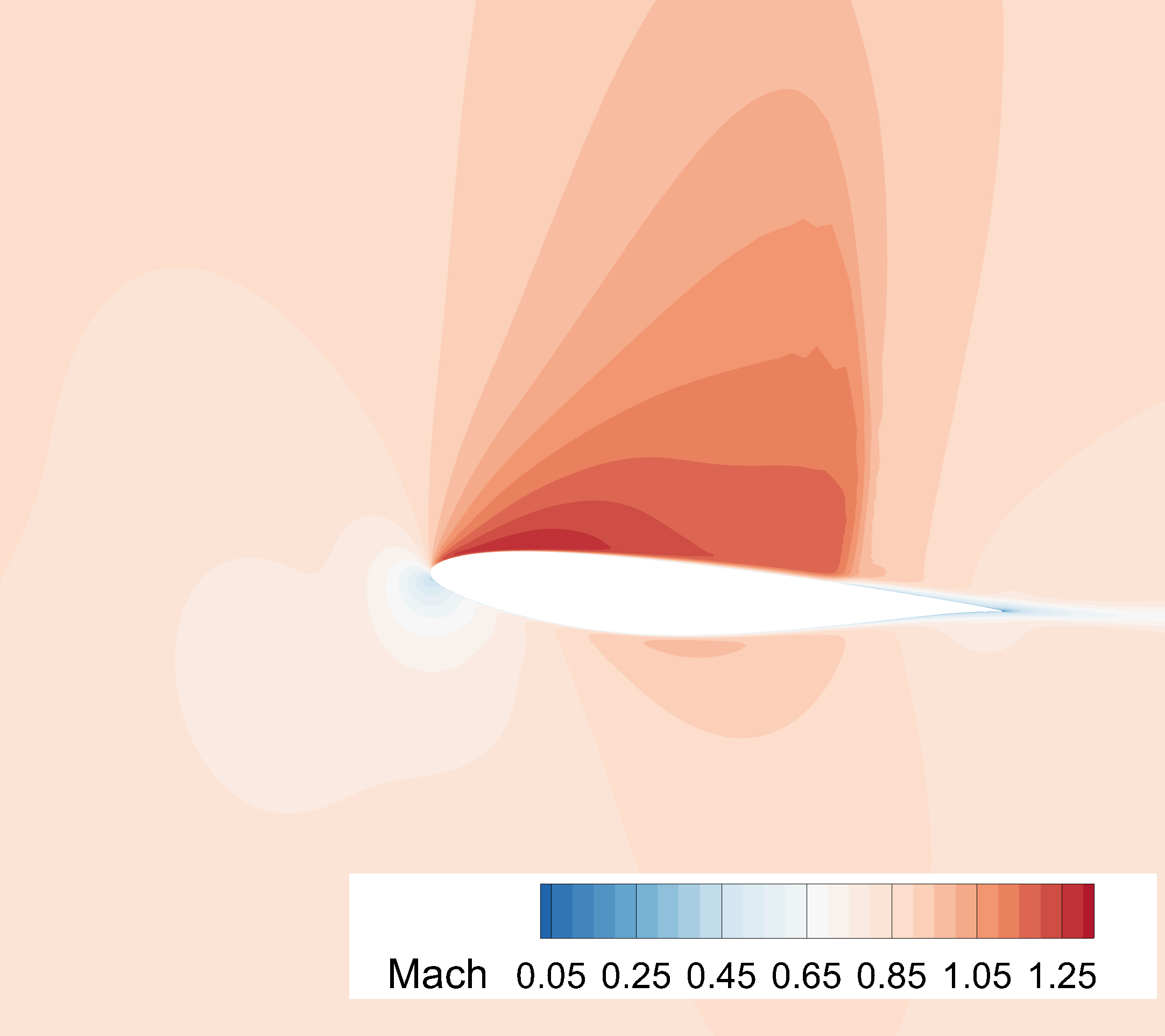}}
  \subfigure[The station 6 ($\eta=0.2828$)]{\includegraphics[width=0.4\textwidth]{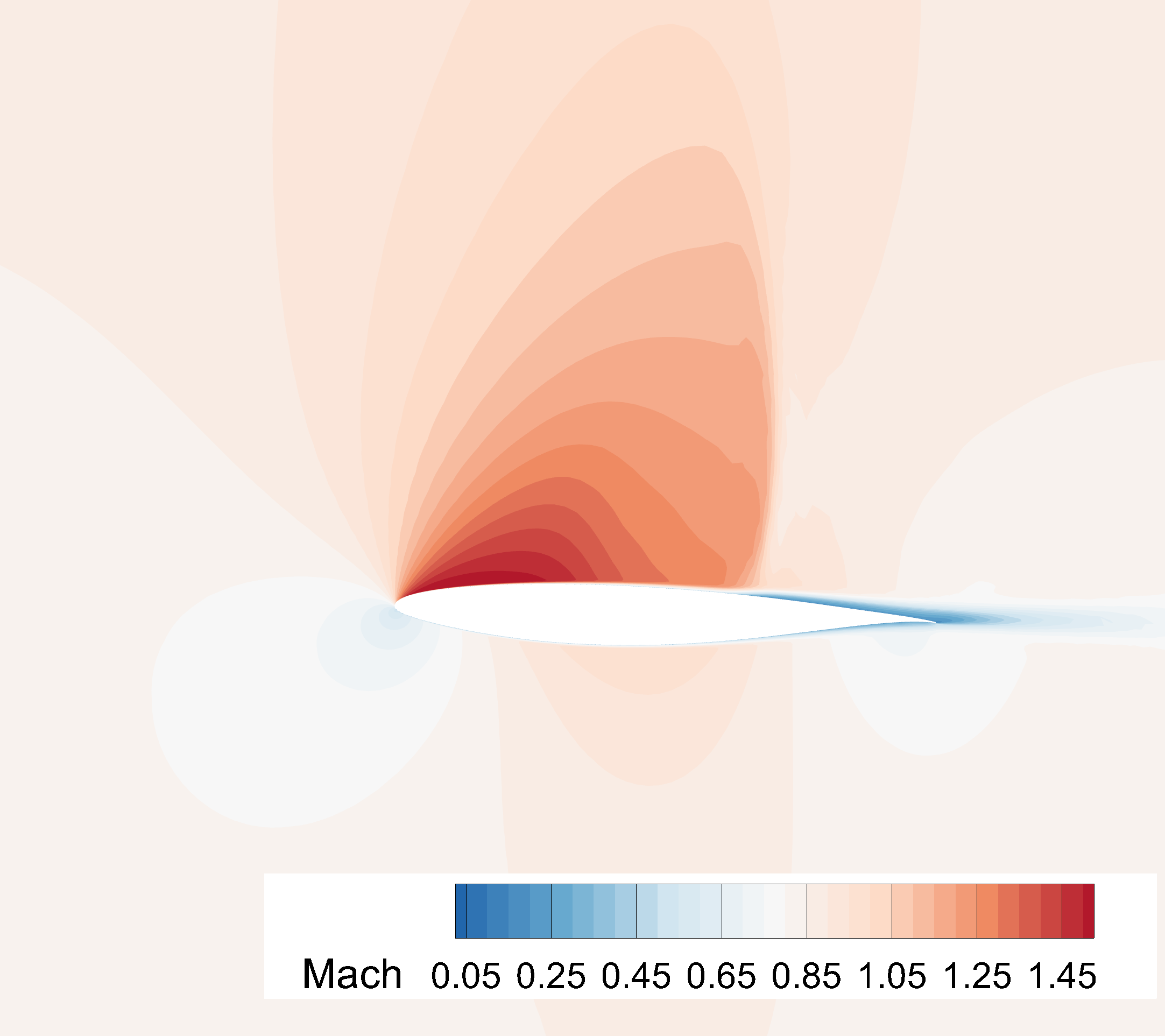}}\\
  \subfigure[The station 8 ($\eta=0.3700$)]{\includegraphics[width=0.4\textwidth]{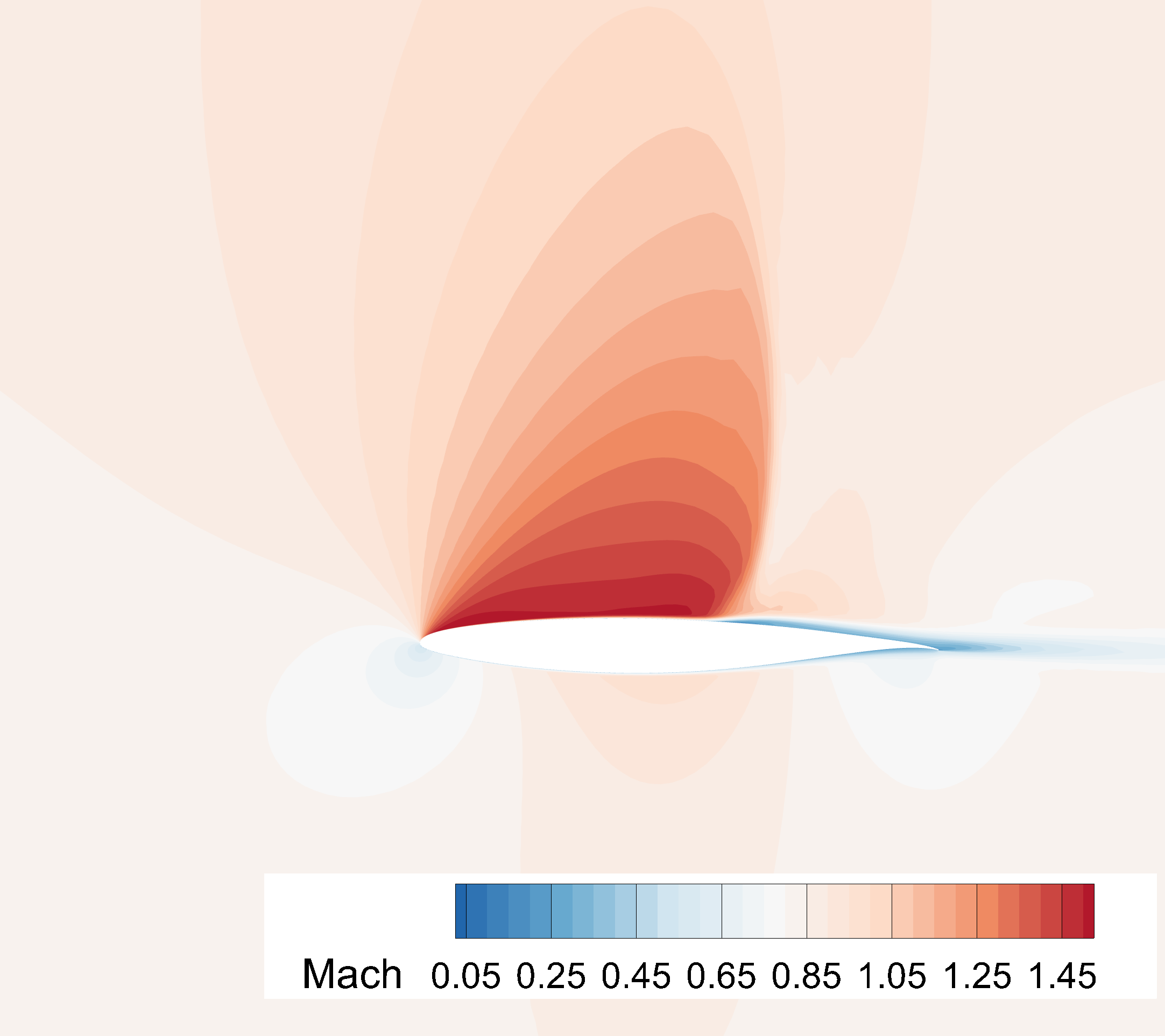}}
  \subfigure[The station 10 ($\eta=0.5024$)]{\includegraphics[width=0.4\textwidth]{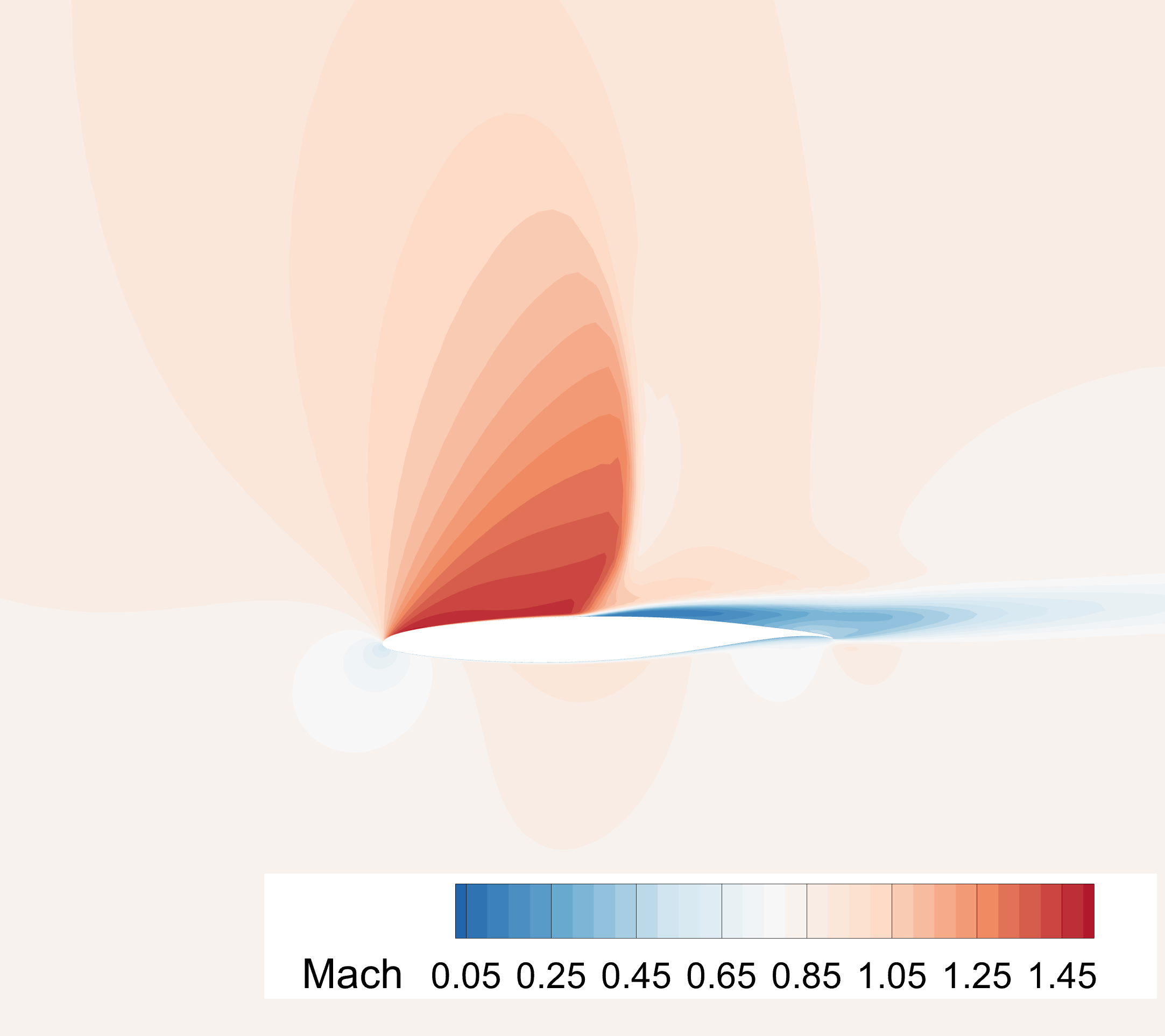}}
  \caption{The contour of Mach number at different spanwise positions at $\alpha=4^\circ$ angle of attack}
  \label{fig:CRMMach}
\end{figure}

\begin{figure}[htb!]
  \centering
  \subfigure[The pressure coefficient distribution]{\includegraphics[width=0.4\textwidth]{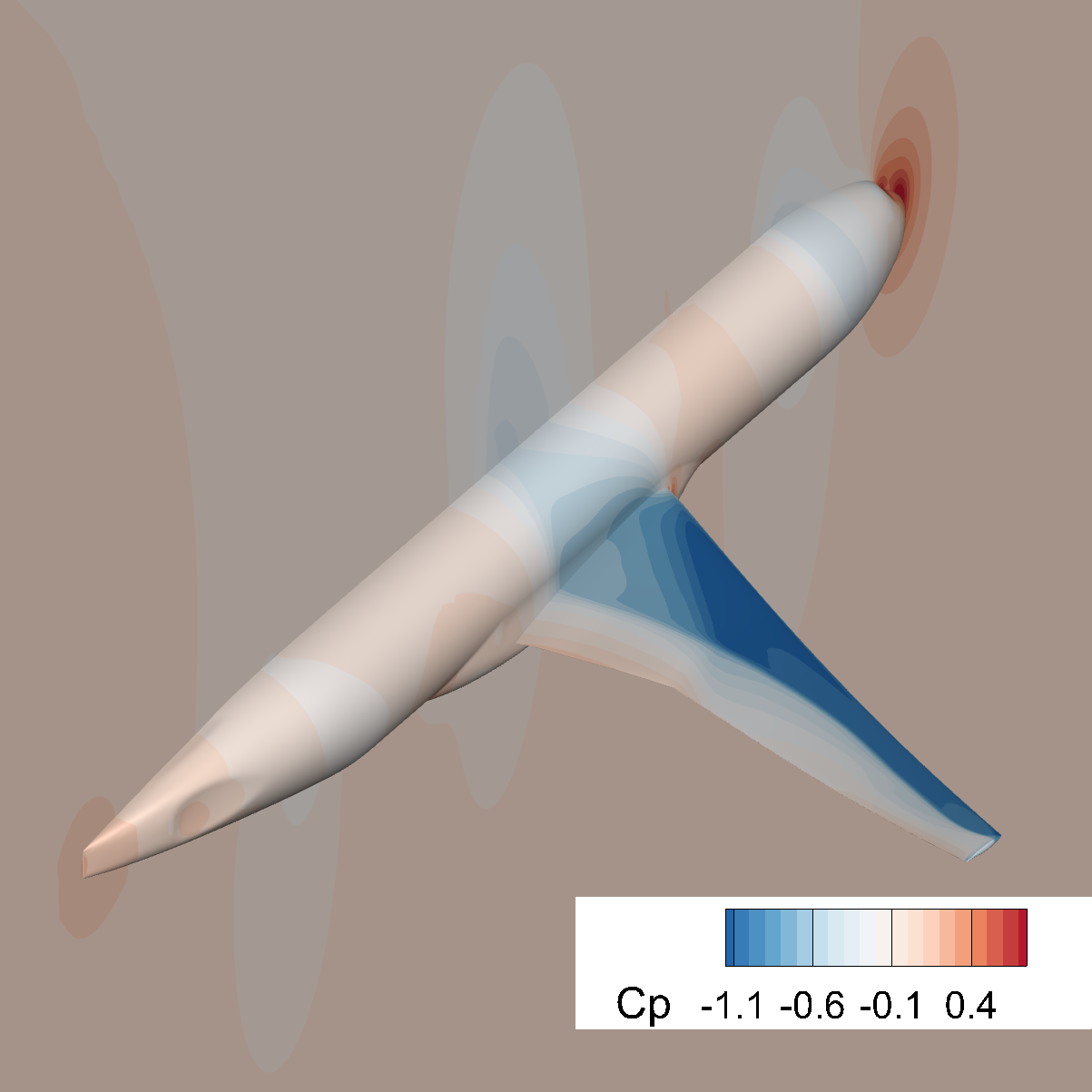}}
  \subfigure[The surface streamlines]{\includegraphics[width=0.4\textwidth]{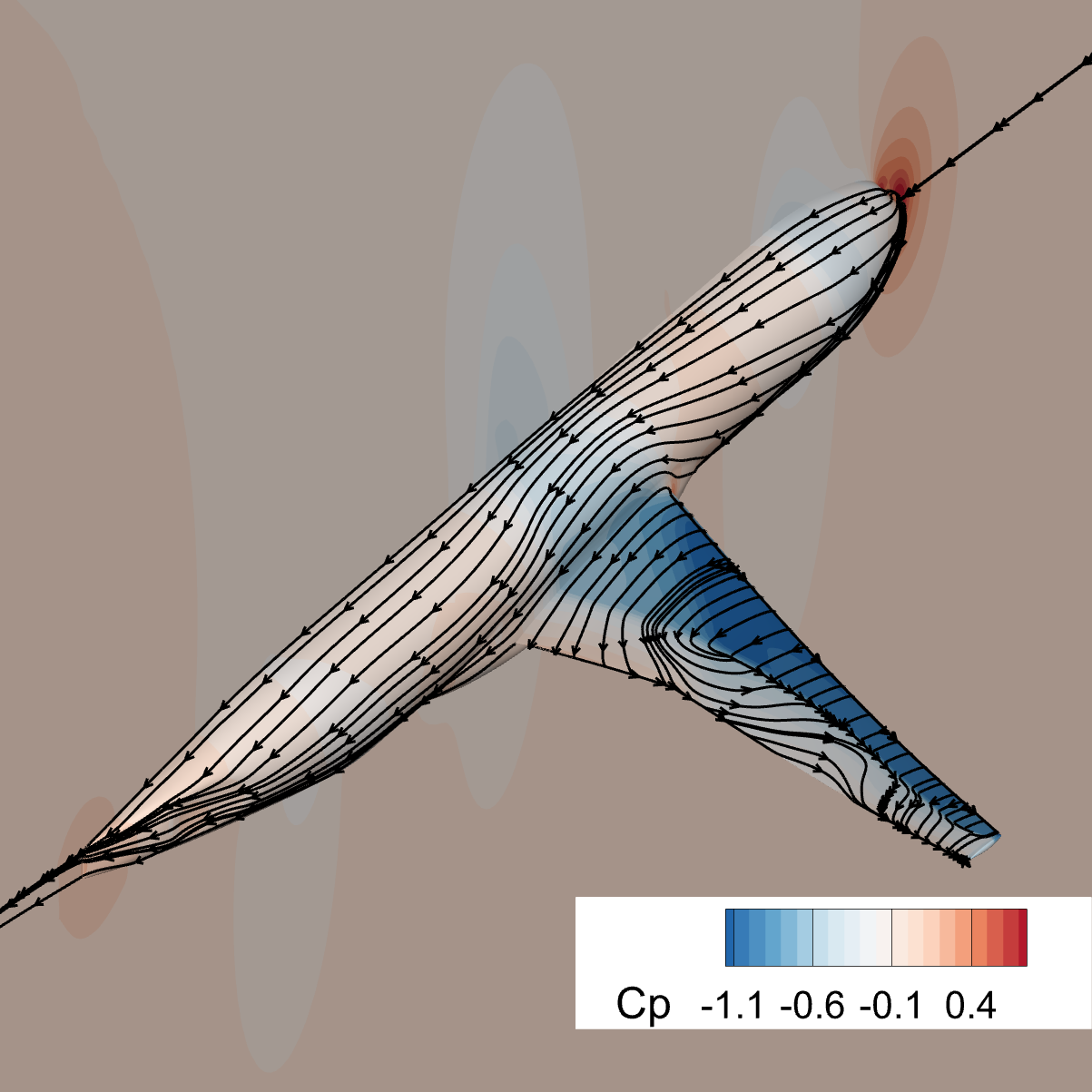}}
  \caption{The pressure coefficient distribution and the surface streamlines at $\alpha=4^\circ$ angle of attack}
  \label{fig:CRMCPStreamline}
\end{figure}

The contour of Mach number at different spanwise positions at $\alpha=4^\circ$ angle of attack is shown in Fig. \ref{fig:CRMMach}. The contour shows the interaction between the shock wave and the boundary layer separation. Moreover, the contour of pressure coefficient and the surface streamlines are shown in Fig. \ref{fig:CRMCPStreamline}, indicating a mass flow separation starts at about 50\% of the chord.
\subsection{NASA Rotor 67}
NASA Rotor 67 is a typical transonic rotor blade, which is widely used in the field of turbomachinery. The geometry database and experimental data can be found in \cite{strazisar1989laser,fottner1990test}. The basic information of the blade is shown in Table \ref{tab:rotor67}.
\begin{table}[htb!]
  \centering
  \caption{Basic information of NASA Rotor 67 \cite{strazisar1989laser}}
  \label{tab:rotor67}
  \begin{tabular}{cc}
    \toprule 
    Parameter & Value \\
    \midrule 
    Number of blades & 22 \\
    Design rotational speed (rpm) & 16043 \\
    Design mass flow rate (kg/s) & 33.25 \\
    Design pressure ratio & 1.63 \\
    Tip speed (m/s) & 429 \\
    Tip clearance at design speed (cm) & 0.1061 \\
    Relative Mach number at tip inlet & 1.38 \\
    Aspect ratio (mean blade height/root axial chord) & 1.56 \\
    Solidity at root & 3.11 \\
    Solidity at tip & 1.29 \\
    Tip diameter at inlet (cm) & 51.4 \\
    Tip diameter at outlet (cm) & 48.5 \\
    Hub-to-tip ratio at inlet & 0.375 \\
    Hub-to-tip ratio at outlet & 0.478 \\
    \bottomrule 
  \end{tabular}
\end{table}

\begin{figure}[htb!]
  \centering
  \subfigure[The mesh at blade surface]{\includegraphics[height=0.35\textwidth]{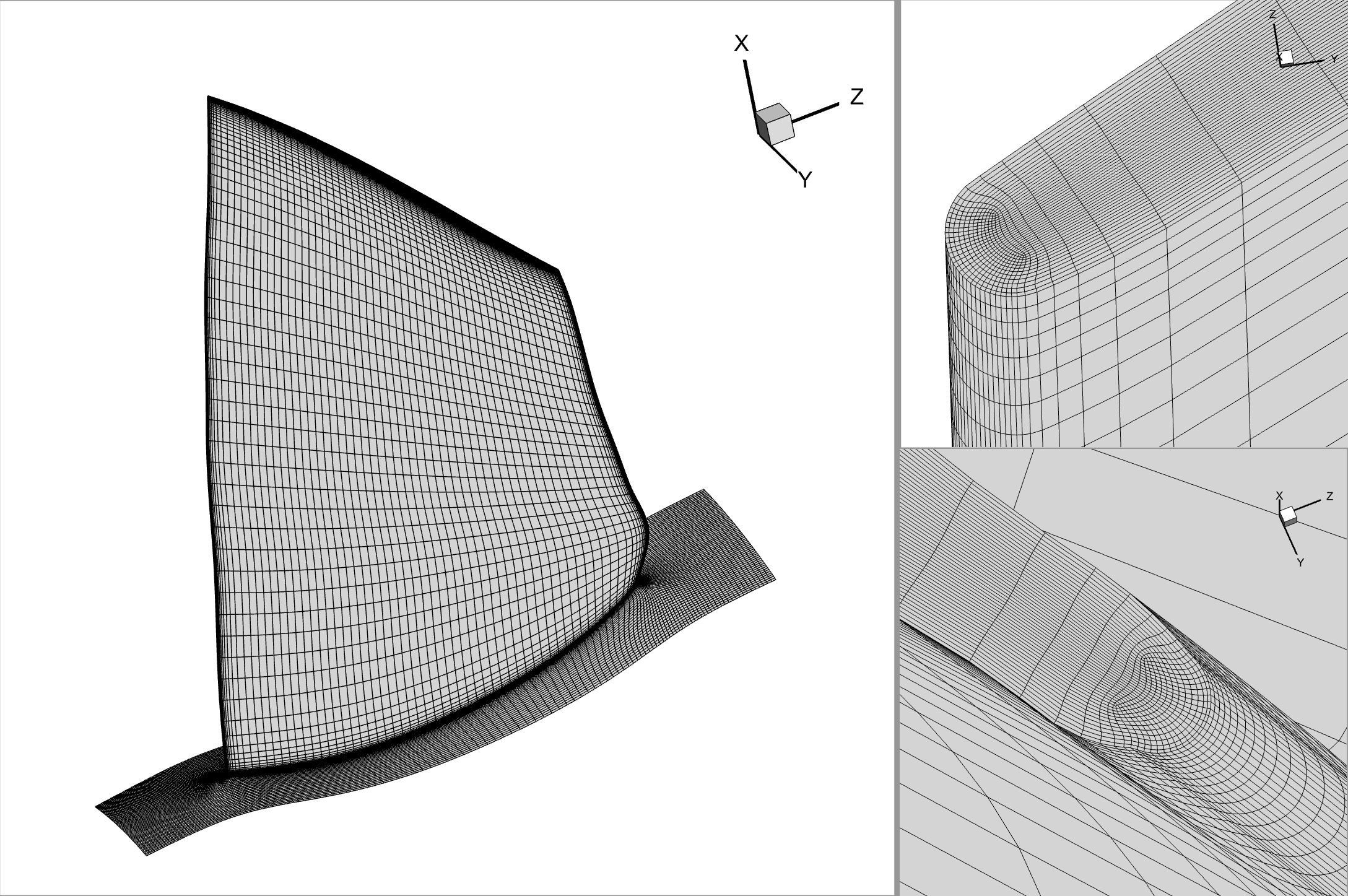}}
  \subfigure[The mesh projection on S2 surface\label{fig:rotor67MeshS2}]{\includegraphics[height=0.35\textwidth]{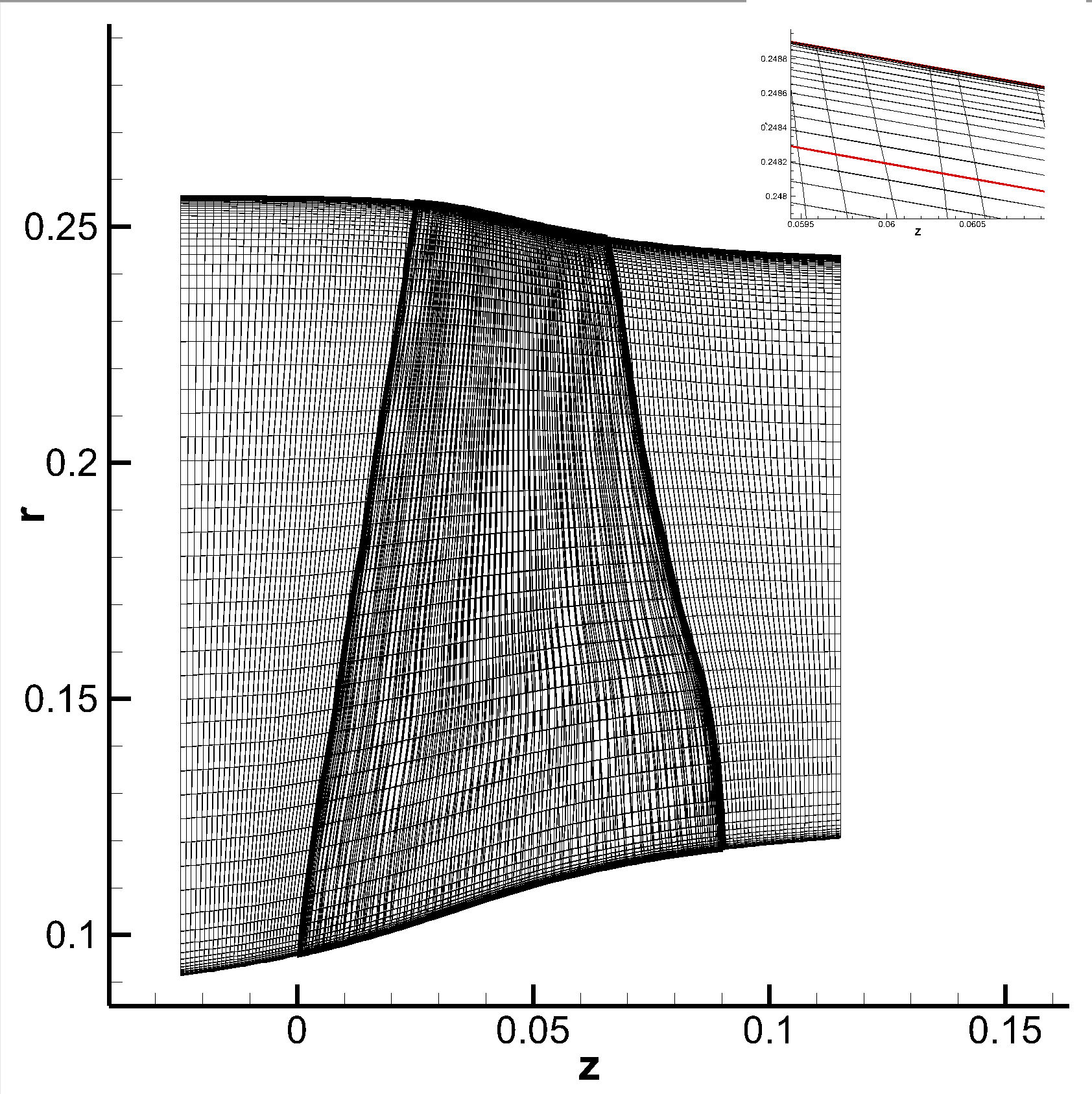}}
  \subfigure[The mesh projection on S1 surface\label{fig:rotor67MeshS1}]{\includegraphics[width=0.7\textwidth]{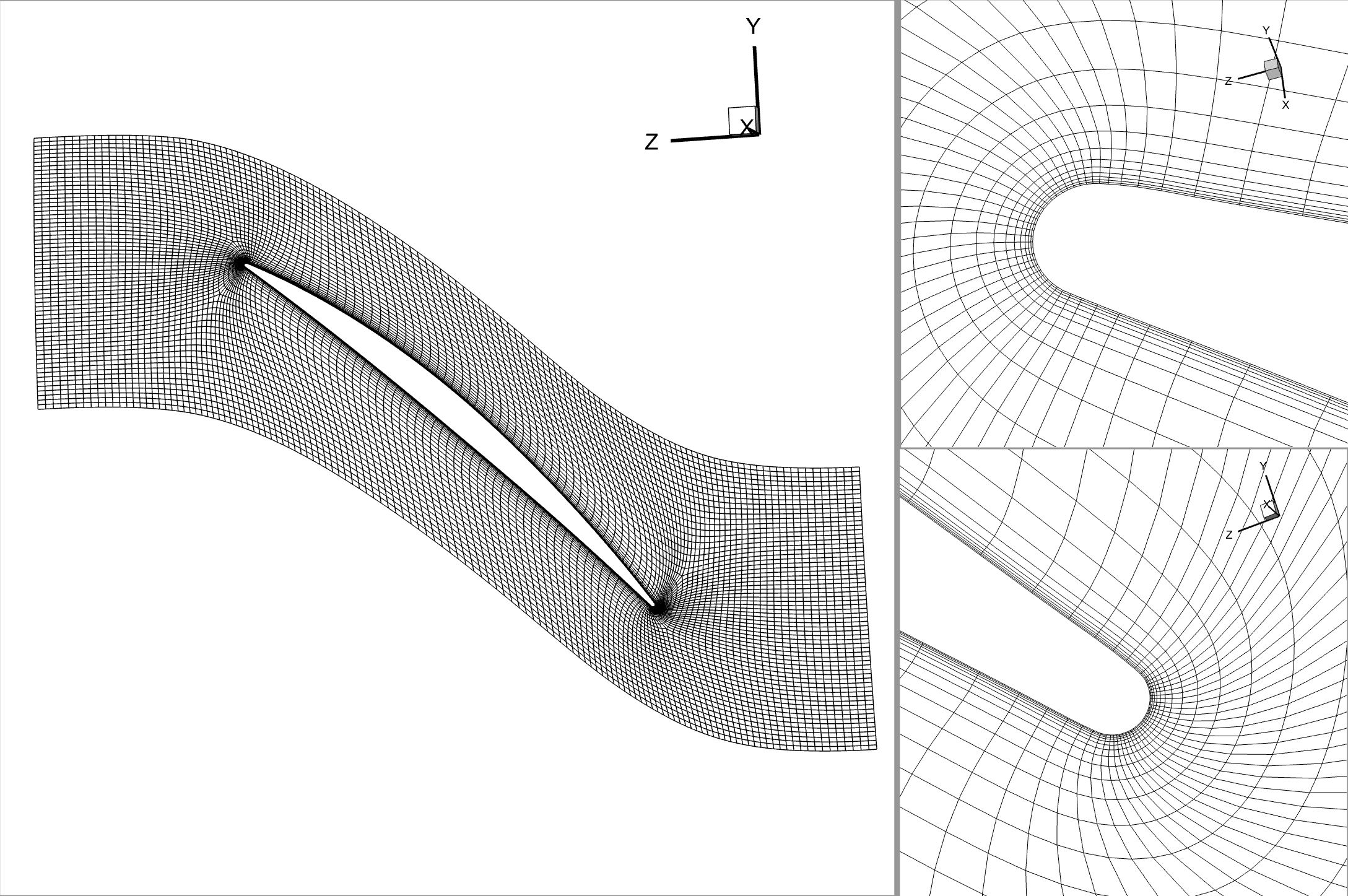}}
  \caption{The computational mesh of NASA Rotor 67}
  \label{fig:rotor67}
\end{figure}

The computational mesh of NASA Rotor 67 is shown in Fig. \ref{fig:rotor67}. The positions of the inlet and outlet are set at the experimental station. The total number of cells is $941,824$. The H-O-H type of topology is used to mesh the blade surface, and the first layer of cells is located at the blade surface with a distance of $10^{-6}$ to the blade surface. As shown in Fig. \ref{fig:rotor67MeshS2}, the tip clearance is set as 0.6 times of the design tip clearance (0.1061 cm) by considering the vena contraction effect, instead of refining the mesh near the blade tip.

At the inlet, a total pressure of 101325 Pa and a total temperature of 288.15 K are specified, with absolute flow direction parallel to the rotation axis. Static pressure is prescribed at the outlet at the shroud, with radial equilibrium determining pressures at other positions. Performance characteristics are computed using outlet pressures ranging from 105,000 Pa to 141,000 Pa. The blade and hub walls rotate at 16043 rpm with adiabatic wall conditions, while periodic boundaries are used in the circumferential direction. The shroud is set as a static wall. For initialization, the lowest pressure case starts with zero velocity and inlet total conditions, while subsequent cases use the converged solution from the previous pressure level.

\begin{figure}[htb!]
  \centering
  \subfigure[Total pressure ratio]{\includegraphics[height=0.4\textwidth]{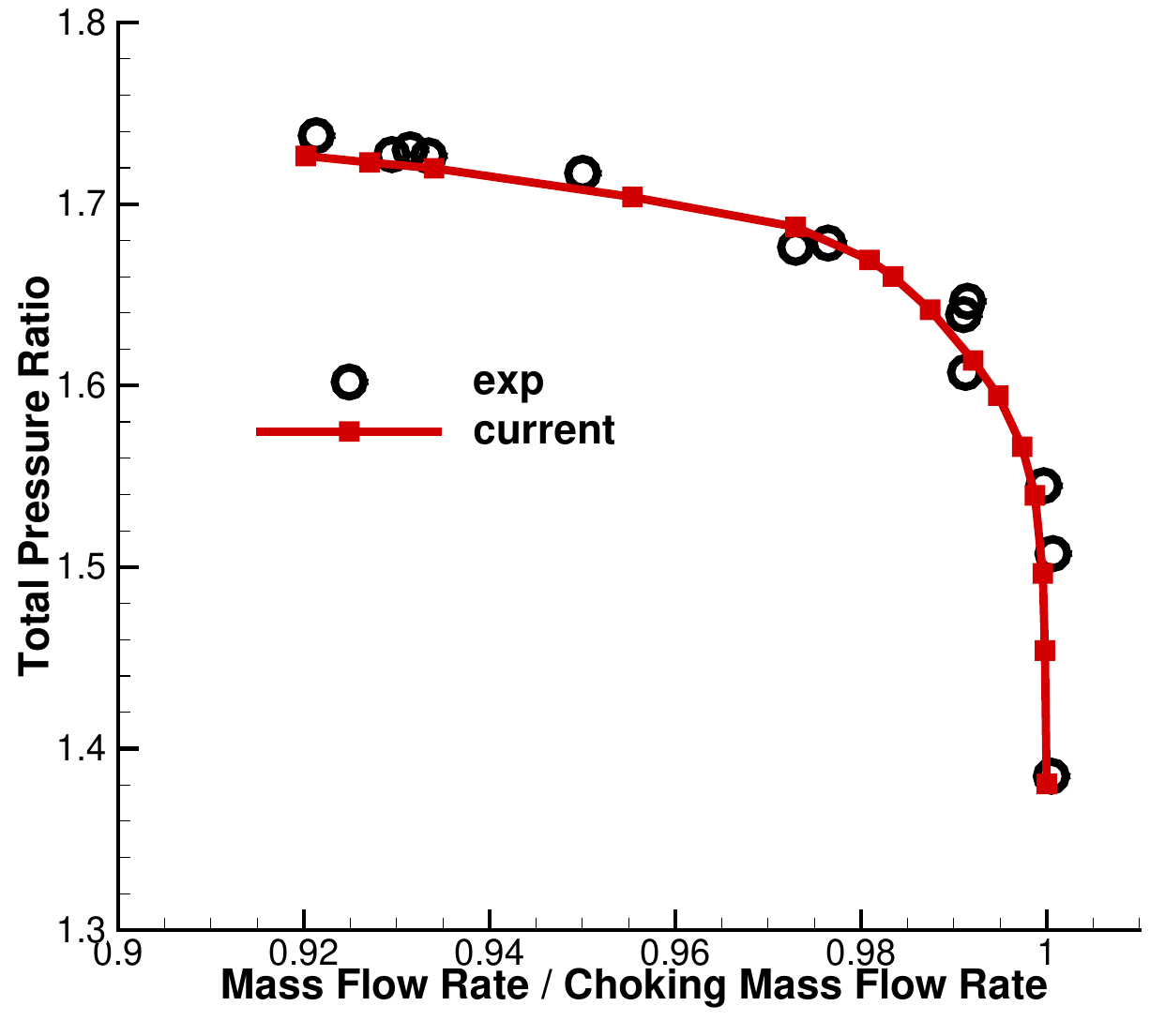}}
  \subfigure[Adiabatic efficiency]{\includegraphics[height=0.4\textwidth]{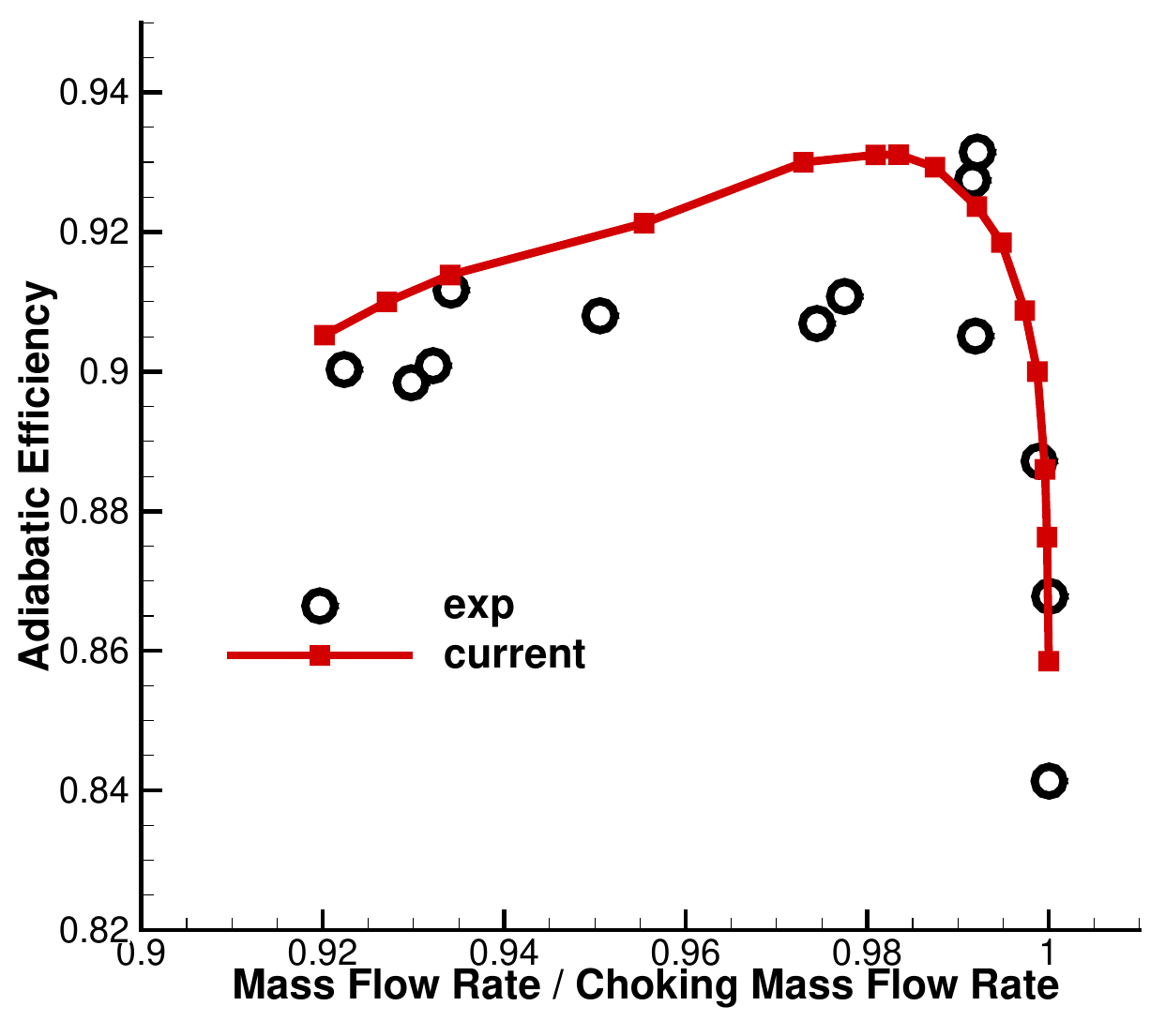}}
  \caption{Performance characteristics of NASA Rotor 67}
  \label{fig:rotor67char}
\end{figure}

The computed choked mass flow rate is 34.67 kg/s, while the corresponding experimental result is 34.96 kg/s, with a relative error of 0.84\%. The computed performance characteristics with normalized mass flow rate as the parameter are shown in Fig. \ref{fig:rotor67char}. The adiabatic efficiency is defined as
\begin{equation*}
  \eta_a = \frac{T_{\text{out,isentropy}}^*-T_{\text{in}}^*}{T_{\text{out}}^*-T_{\text{in}}^*} = \frac{(P_{\text{out}}^*/P_{\text{in}}^*)^{(\gamma-1)/\gamma} -1}{P_{\text{out}}^*/P_{\text{in}}^*-1},
\end{equation*}
where $T_{\text{out}}^*$ and $T_{\text{in}}^*$ are the outlet and inlet total temperature respectively, $T_{\text{out,isentropy}}^*$ is the outlet total temperature that would be achieved through an isentropic process starting from the inlet total pressure and ending at the outlet total pressure, $P_{\text{out}}^*$ and $P_{\text{in}}^*$ are the outlet and inlet total pressure respectively, and $\gamma$ is the specific heat ratio.
The peak efficiency and near-stall points can be well captured and agree with the experimental data.

\begin{figure}[htb!]
  \centering
  \subfigure[The convergence history of mass flow rate]{\includegraphics[width=0.3\textwidth]{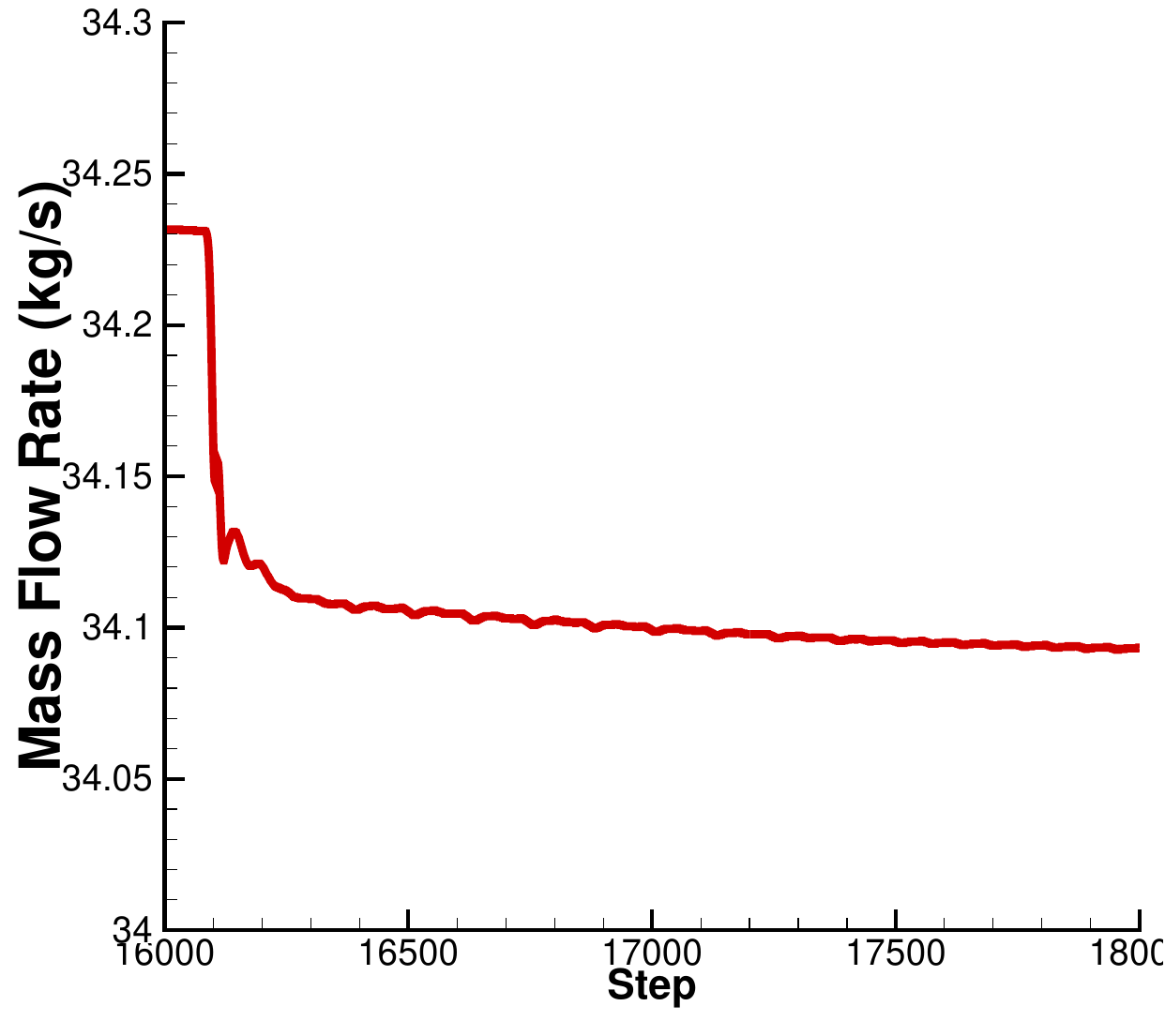}}
  \subfigure[The convergence history of total pressure ratio]{\includegraphics[width=0.3\textwidth]{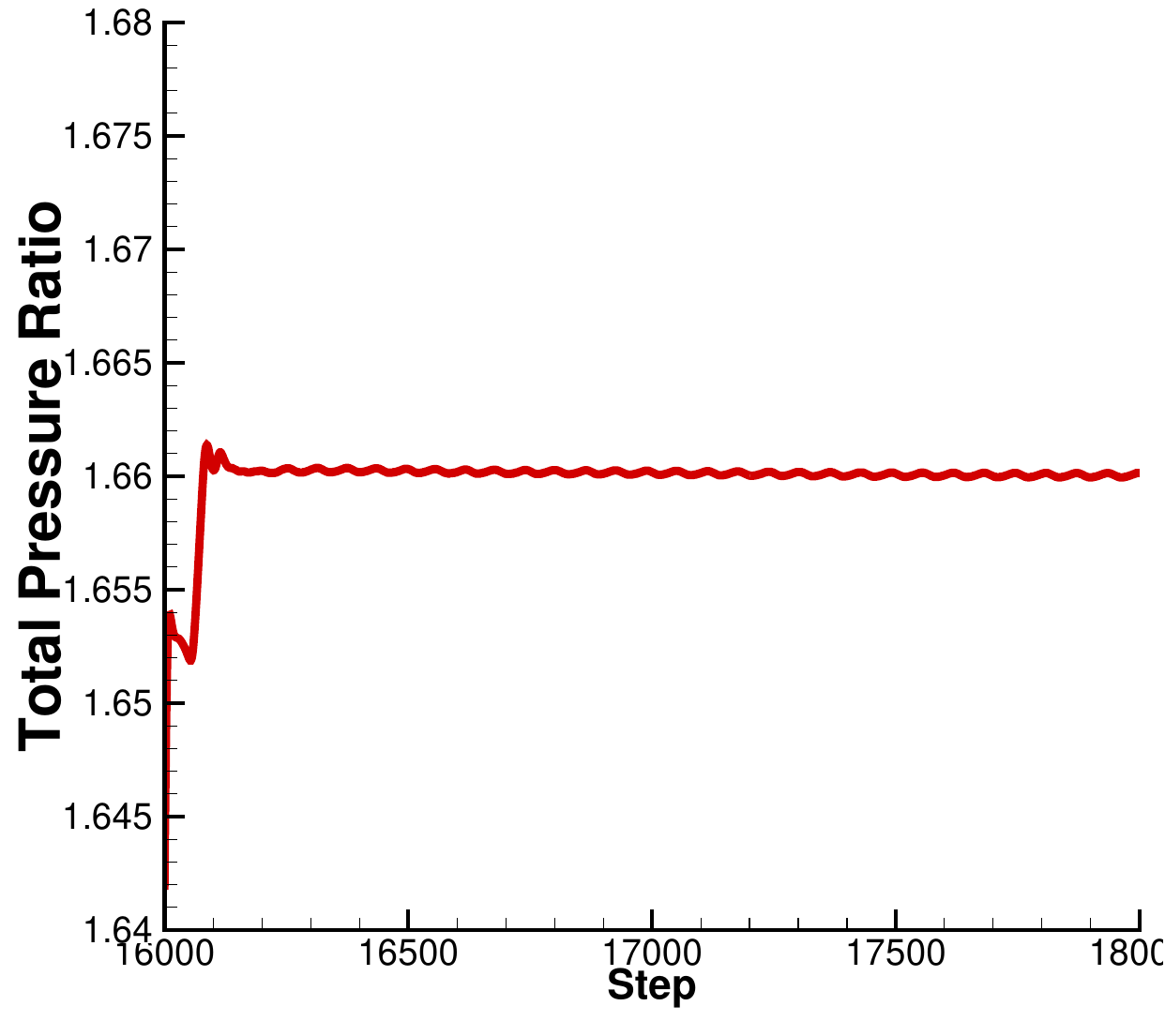}}
  \subfigure[The convergence history of adiabatic efficiency]{\includegraphics[width=0.3\textwidth]{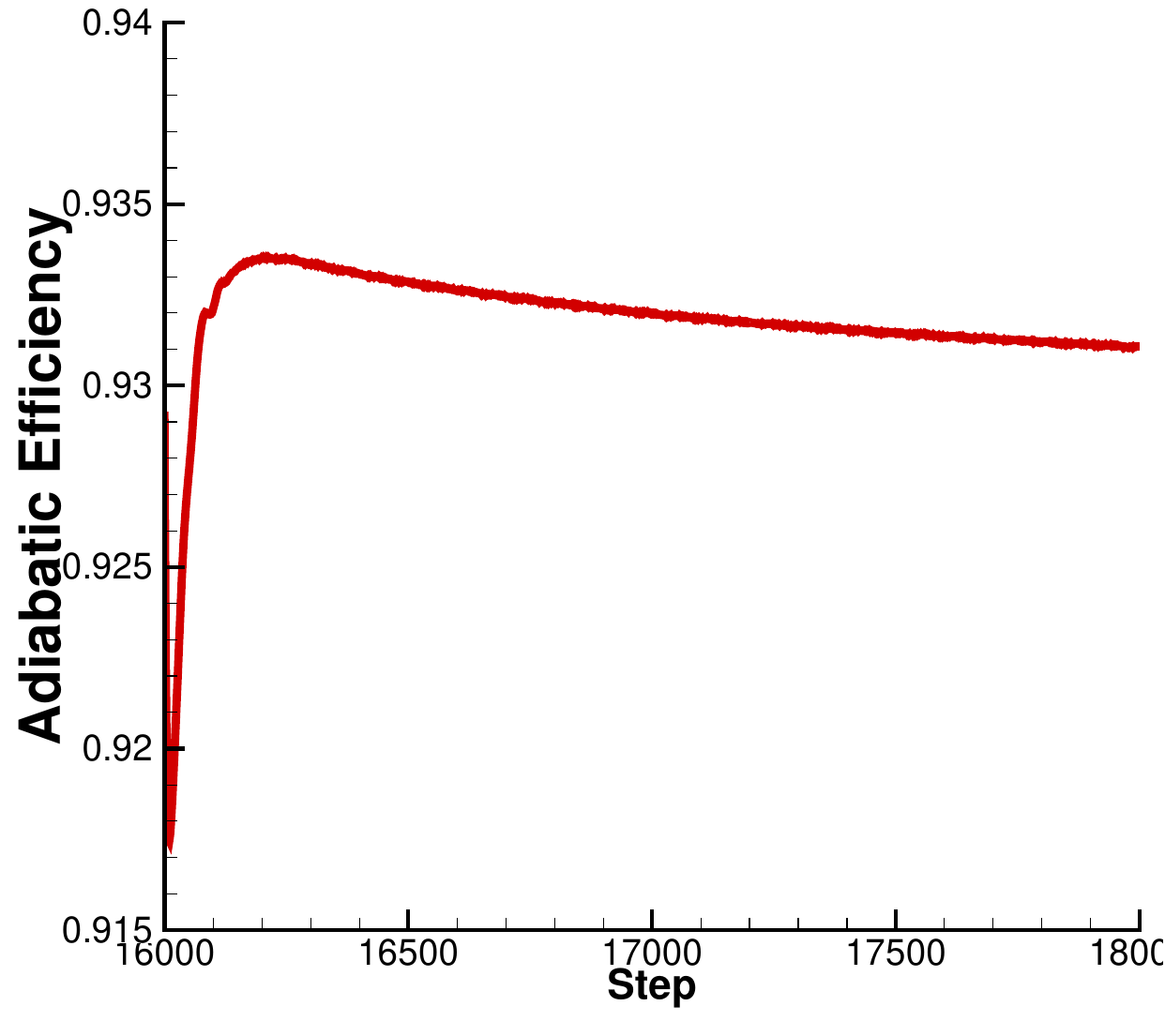}}
  \caption{The convergence history of NASA Rotor 67 at peak efficiency point}
  \label{fig:rotor67convergence}
\end{figure}

The computation time for 1000 steps is about 1506s with 128 cores. For each working point, 2000 steps are run to reach the convergence. Fig. \ref{fig:rotor67convergence} shows the convergence history of mass flow rate, total pressure ratio, and adiabatic efficiency at the peak efficiency point. The CFL number is an exponential function increasing from 1 to 100 during the first 100 steps, and remains at 100 afterwards. The results show that the performance parameters converge after approximately 300 steps.

\begin{figure}[htb!]
  \centering
  \subfigure[90\% span relative Mach number contours]{
    \raisebox{-0.5\height}{\includegraphics[width=0.22\textwidth]{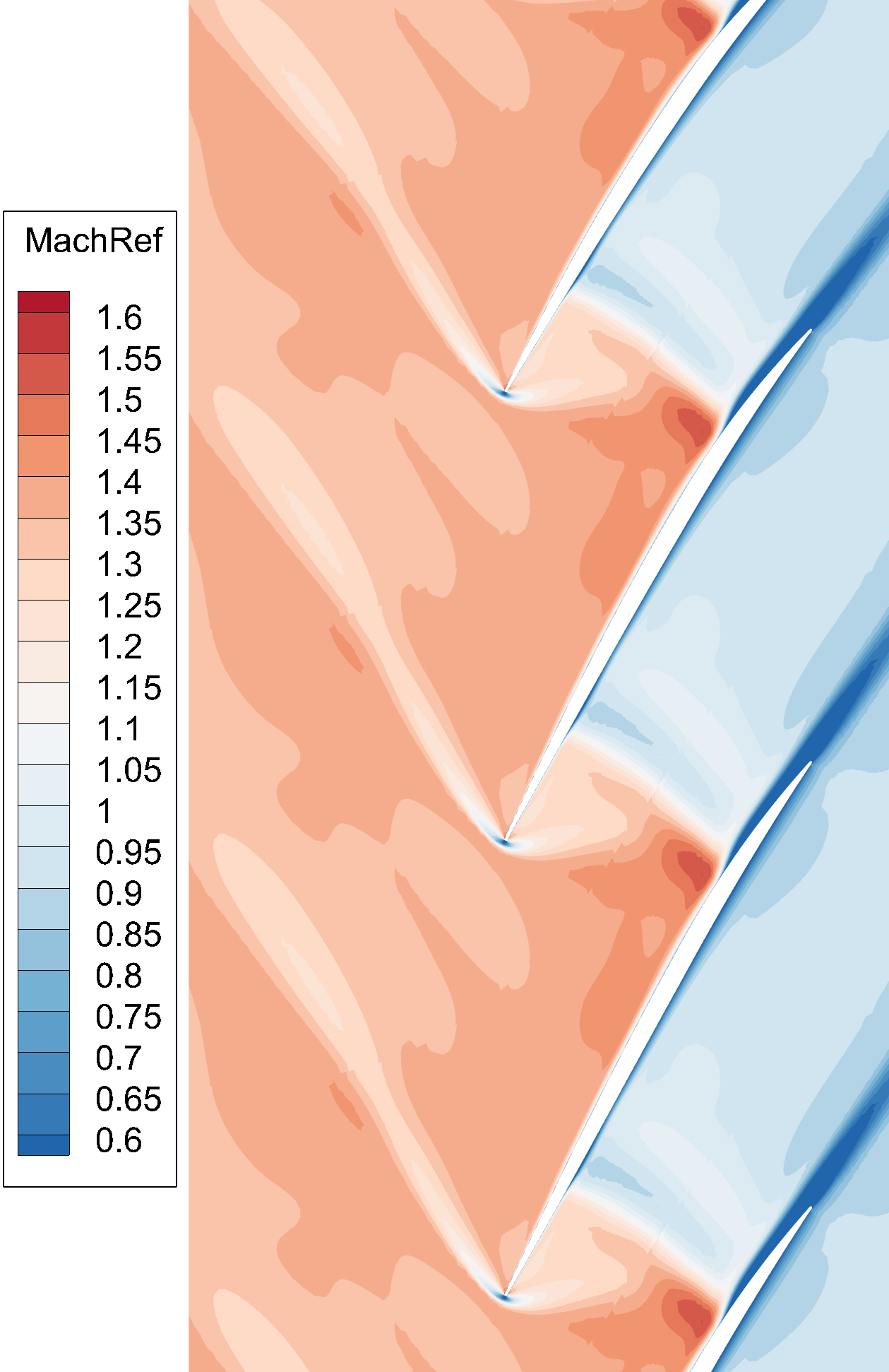}}
    \raisebox{-0.5\height}{\includegraphics[width=0.22\textwidth]{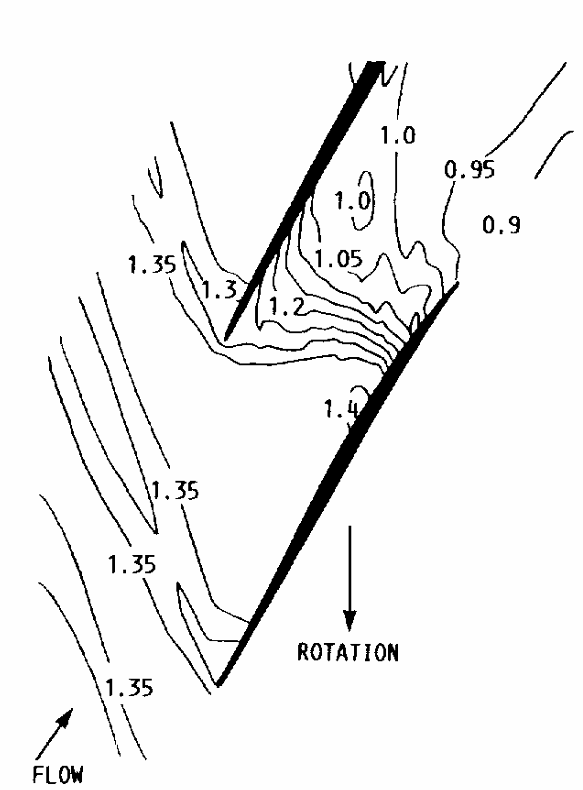}} \quad
    \raisebox{-0.5\height}{\includegraphics[width=0.22\textwidth]{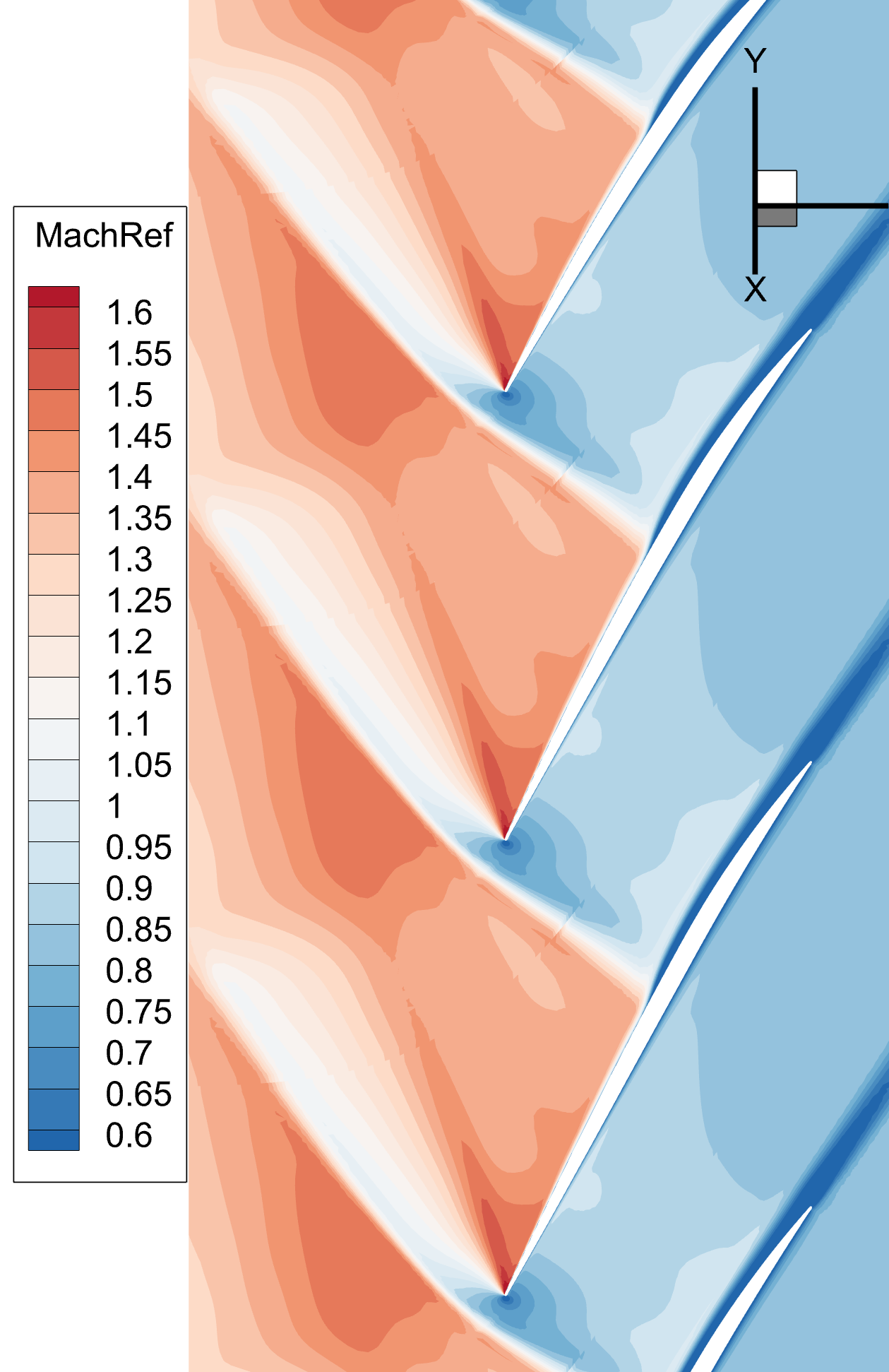}}
    \raisebox{-0.5\height}{\includegraphics[width=0.22\textwidth]{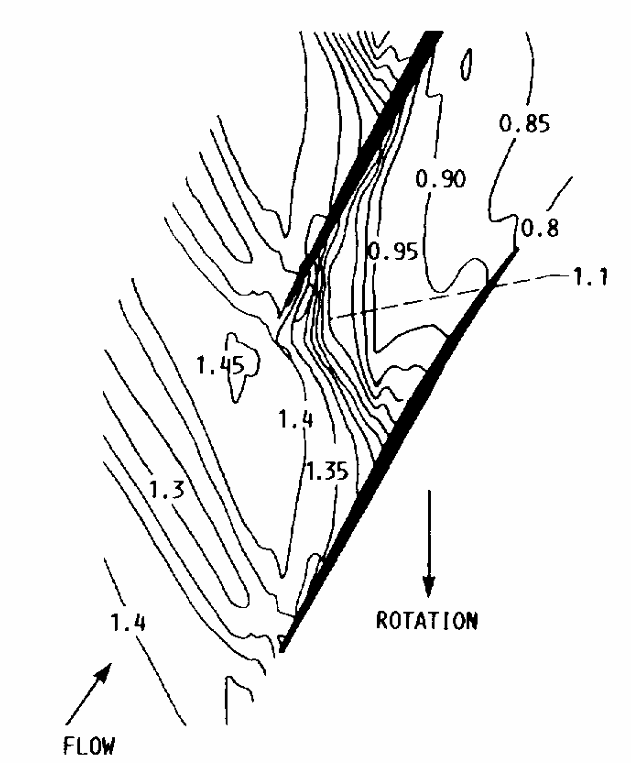}}
    }
    \subfigure[70\% span relative Mach number contours]{
    \raisebox{-0.5\height}{\includegraphics[width=0.22\textwidth]{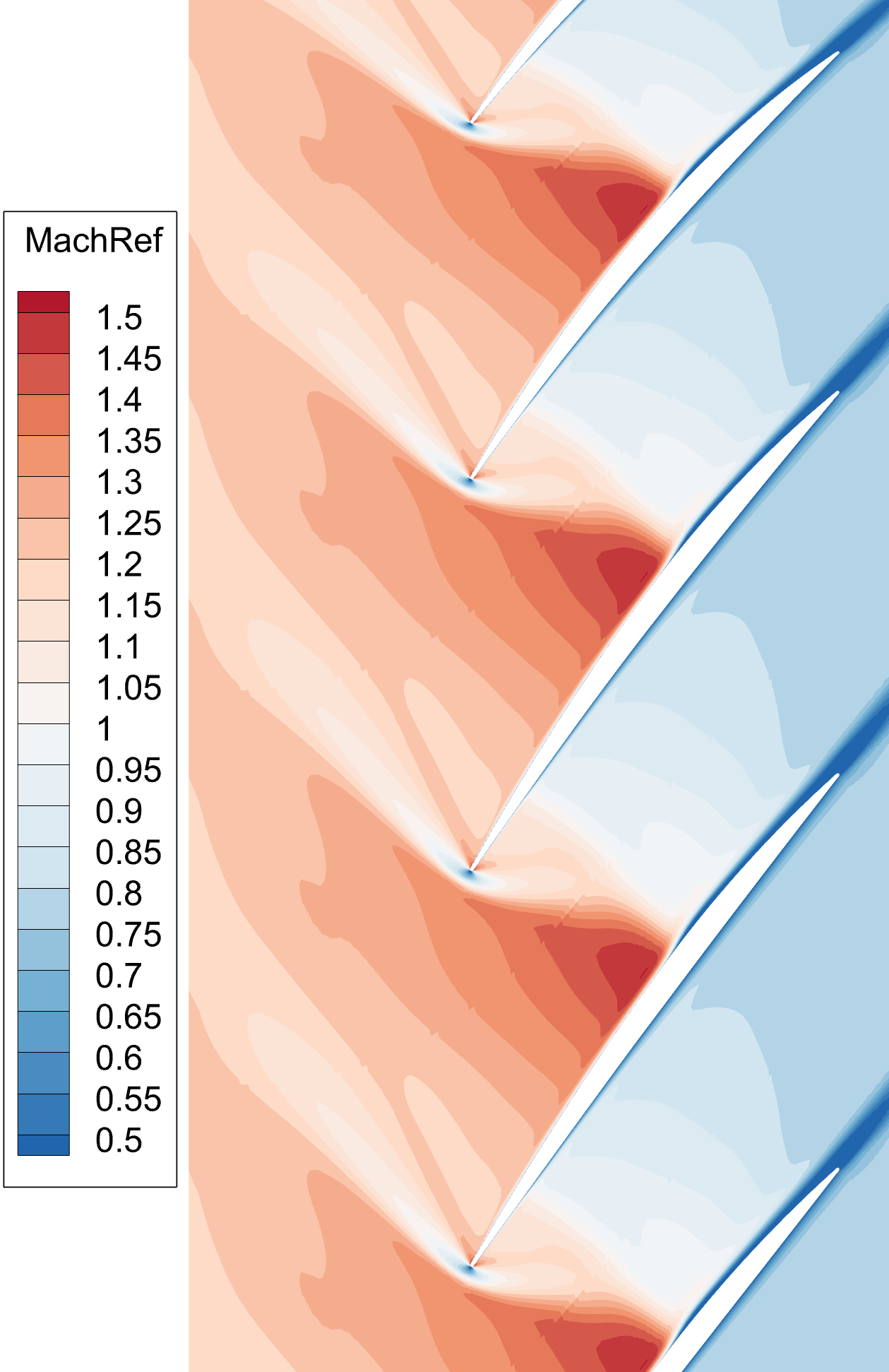}}
    \raisebox{-0.5\height}{\includegraphics[width=0.22\textwidth]{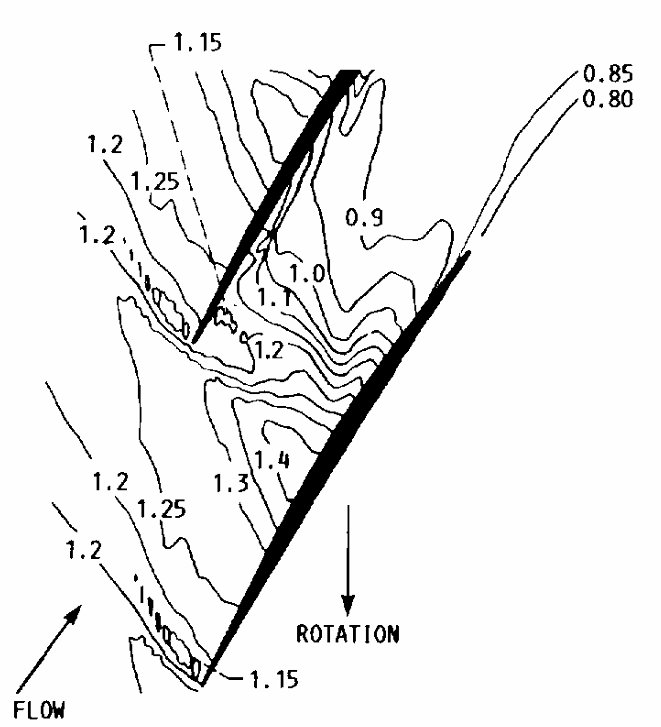}} \quad
    \raisebox{-0.5\height}{\includegraphics[width=0.22\textwidth]{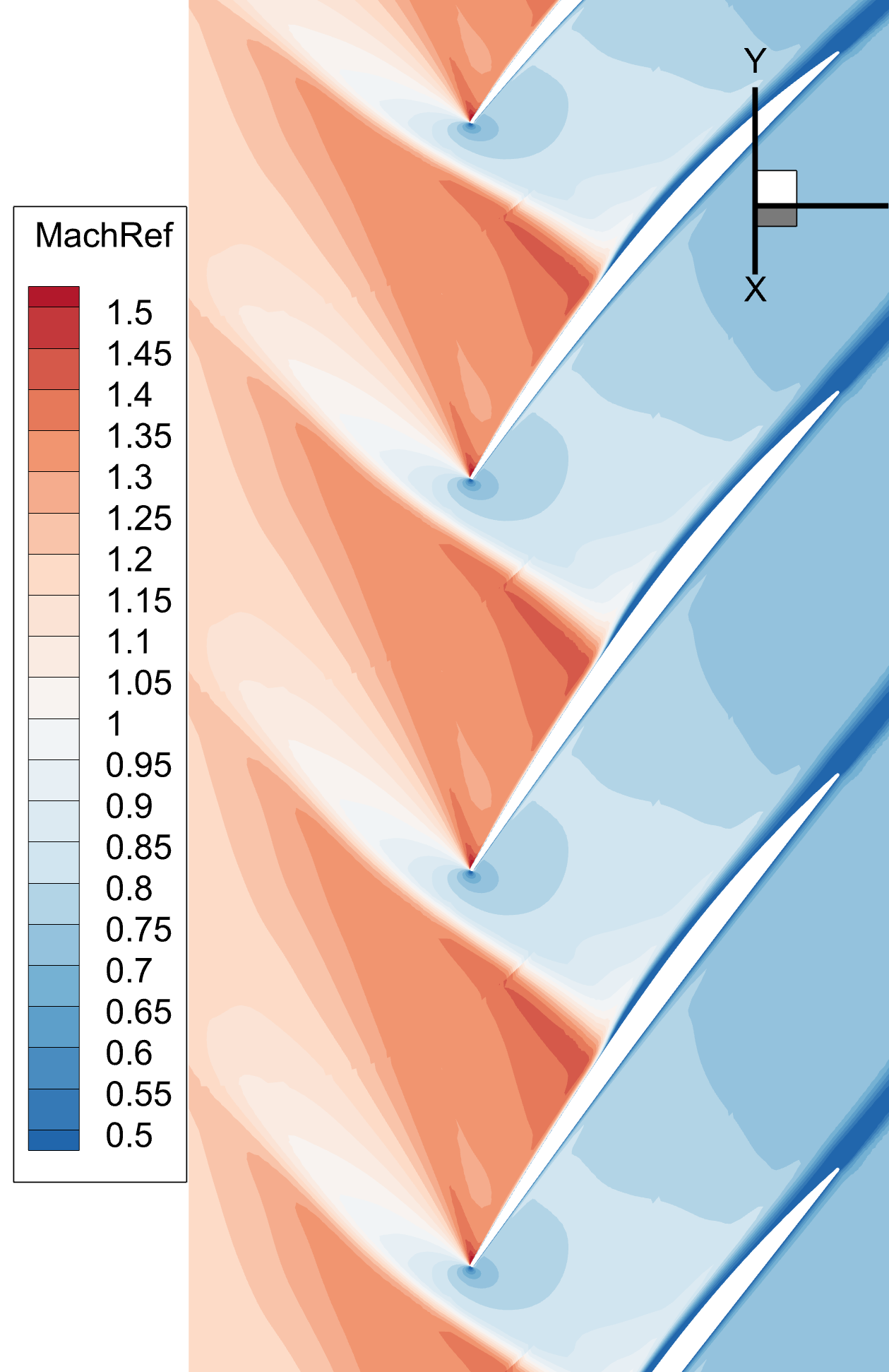}}
    \raisebox{-0.5\height}{\includegraphics[width=0.22\textwidth]{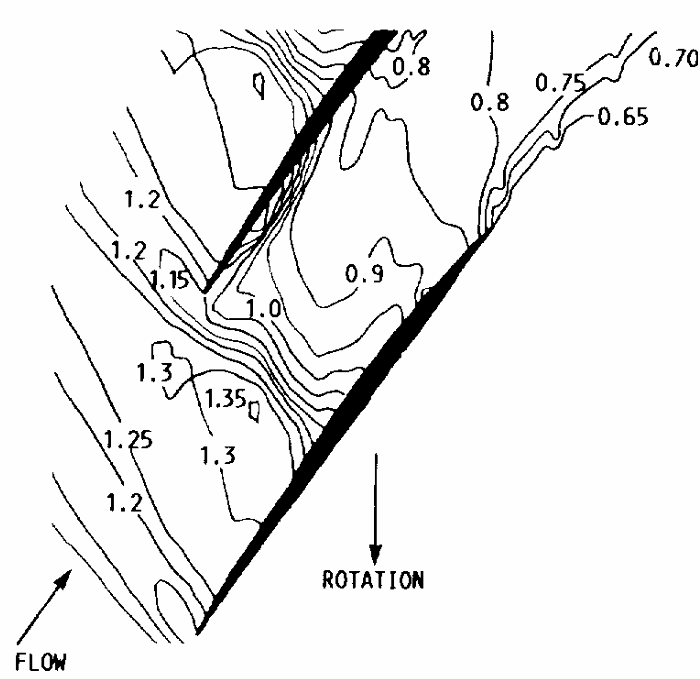}}
    }
    \subfigure[30\% span relative Mach number contours]{
    \raisebox{-0.5\height}{\includegraphics[width=0.22\textwidth]{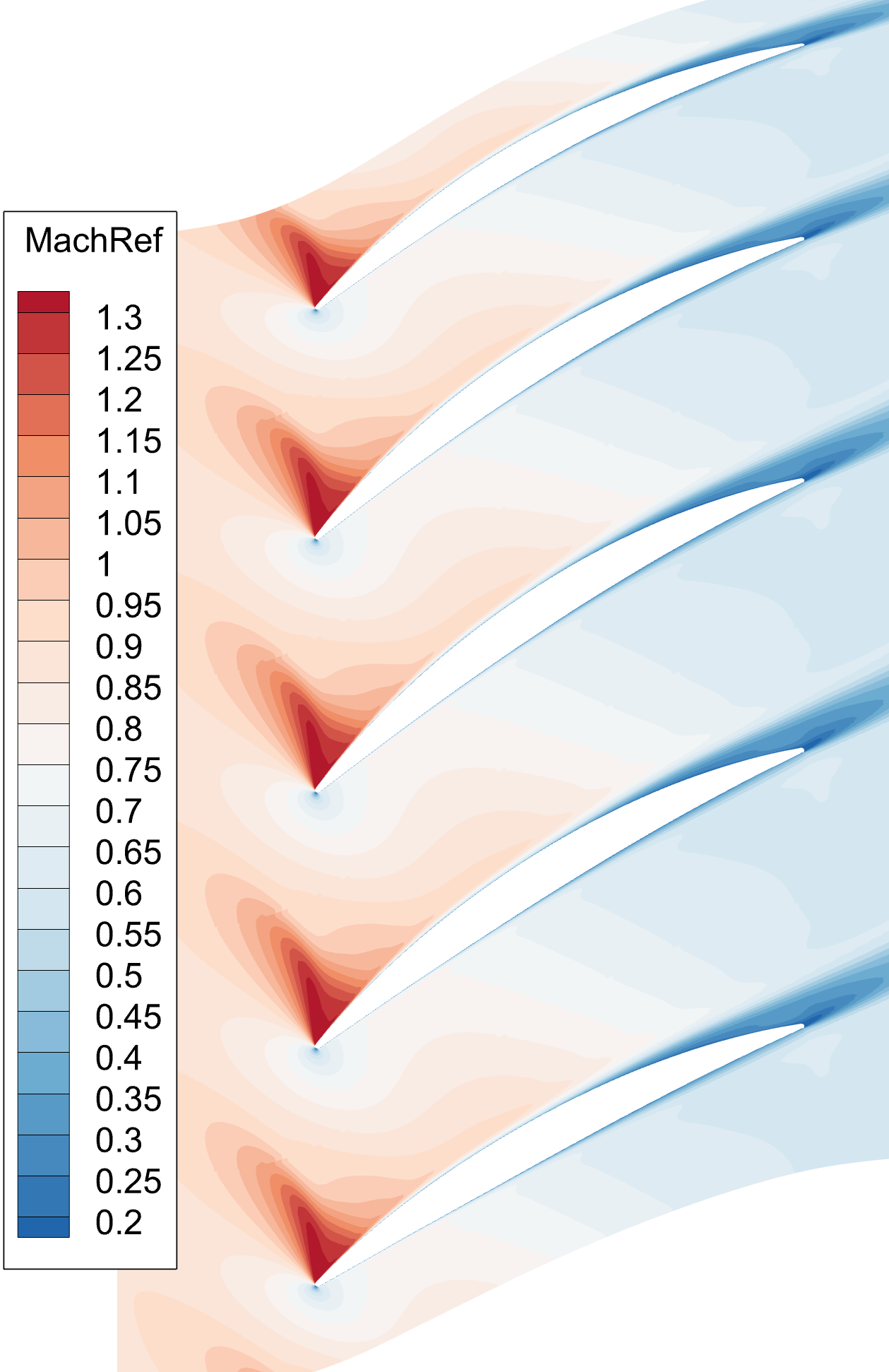}}
    \raisebox{-0.5\height}{\includegraphics[width=0.22\textwidth]{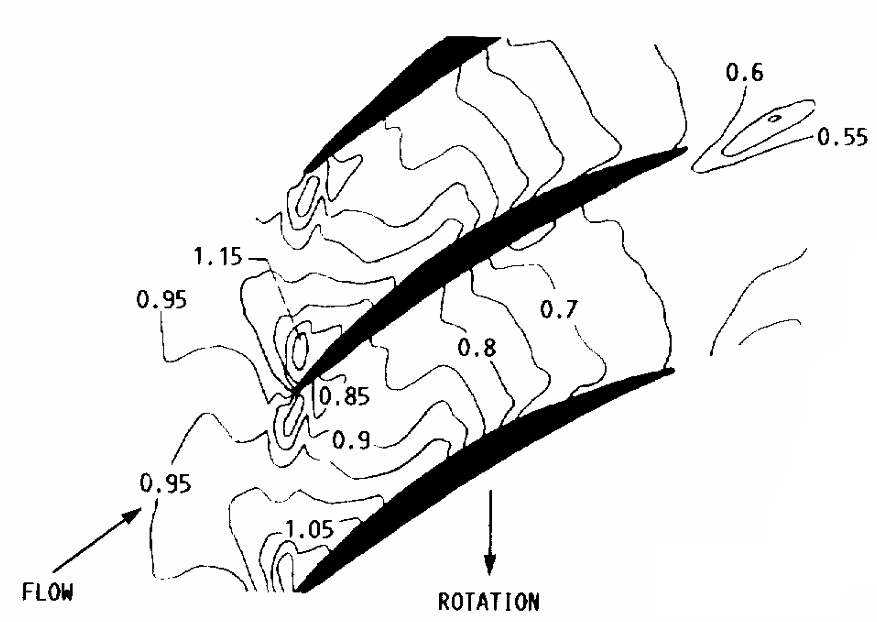}} \quad
    \raisebox{-0.5\height}{\includegraphics[width=0.22\textwidth]{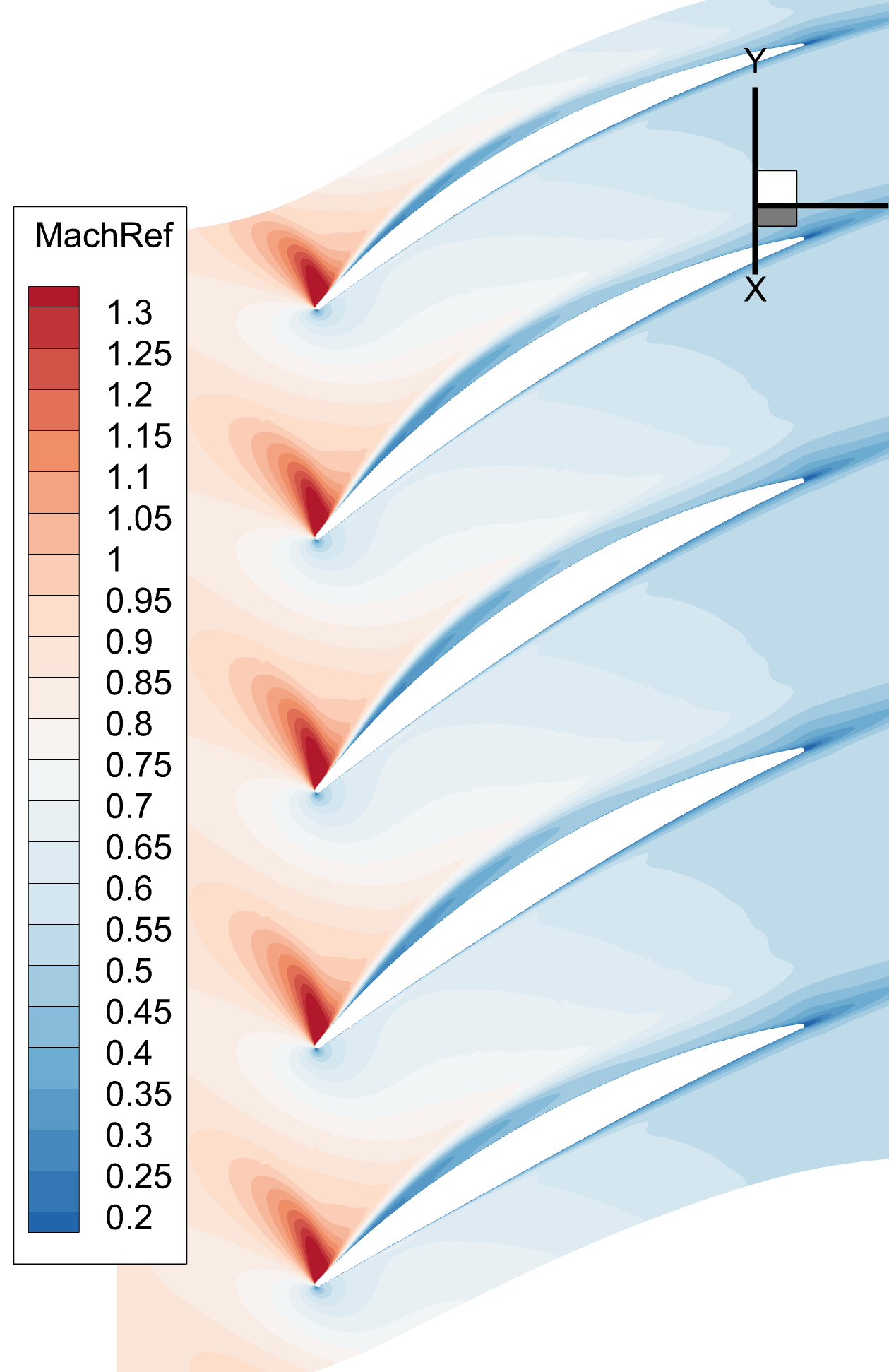}}
    \raisebox{-0.5\height}{\includegraphics[width=0.22\textwidth]{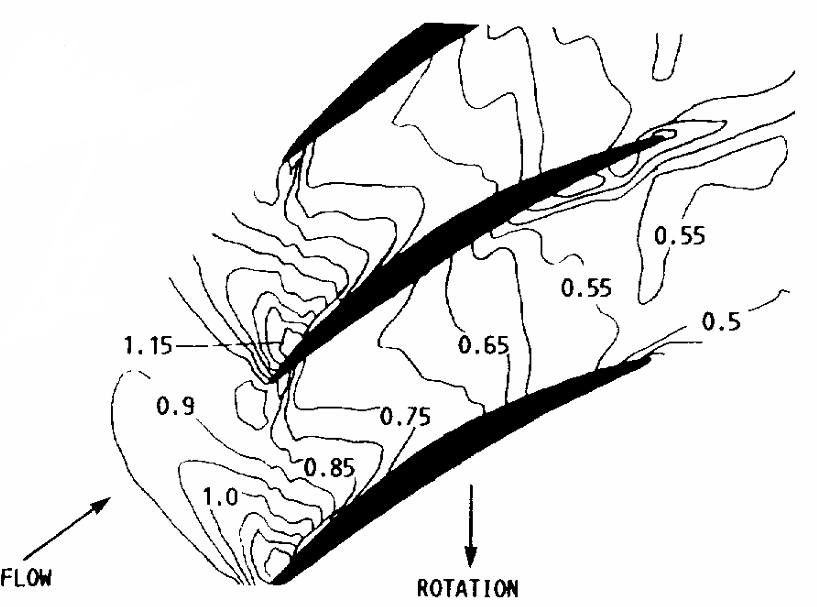}}
    }
  \caption{Comparison of relative Mach number contours for NASA Rotor 67: numerical results and experimental contours \cite{strazisar1989laser} at peak efficiency point (left two) and near stall point (right two)}
  \label{fig:rotor67machContour}
\end{figure}

\begin{figure}[htb!]
  \centering
  \subfigure[Total pressure at peak efficiency]{\includegraphics[width=0.3\textwidth]{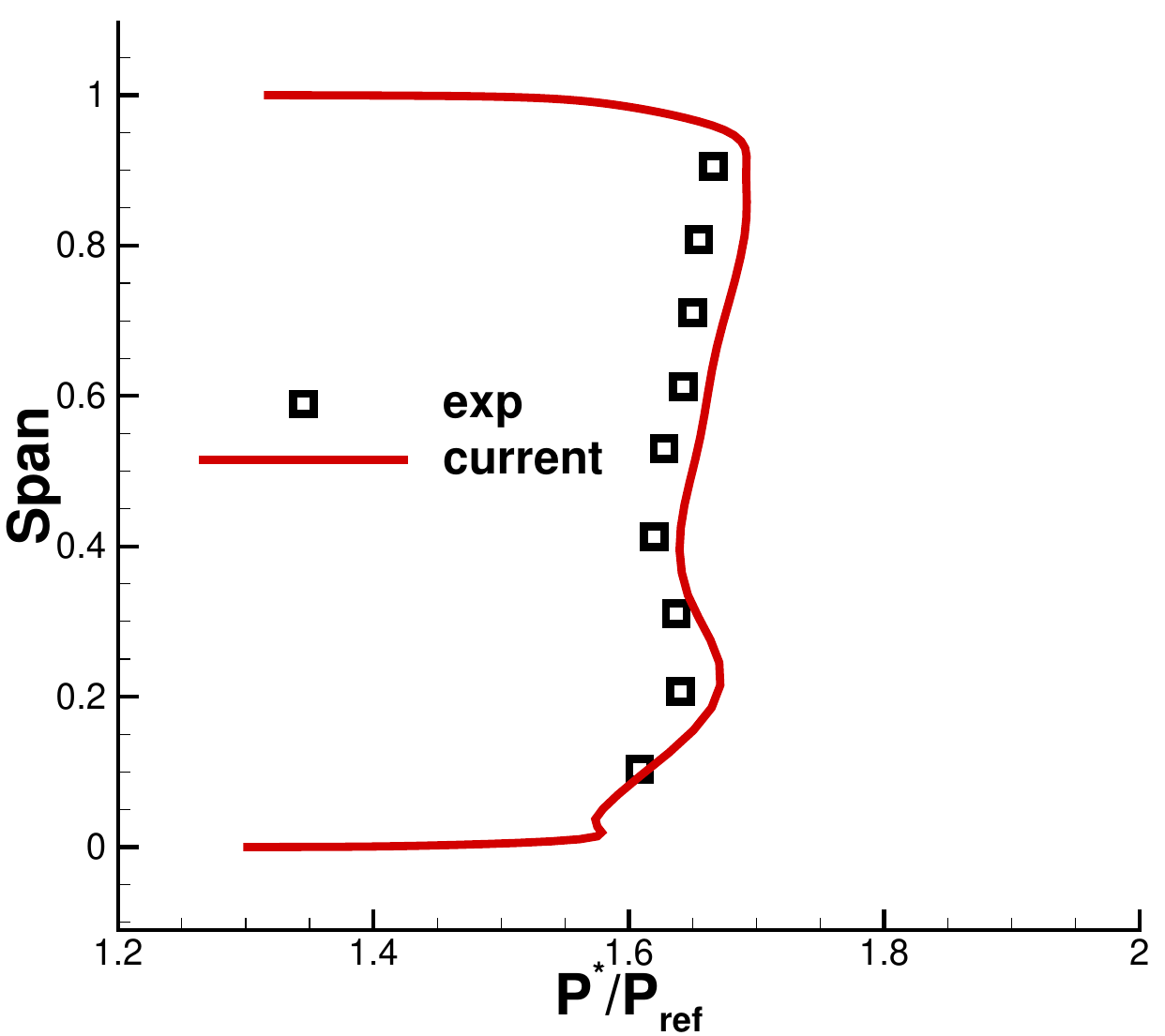} }\quad
  \subfigure[Total temperature at peak efficiency]{\includegraphics[width=0.3\textwidth]{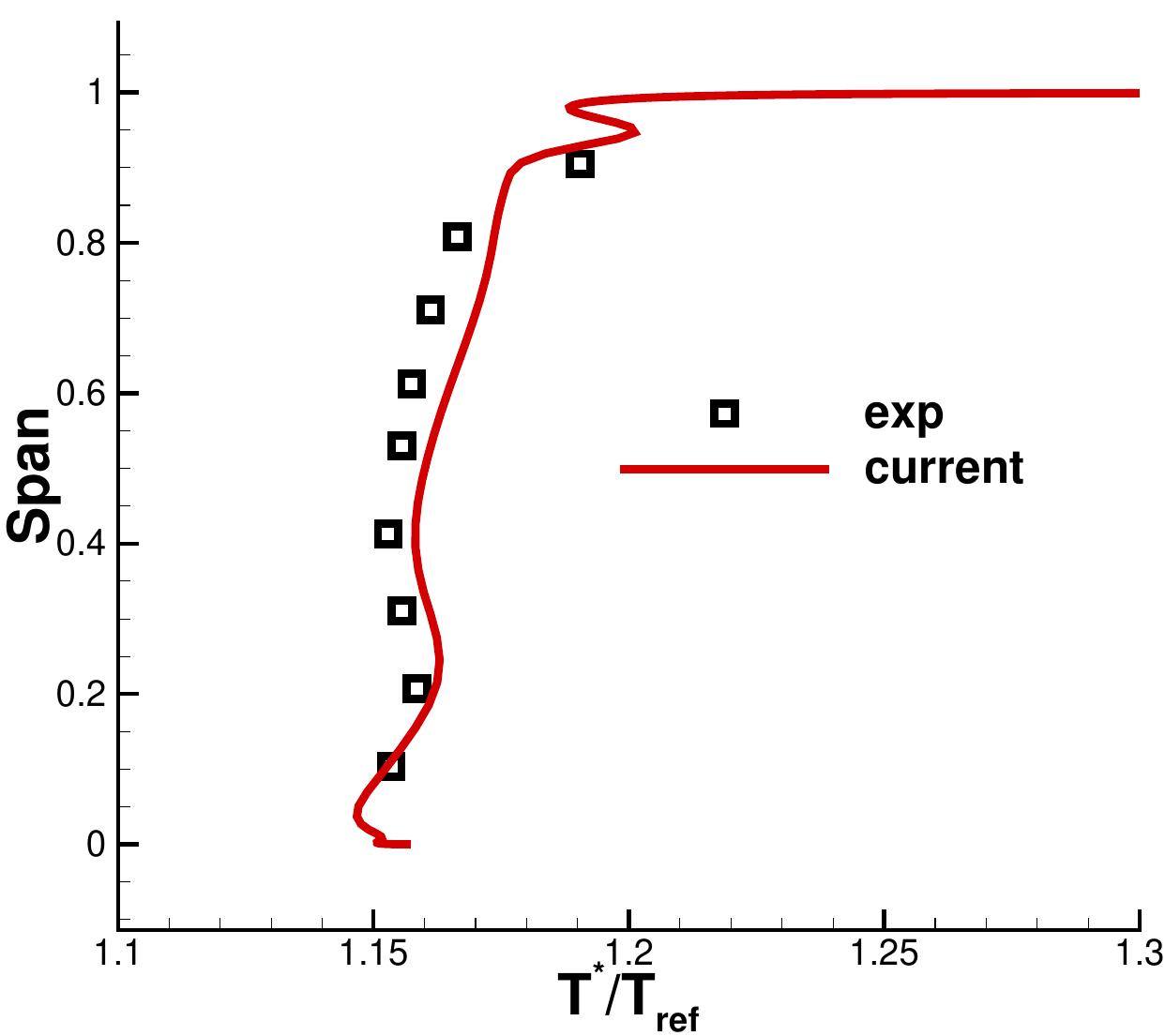}}\quad
  \subfigure[Flow angle at peak efficiency]{\includegraphics[width=0.3\textwidth]{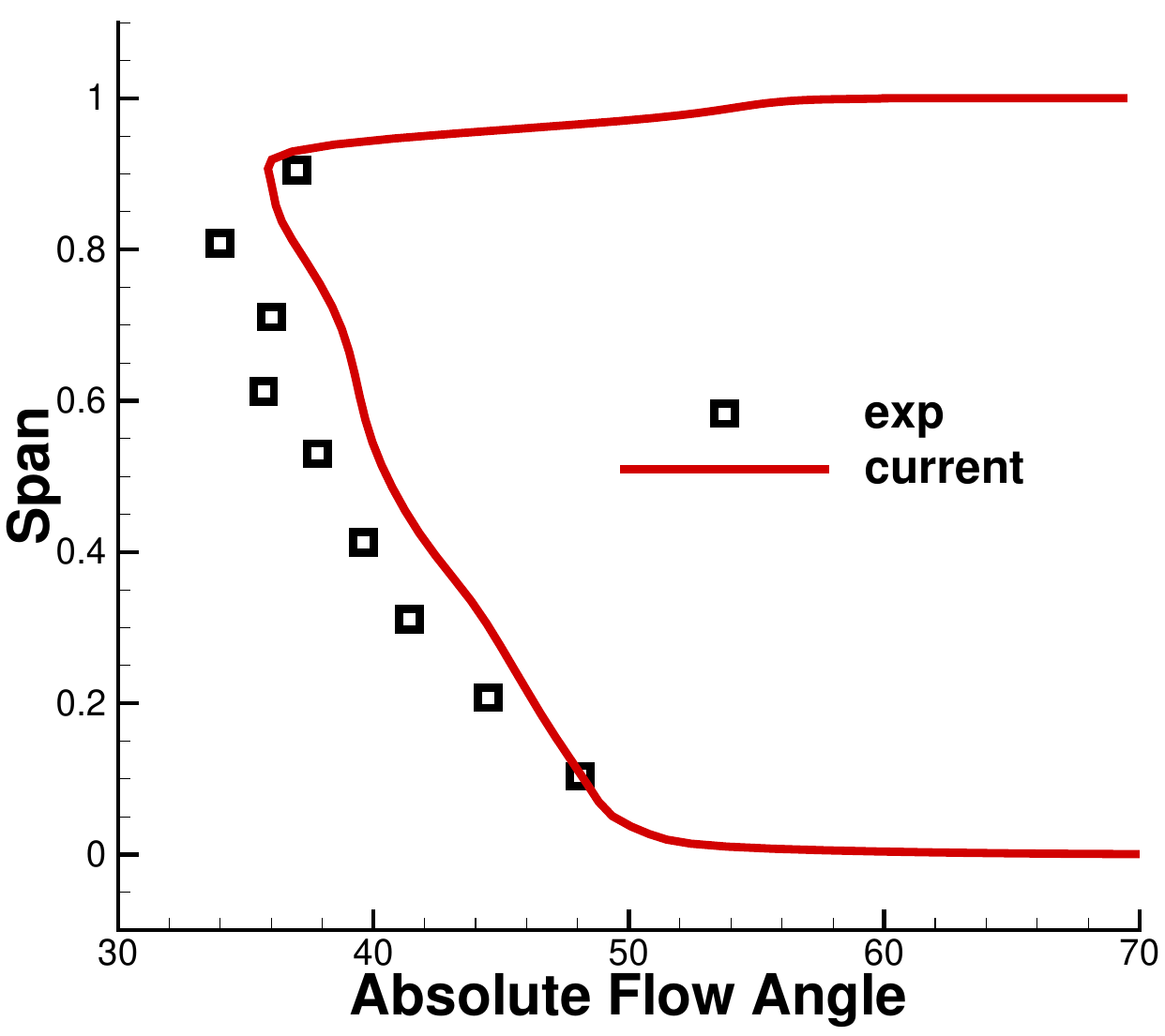}}
  \subfigure[Total pressure at near stall]{\includegraphics[width=0.3\textwidth]{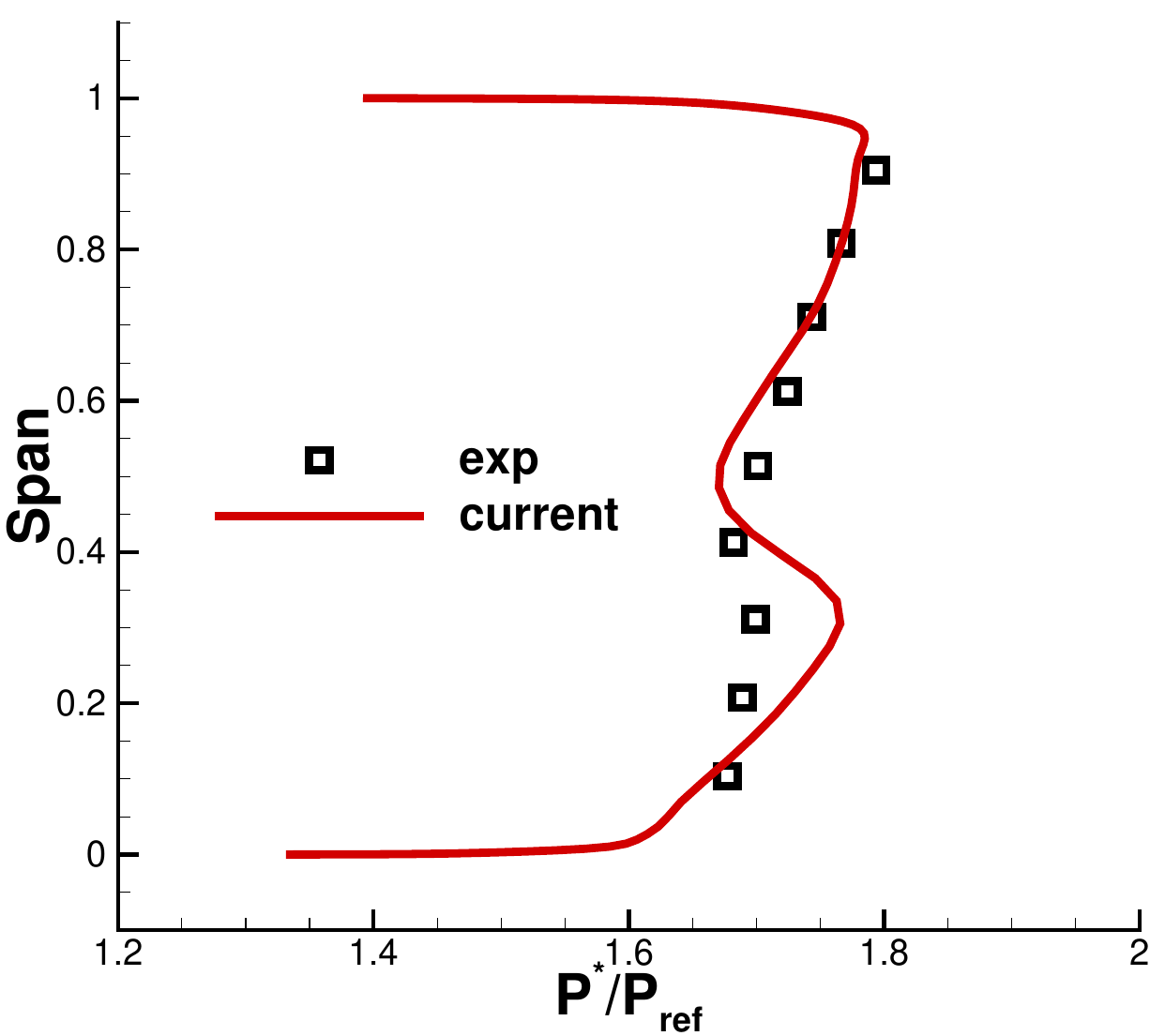}}\quad
  \subfigure[Total temperature at near stall]{\includegraphics[width=0.3\textwidth]{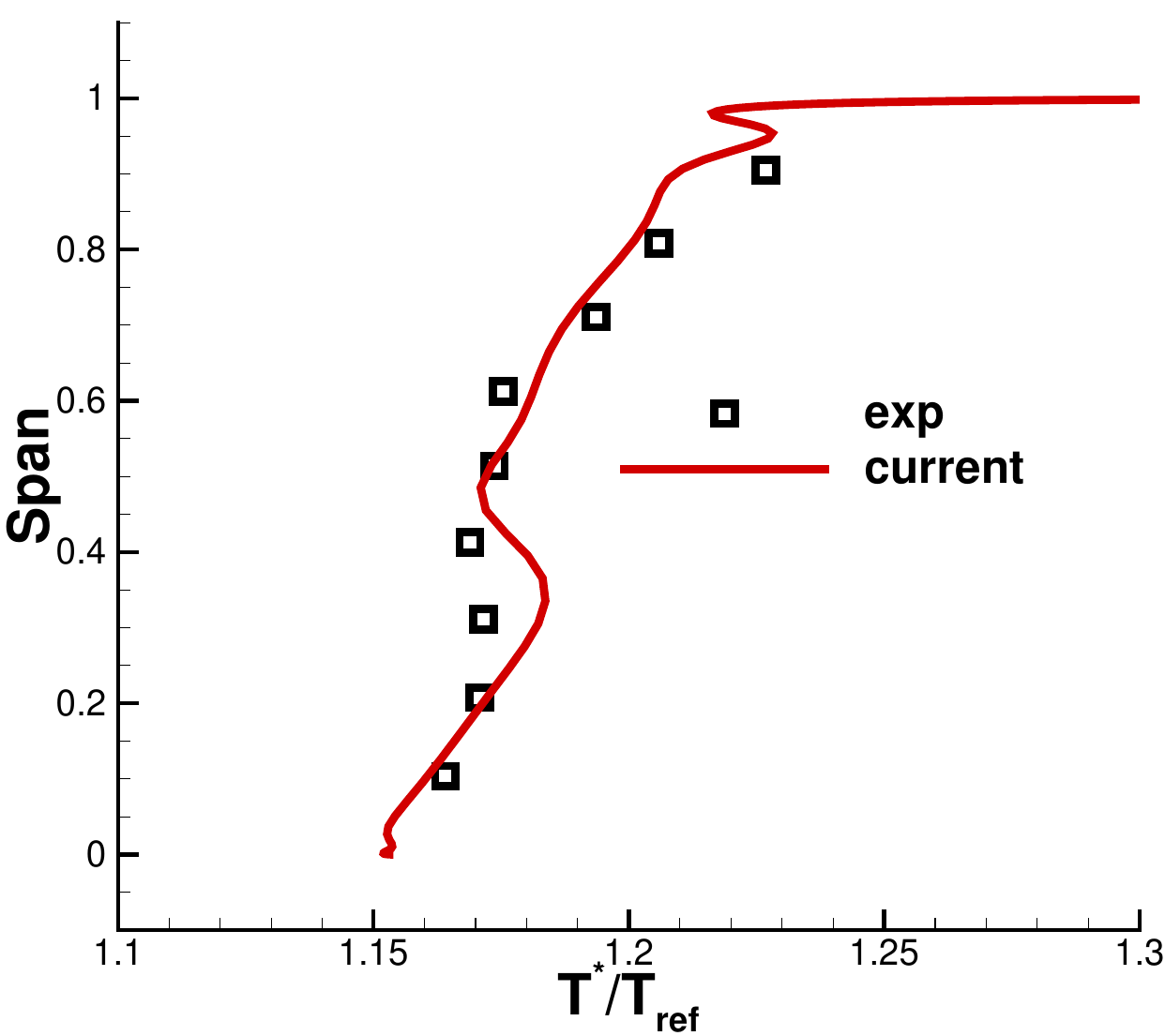}}\quad
  \subfigure[Flow angle at near stall]{\includegraphics[width=0.3\textwidth]{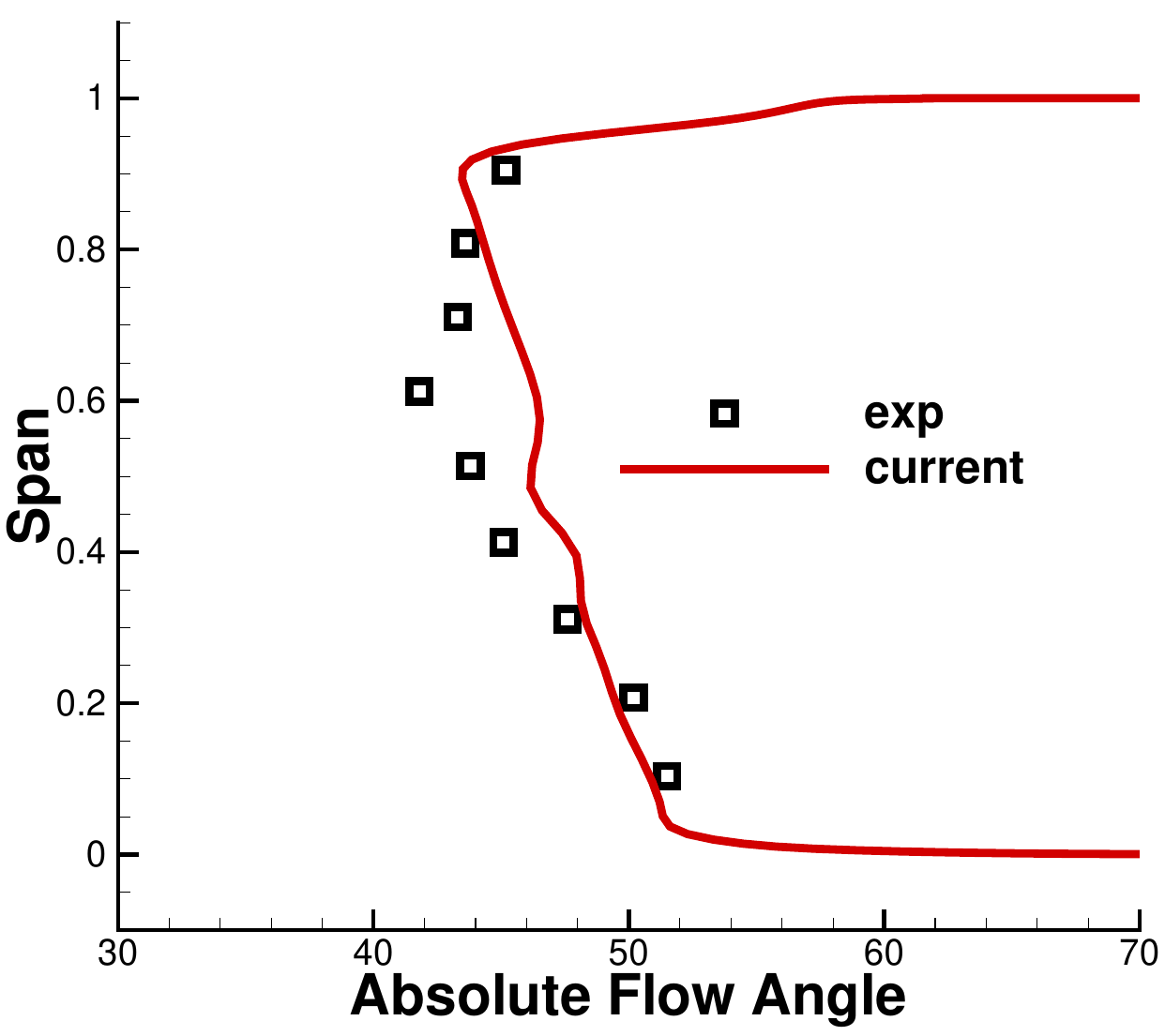}}
  \caption{comparison of circumferentially averaged outlet parameters of NASA Rotor 67 between computed results and experimental data \cite{strazisar1989laser}}
  \label{fig:rotor67circAvg}
\end{figure}

The comparison of relative Mach number contours between computed results and experimental data at the peak efficiency point and near the stall point is shown in Fig. \ref{fig:rotor67machContour} at different spanwise positions. The computed results can capture the position of the shock wave from the experimental data. However, at the near stall point, at 30\% span, the computed results show flow separation on the suction surface. In contrast, the experimental data show separation on the pressure surface, indicating that the computed results have a larger angle of attack corresponding to a lower axial velocity of the incoming flow.
The comparison of circumferentially averaged outlet parameters between computed results and experimental data at the peak efficiency point and near the stall point is shown in Fig. \ref{fig:rotor67circAvg}. The results of total pressure and total temperature are normalized by the inlet total pressure and total temperature, respectively. At the peak efficiency point, the computed results show good agreement with the experimental data. However, at the near stall point, the computed results show higher total pressure and total temperature than the experimental data between 20\% and 40\% span, which agrees with the result of relative Mach number contours at 30\% span. The higher total pressure and total temperature indicate more work input per unit mass flow due to the lower mass flow rate in this region.

\subsection{NASA Rotor 37}
NASA Rotor 37 was initially designed for an experimental study of a four-stage axial compressor. The design parameters of this four-stage compressor were chosen to be typical of the inlet stage of an aircraft engine's high-pressure compressor, with experimental results available in reference \cite{reid1978design}. Later, NASA conducted more detailed measurements on one of the rotors in isolation (without inlet guide vanes and outlet stator) and named it Rotor 37. In 1994, the International Gas Turbine Institute of the American Society of Mechanical Engineers (ASME/IGTI) used NASA Rotor 37 as a blind test case to evaluate turbomachinery CFD codes \cite{wisler1994rotor}, aiming to assess the capability of CFD programs in predicting compressor flow fields at that time. NASA Rotor 37 has attracted considerable numerical simulation studies \cite{arima1999numerical,suder1996experimental,chima1998calculation,hah1999development} not only because of its comprehensive experimental data and simple isolated rotor flow with accurately definable boundary conditions allowing fair and detailed evaluation of turbomachinery CFD, but also due to its representativeness among transonic rotors making it essential to understand its flow structure and loss mechanisms. The basic geometric and aerodynamic parameters of NASA Rotor 37 are shown in Table \ref{tab:rotor37}.
\begin{table}[htb!]
  \centering
  \caption{Basic information of NASA Rotor 37 \cite{wisler1994rotor}}
  \label{tab:rotor37}
  \begin{tabular}{cc}
    \toprule
    Parameter & Value \\
    \midrule
    Number of blades & 36 \\
    Design rotational speed (rpm) & 17188.7 \\
    Design mass flow rate (kg/s) & 20.19 \\
    Design pressure ratio & 2.106 \\
    Tip speed (m/s) & 454.14 \\
    Tip clearance at design speed (cm) & 0.0356 \\
    Relative Mach number at tip inlet & 1.48 \\
    Relative Mach number at hub inlet & 1.13 \\
    Aspect ratio (mean blade height/root axial chord) & 1.19 \\
    Solidity at tip & 1.29 \\
    Hub-to-tip ratio at inlet & 0.7 \\
    \bottomrule
  \end{tabular}
\end{table}

\begin{figure}[htb!]
  \centering
  \subfigure[The mesh projection on S2 surface\label{fig:rotor37MeshS2}]{\includegraphics[height=0.35\textwidth]{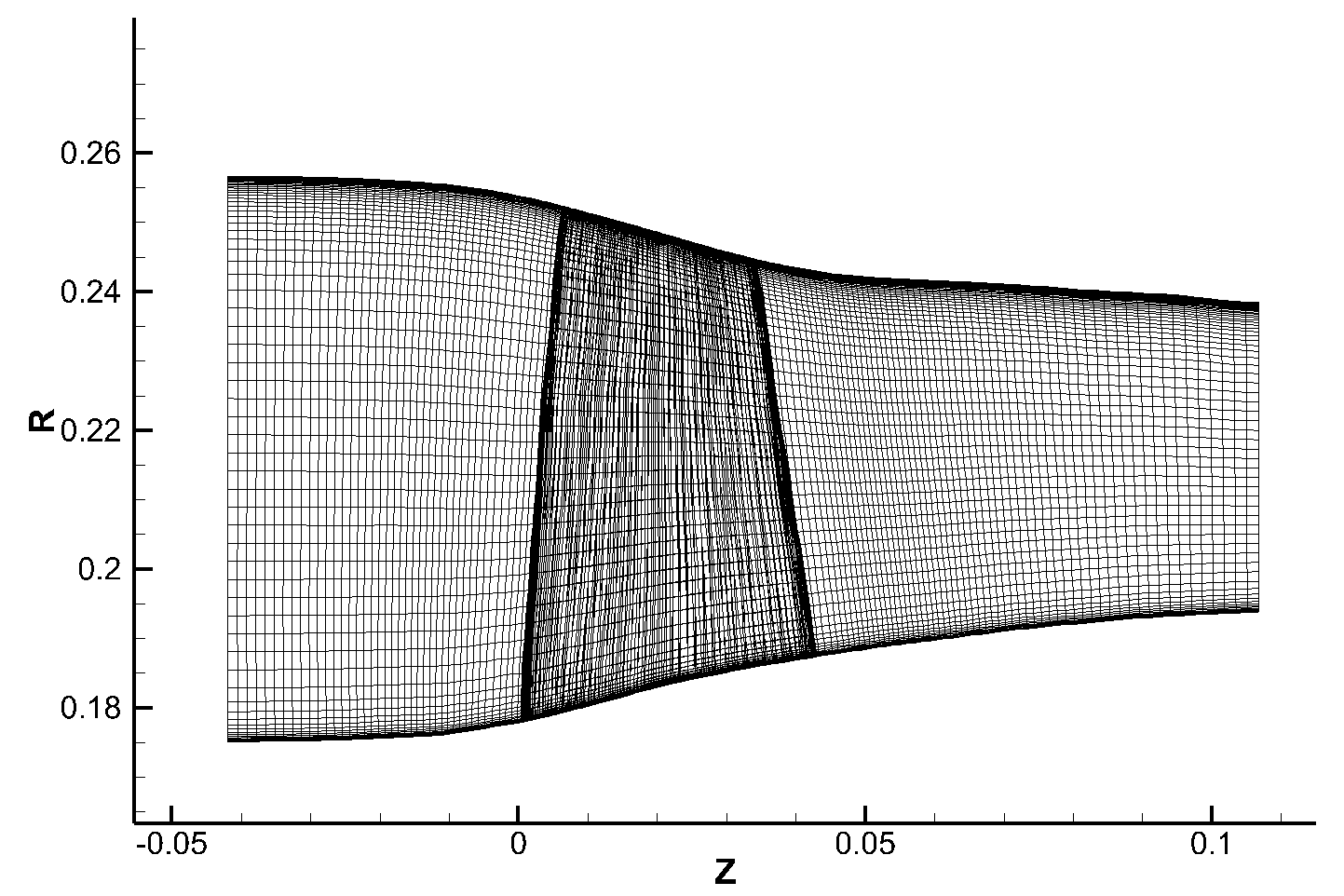}}
  \subfigure[The mesh projection on S1 surface\label{fig:rotor37MeshS1}]{\includegraphics[height=0.35\textwidth]{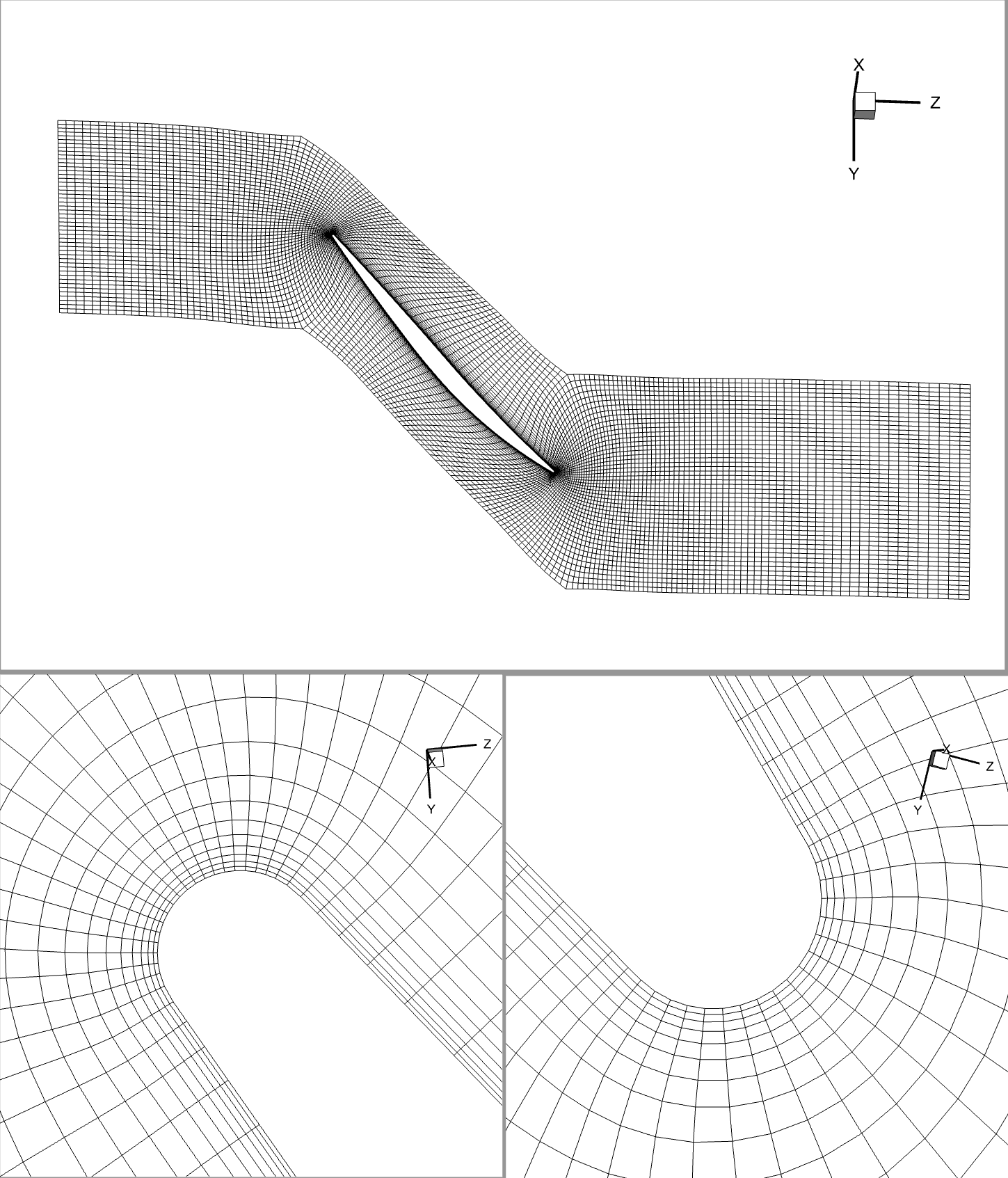}}
  \caption{The computational mesh of NASA Rotor 37}
  \label{fig:rotor37}
\end{figure}

The computational mesh of NASA Rotor 37 is shown in Fig. \ref{fig:rotor37}. The positions of the inlet and outlet are set at the experimental station. The total number of cells is $972,800$. The H-O-H type of topology is used to mesh the blade surface, and the first layer of cells is located at the blade surface with a distance of $10^{-6}$ to the blade surface. As in the case of NASA Rotor 67, the tip clearance is set as 0.6 times the design tip clearance (0.0356 cm) by considering the vena contraction effect, instead of refining the mesh near the tip of the blade.

\begin{figure}[htb!]
  \centering
  \subfigure[Total pressure ratio]{\includegraphics[width=0.4\textwidth]{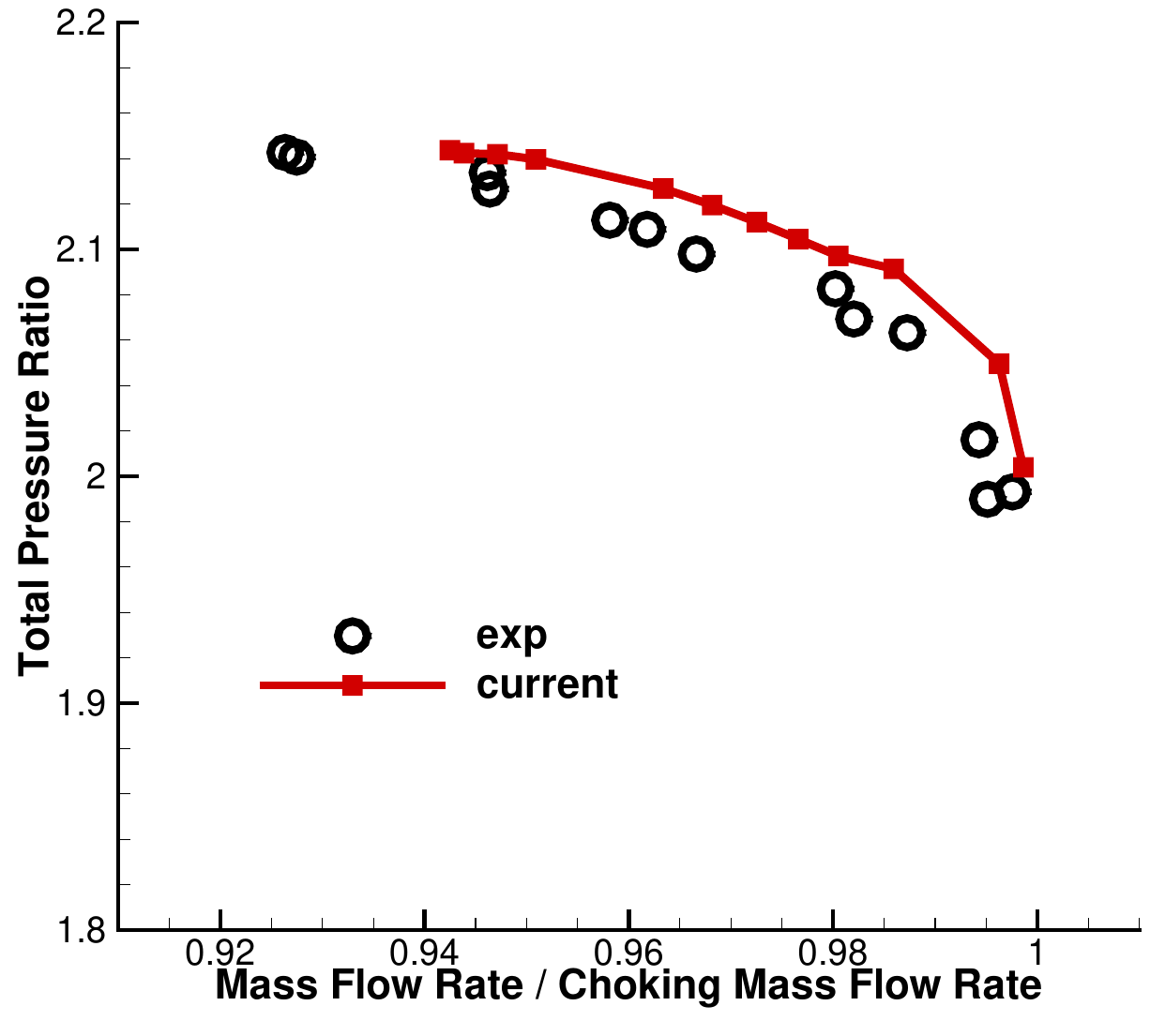}}
  \subfigure[Adiabatic efficiency]{\includegraphics[width=0.4\textwidth]{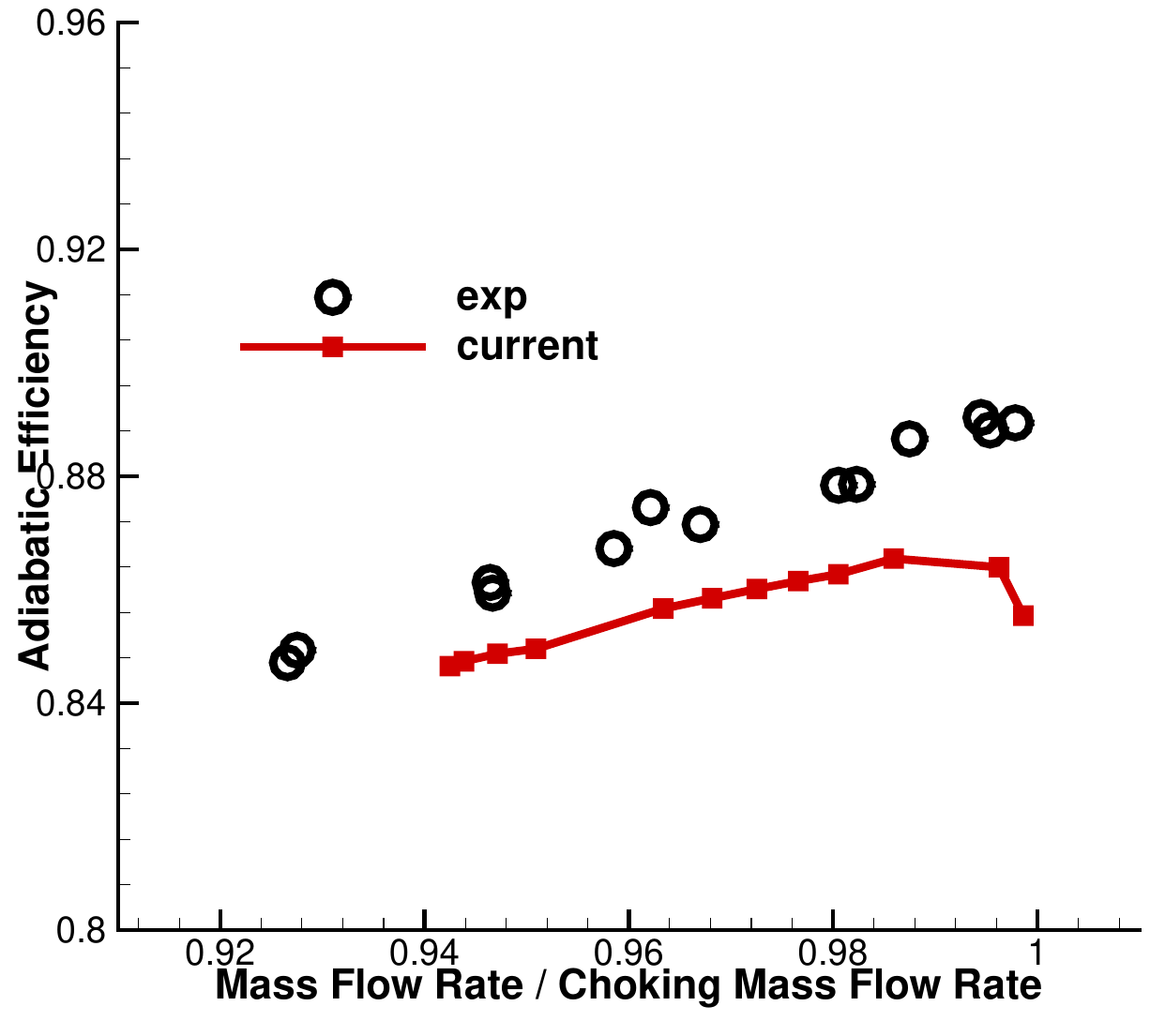}}
  \caption{Performance characteristics of NASA Rotor 37}
  \label{fig:rotor37char}
\end{figure}

\begin{figure}[htb!]
  \centering
  \subfigure[The convergence history of mass flow rate]{\includegraphics[width=0.35\textwidth]{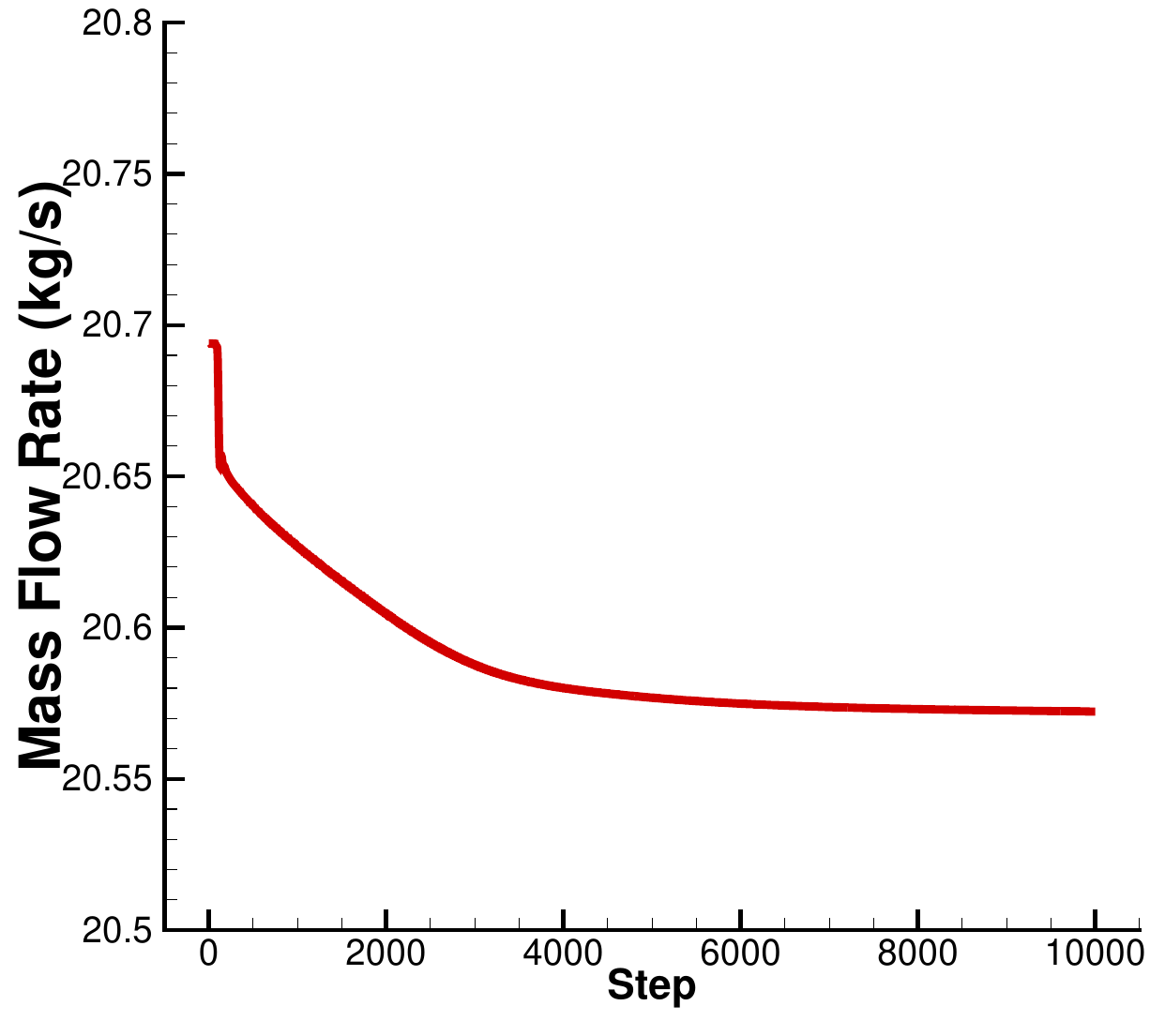}}
  \subfigure[The convergence history of total pressure ratio]{\includegraphics[width=0.3\textwidth]{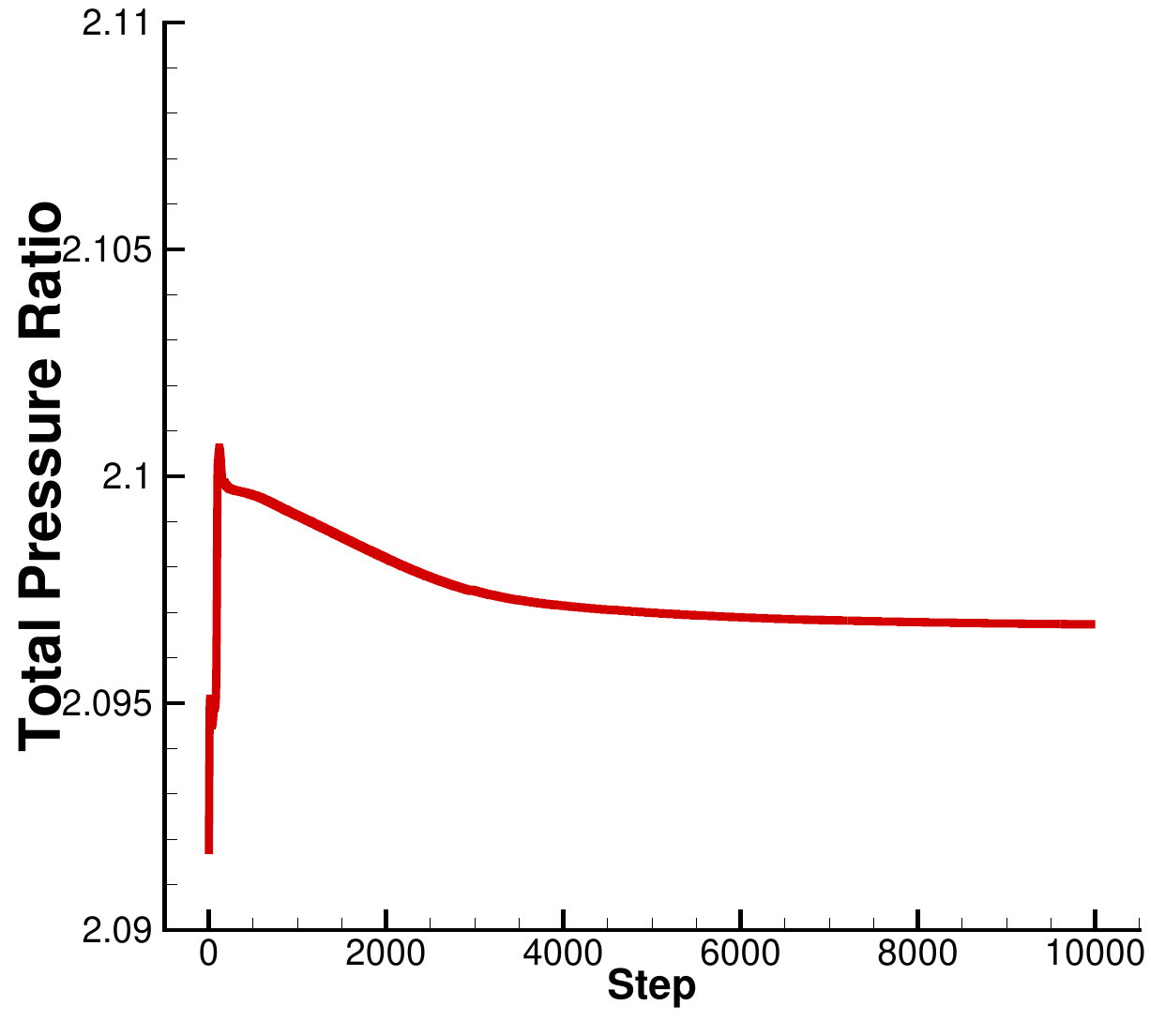}}
  \subfigure[The convergence history of adiabatic efficiency]{\includegraphics[width=0.3\textwidth]{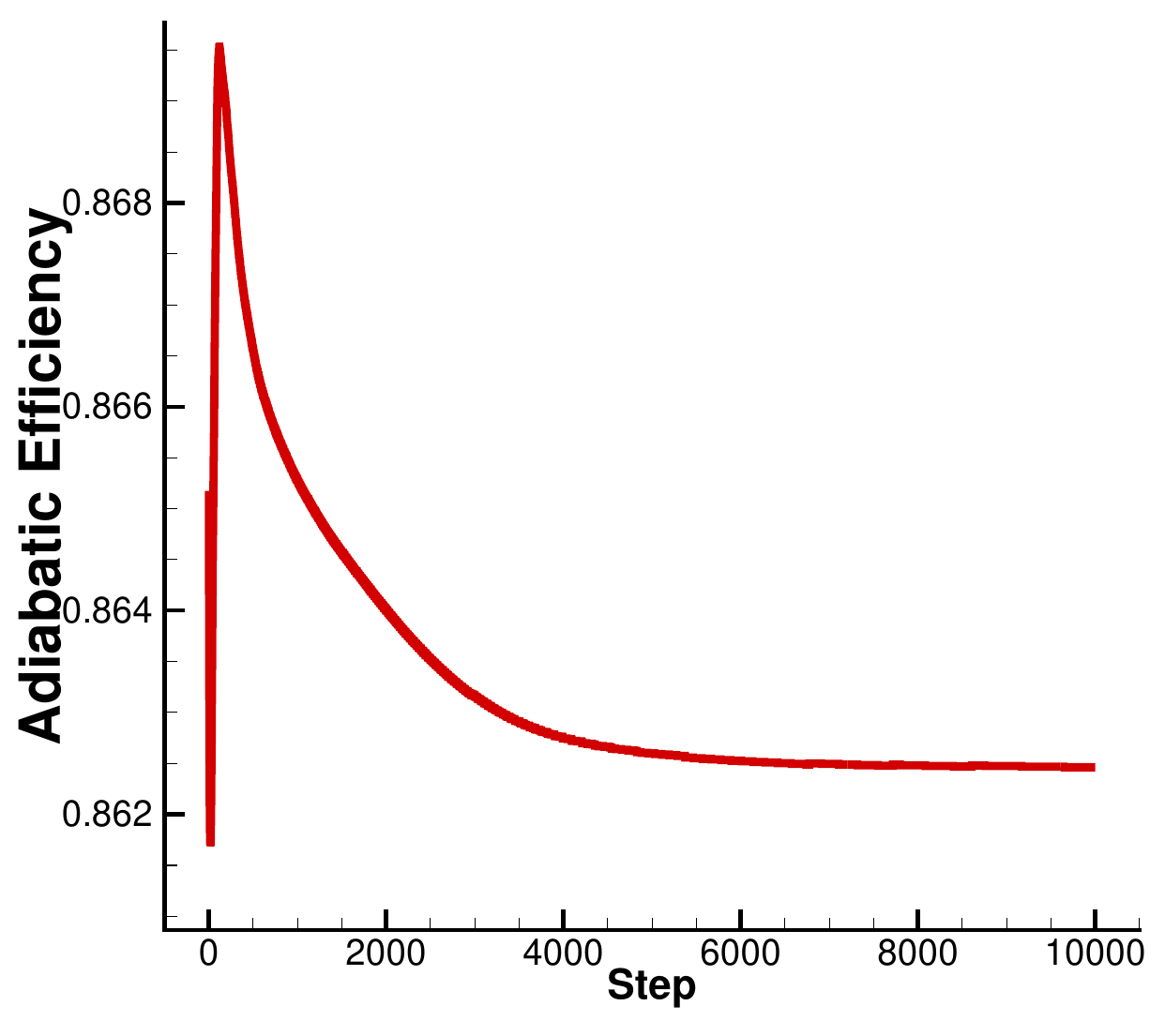}}
  \caption{The convergence history of NASA Rotor 37 at 98\% choked mass flow rate point}
  \label{fig:rotor37convergence}
\end{figure}

The boundary conditions and numerical settings are the same as those of NASA Rotor 67. The computed choked mass flow rate is 20.96 kg/s, while the corresponding experimental result is $20.93\pm 0.14$ kg/s, which falls within the experimental error band. The computed performance characteristics with normalized mass flow rate as the parameter are shown in Fig. \ref{fig:rotor37char}. The comparison shows that our computed efficiency is slightly lower than the experimental value, and the stall margin is smaller than the experimental result.
The convergence history at the 98\% choked mass flow rate point of mass flow rate, total pressure ratio, and adiabatic efficiency is shown in Fig. \ref{fig:rotor37convergence}. The results show that the performance parameters converge after approximately 4000 steps. It is also noted that significant variations only occur in the first 500 steps, while the subsequent convergence process is relatively smooth with variations within 0.5\%.

\begin{figure}[htb!]
  \centering
  \subfigure[90\% span]{\includegraphics[width=0.3\textwidth]{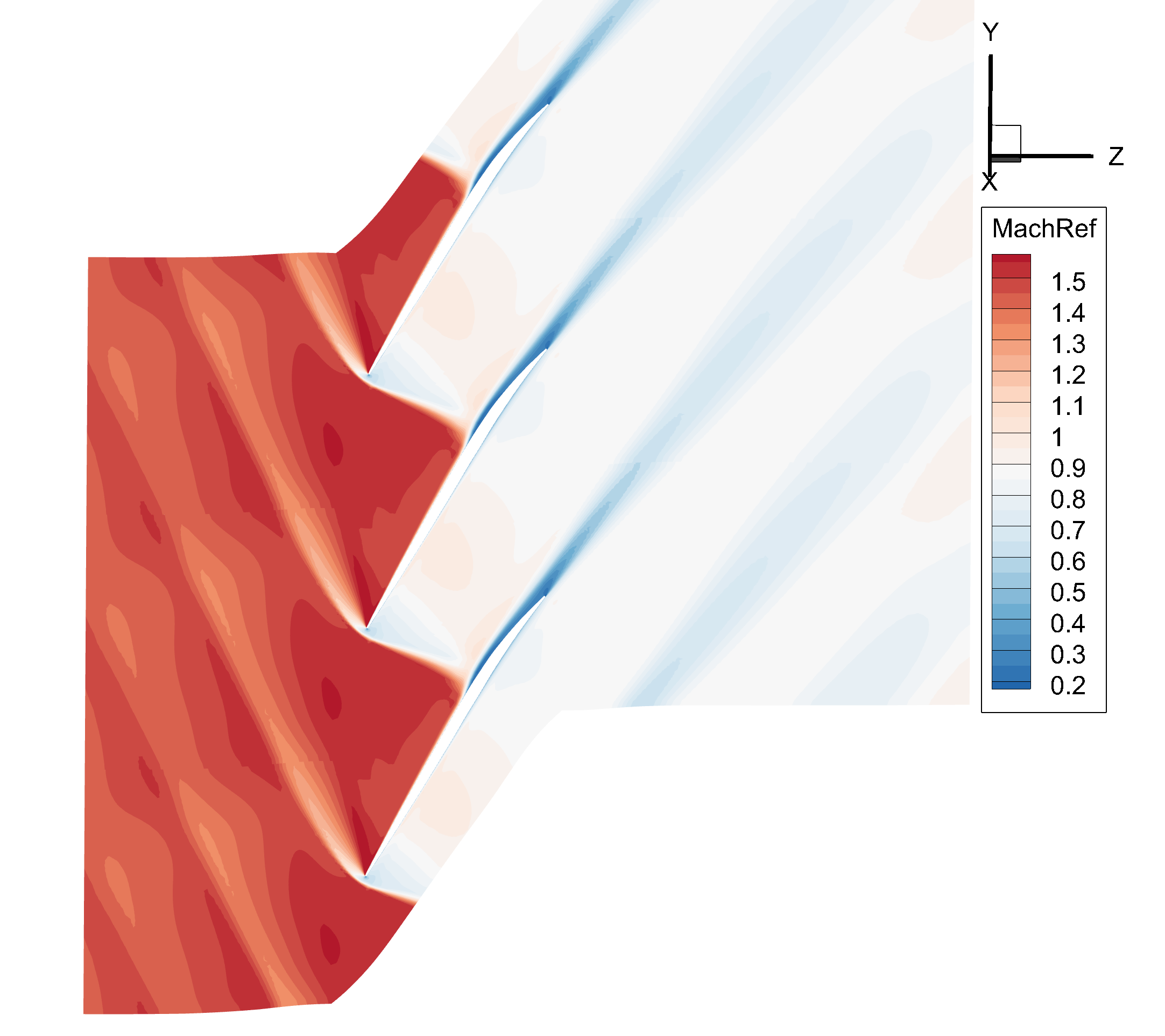}}
  \subfigure[70\% span]{\includegraphics[width=0.3\textwidth]{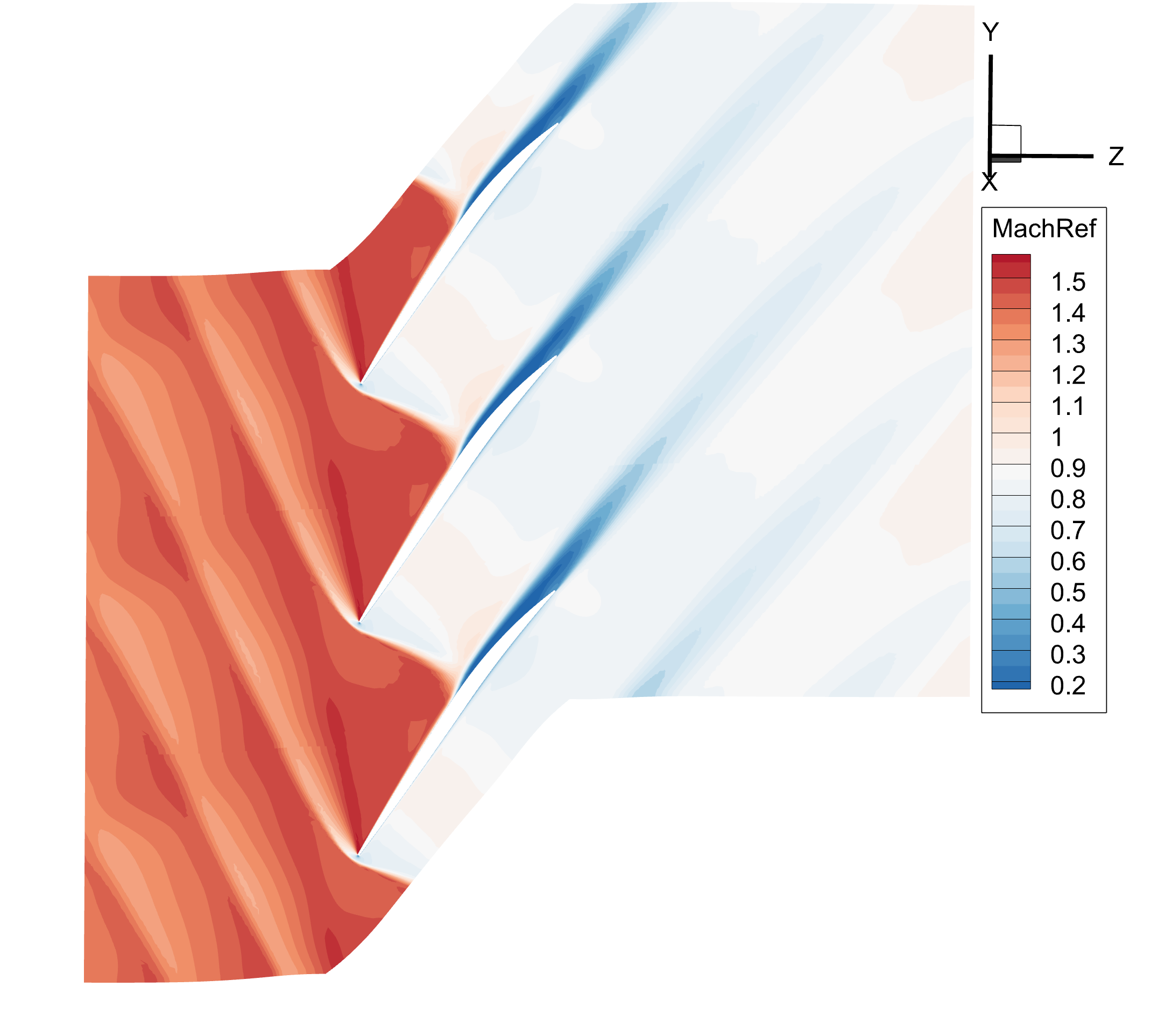}}
  \subfigure[30\% span]{\includegraphics[width=0.3\textwidth]{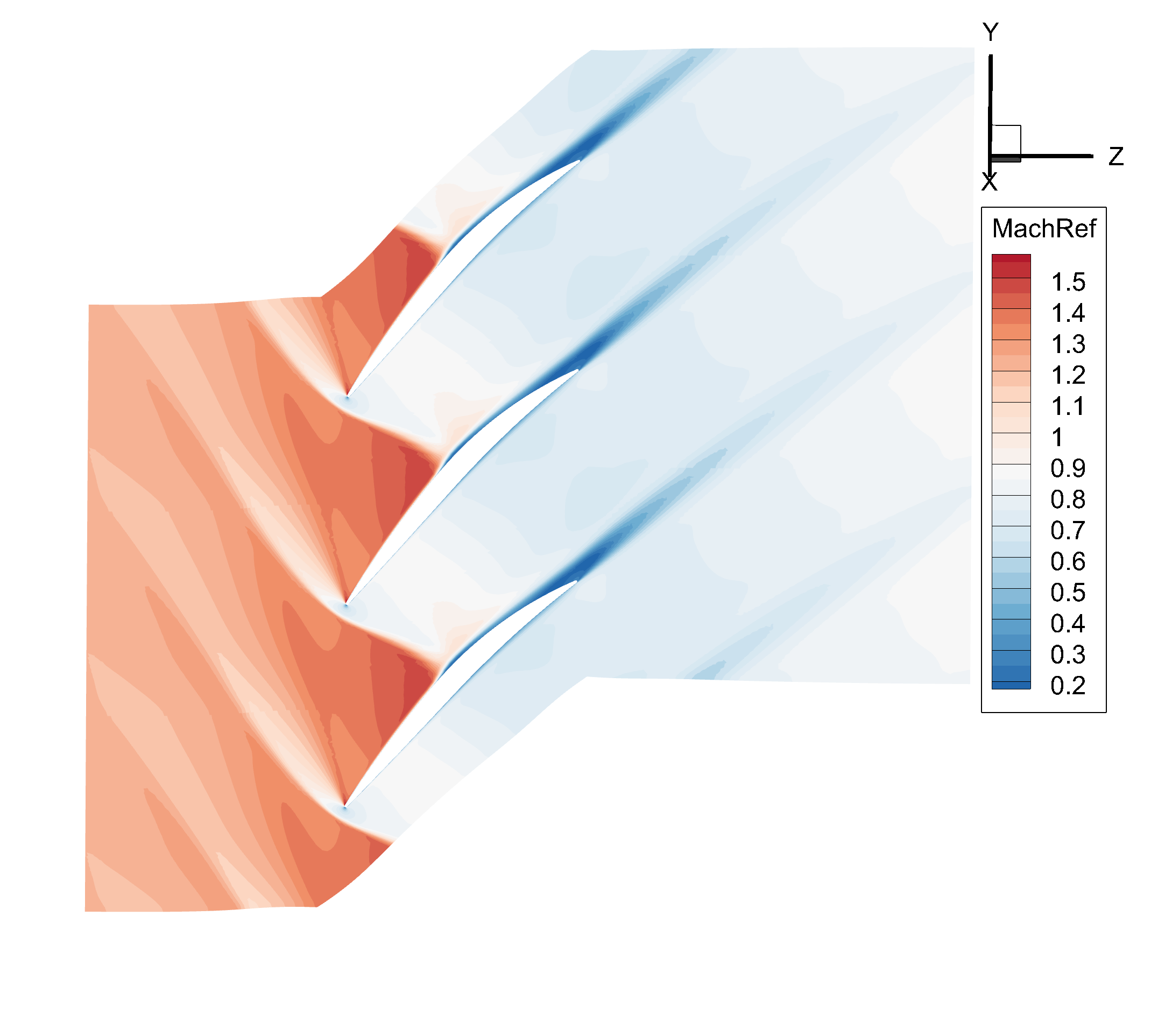}}
  \caption{The relative Mach number contours of NASA Rotor 37 at 98\% choked mass flow rate}
  \label{fig:rotor37machContour}
\end{figure}

\begin{figure}[htb!]
  \centering
  \subfigure[Total pressure]{\includegraphics[width=0.4\textwidth]{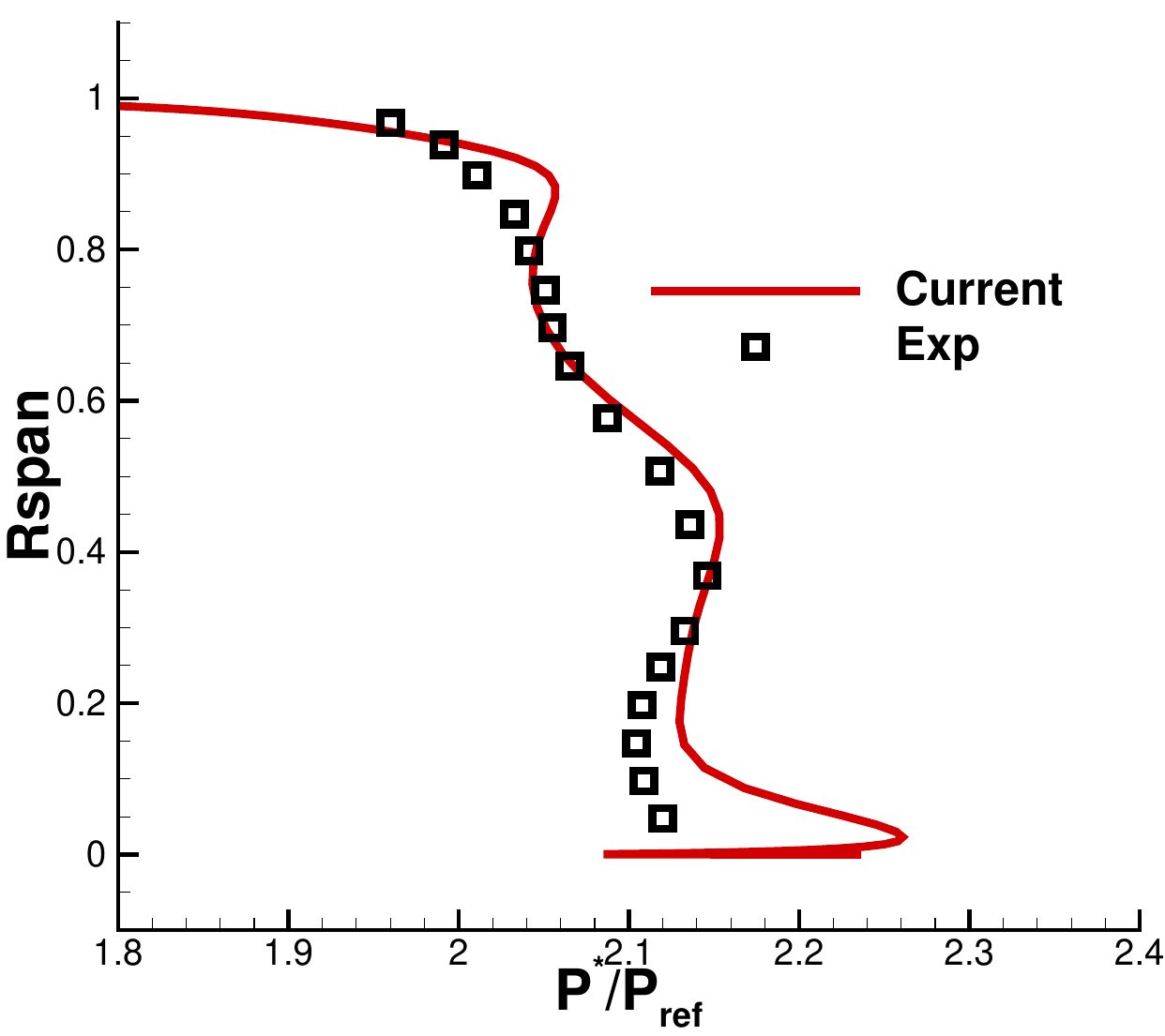}}
  \subfigure[Total temperature]{\includegraphics[width=0.4\textwidth]{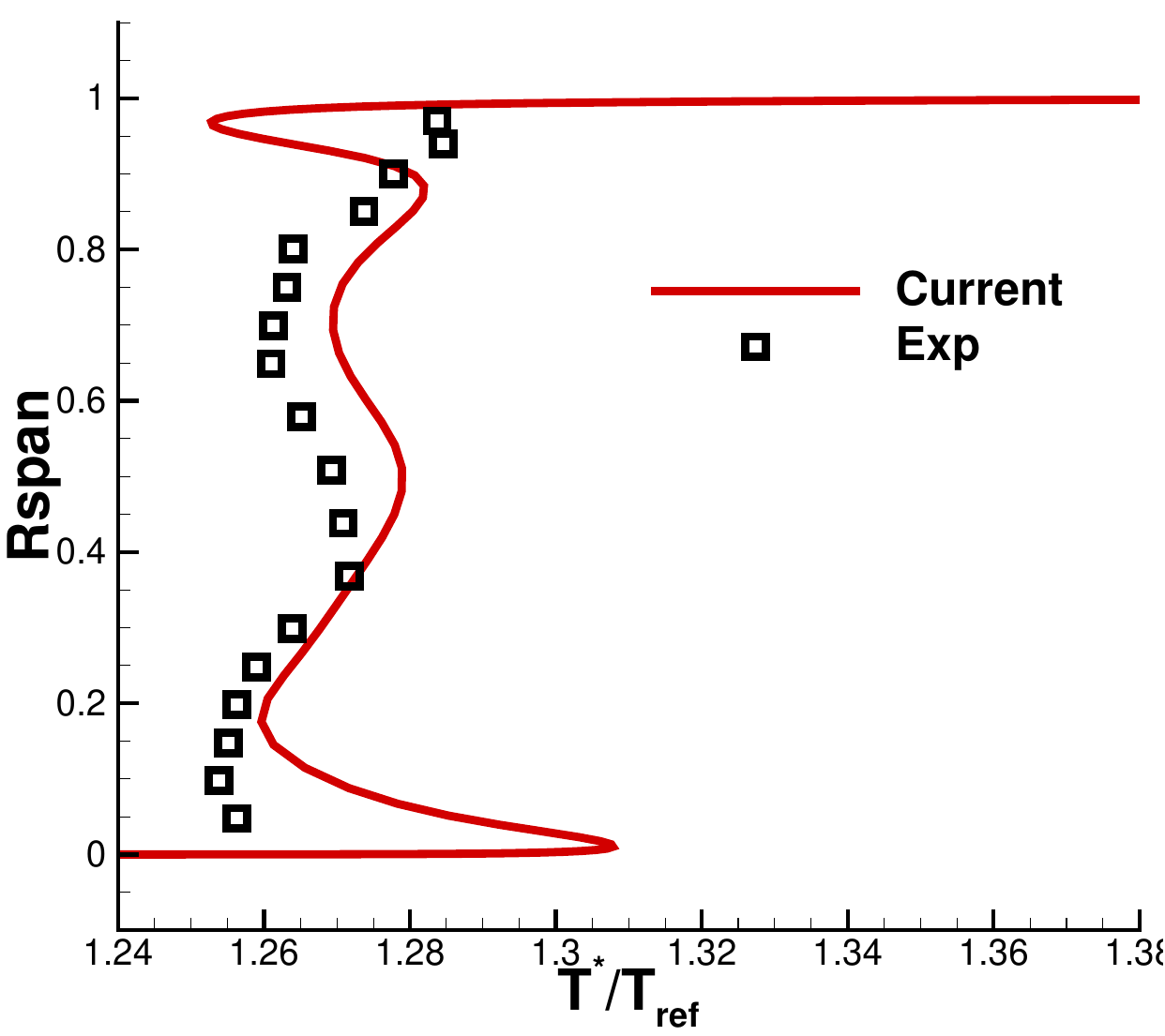}}
  \subfigure[Adiabatic efficiency]{\includegraphics[width=0.4\textwidth]{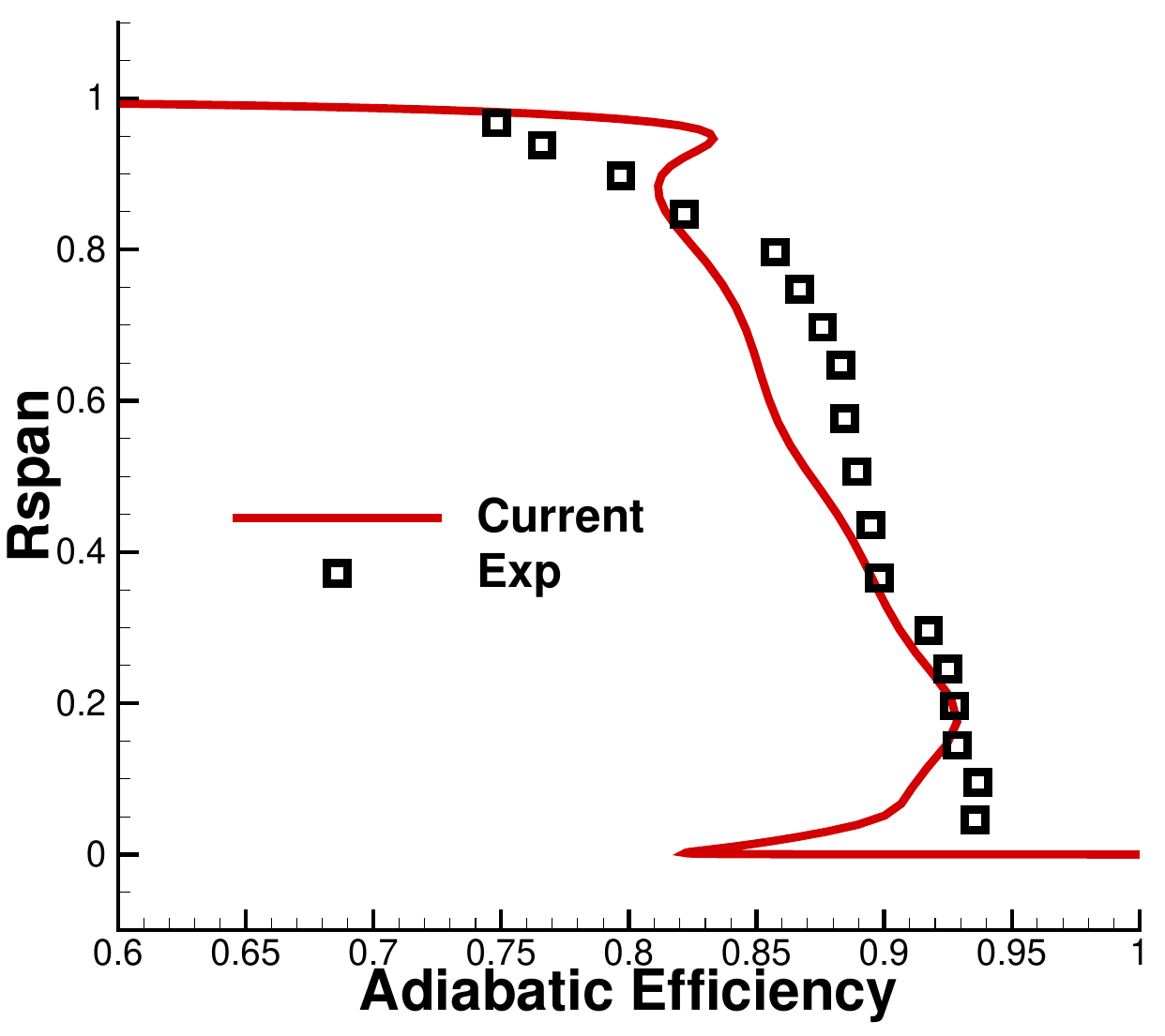}}
  \subfigure[Flow angle]{\includegraphics[width=0.4\textwidth]{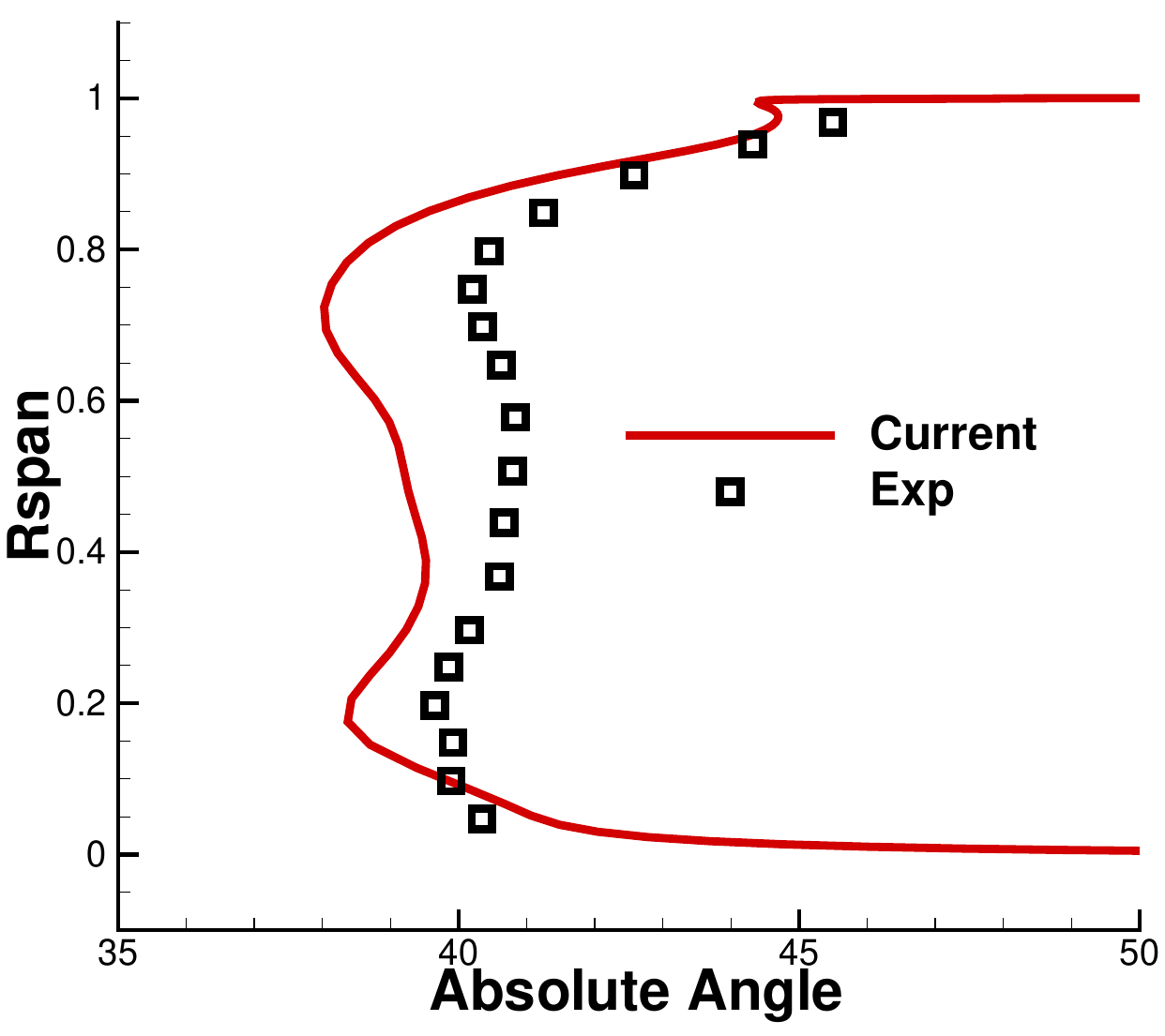}}
  \caption{The circumferentially averaged outlet parameters of NASA Rotor 37 at 98\% choked mass flow rate}
  \label{fig:rotor37circAvg}
\end{figure}

The relative Mach number contours of NASA Rotor 37 at 98\% choked mass flow rate are shown in Fig. \ref{fig:rotor37machContour}. The contours show the flow structure of NASA Rotor 37 at 98\% choked mass flow rate.
Fig. \ref{fig:rotor37circAvg} shows the circumferentially averaged outlet parameters (total pressure, total temperature, adiabatic efficiency, and flow angle) of NASA Rotor 37 at 98\% choked mass flow rate. At 98\% choked mass flow rate, the computed total temperature and total pressure under the 30\% span are higher than the experimental values, which is related to the boundary condition settings. The entire hub is set as a rotating wall, while in the experiment, only the rotor section of the hub is rotating, resulting in higher computational results. Between 40\% and 80\% span, the total pressure matches well with the experiment, but the total temperature is slightly higher than the experimental value, thus corresponding to lower efficiency. Meanwhile, the computed absolute flow angle is slightly smaller than the experimental value, corresponding to larger separation and higher losses.
\section{Conclusion}
In this paper, an implicit GMRES solver based on the gas-kinetic scheme was developed. First, the gas-kinetic scheme was extended to accommodate a rotating frame of reference. Subsequently, the implicit GMRES solver was formulated, with the Jacobian matrix derived using the kinetic flux vector splitting (KFVS) approximation for inviscid fluxes and the thin-layer approximation for viscous fluxes. To address turbulent flow, the shear-stress transport turbulence model was employed. The solver was validated through four test cases, which included external flows (Onera M6 wind and CRM Wing-Body Buffet Study) and internal flows (NASA Rotor 67 and NASA Rotor 37). The results demonstrated that the solver effectively captures the characteristics of these test cases. For most of the test cases, the convergence is achieved within 1000 steps. For the external flow test cases, our solver can accurately capture the position of the shock wave and the separation region. For the internal flow test cases, our solver can give a good prediction of the performance parameters.
In the future, to enhance the solver's capability for practical engineering problems, the implicit GMRES solver will be extended to handle unsteady flow problems using the dual time-stepping method. Additionally, more advanced turbulence models, such as Detached Eddy Simulation (DES) and Large Eddy Simulation (LES), will be implemented to better address turbulent flow simulation.

\section*{Acknowledgements}
We thank Mr. Junzhe Cao and Mr. Yuze Zhu for their assistance in mesh generation and discussions regarding the test cases, as well as Dr. Yajun Zhu for the Jacobian matrix derivation. This work was supported by the National Key R\&D Program of China (Grant No. 2022YFA1004500), the National Natural Science Foundation of China (Nos. 12172316, 92371107, 12302378 and 92371201), the Hong Kong Research Grant Council (Nos. 16208021, 16301222, and 16208324) and the Funding of National Key Laboratory of Computational Physics, and the Natural Science Basic Research Plan in Shaanxi Province of China (No. 2025SYS-SYSZD-070).

\bibliography{mybibfile}

\appendix

\section{Jacobian Matrix for Inviscid Flux and Viscous Flux}
\label{flux_jacobian}
\subsection{Jacobian matrix for inviscid flux}
At the cell face, consider the approach of flux calculation by using the moments of relative velocity (Eq. (\ref{fluxTransform}) and Eq. (\ref{fluxIntegralW})). The Jacobian matrix of the inviscid flux is given by
\begin{equation*}
  \frac{\partial \mathbb{F}}{\partial \mathbf{W}_{l,r}} = \frac{\partial \mathbb{F}}{\partial \mathbb{F}^\prime}\frac{\partial \mathbb{F}^\prime}{\partial \mathbf{W}^\prime_{l,r}}\frac{\partial \mathbf{W}^\prime_{l,r}}{\partial \mathbf{W}_{l,r}},
\end{equation*}
where $\mathbb{F}^\prime$ and $\mathbf{W}^\prime_{l,r}$ are the flux and the conservative variables of relative velocity, respectively. According to Eq. (\ref{fluxTransform}), the Jacobian matrix of $\mathbb{F}$ with respect to $\mathbb{F}^\prime$ is given by
\begin{equation*}
  \frac{\partial \mathbb{F}}{\partial \mathbb{F}^\prime} = \left[ \begin{matrix}
    1 & 0 & 0 & 0 & 0\\
 U_1& 1 & 0 & 0 & 0\\
 U_2& 0 & 1 & 0 & 0\\
 U_3& 0 & 0 & 1 & 0\\
    \frac{1}{2}(U_1^2+U_2^2+U_3^2)& U_1 & U_2 & U_3 & 1
  \end{matrix}\right],
\end{equation*}
where $\mathbf{U}=(U_1,U_2,U_3)$ is the velocity of mesh motion.
And the Jacobian matrix of $\mathbf{W}^\prime=[\rho,\rho W_1,\rho W_2,\rho W_3,p/(\gamma-1)+1/2\rho(W_1^2+W_2^2+W_3^2)]^T$ with respect to $\mathbf{W}=[\rho,\rho V_1,\rho V_2,\rho V_3,p/(\gamma-1)+1/2\rho(V_1^2+V_2^2+V_3^2)]^T$ is given by
\begin{equation*}
  \frac{\partial \mathbf{W}^\prime}{\partial \mathbf{W}} = \left[ \begin{matrix}
    1 & 0 & 0 & 0 & 0\\
 -U_1& 1 & 0 & 0 & 0\\
 -U_2& 0 & 1 & 0 & 0\\
 -U_3& 0 & 0 & 1 & 0\\
    \frac{1}{2}(U_1^2+U_2^2+U_3^2)& -U_1 & -U_2 & -U_3 & 1
  \end{matrix}\right].
\end{equation*}
The flux Jacobian matrix is given by
\begin{equation*}
  \frac{\partial \mathbb{F}^\prime}{\partial \mathbf{W}^\prime} =  T^{-1}\frac{\partial \mathbf{F}_{\text{KFVS}}}{\partial \tilde{\mathbf{W}}}T,
\end{equation*}
where $T$ is the transformation matrix from local coordinate to global coordinate. The flux $\mathbf{F}_{\text{KFVS}}$ is given by
\begin{equation*}
  \mathbf{F}_{\text{KFVS}}=\mathbf{F}_{\text{KFVS}}^l+\mathbf{F}_{\text{KFVS}}^r,
\end{equation*}
where $\mathbf{F}_{\text{KFVS}}^l$ and $\mathbf{F}_{\text{KFVS}}^r$ are calculated by the left and right cells with the same calculation method
\begin{equation*}
  \mathbf{F}_{\text{KFVS}}=\left[ \begin{aligned}
    \rho <w_1^1> \\
    \rho <w_1^2> \\
    \rho W_2<w_1^1> \\
    \rho W_3<w_1^1> \\
    \frac{1}{2}\rho [<w_1^3>+<w_1^1>\mathcal{M}_2]
  \end{aligned} \right],
\end{equation*}
where $\mathcal{M}_2=W_2^2+W_3^2+(K+2)/2\lambda$, and for $\mathbf{F}_{\text{KFVS}}^l$ the half integration $<w_1^n>_{>0}$ is used and for $\mathbf{F}_{\text{KFVS}}^r$ the half integration $<w_1^n>_{<0}$ is employed.
The flux Jacobian matix about $\mathbf{q}=(\rho,W_1,W_2,W_3,\lambda)^T$ gives
\begin{equation*}
  \frac{\partial \mathbf{F}_{\text{KFVS}}}{\partial \mathbf{q}}=\left[
    \begin{matrix}
 <w_1^1>&\rho<w_0^1> &0&0&\rho \mathcal{L}_1 \\
 <w_1^2>&2\rho<w_1^1> &0&0&\rho \mathcal{L}_2 \\
 W_2<w_1^1>&\rho W_2<w_0^1> &\rho<w_1^1>&0&\rho W_2\mathcal{L}_1 \\
 W_3<w_1^1>&\rho W_3<w_0^1> &0&\rho<w_1^1>&\rho W_3\mathcal{L}_1 \\
     \mathcal{D}_1 & \mathcal{D}_2 &\rho W_1<w_1^1> &\rho W_2<w_1^1> &\frac{1}{2}\rho [\mathcal{L}_3+\mathcal{L}_1\mathcal{M}_2],
    \end{matrix}
  \right]
\end{equation*}
where
\begin{equation*}
  \begin{aligned}
    \mathcal{L}_1&=\frac{\partial <w_1^1>}{\partial \lambda }=-\frac{1}{2\lambda}[<w_1^1>-W_1<w_0^1>], \\
    \mathcal{L}_2&=\frac{\partial <w_1^2>}{\partial \lambda }=-\frac{1}{2\lambda^2}<w_1^0>, \\
    \mathcal{L}_3&=\frac{\partial <w_1^3>}{\partial \lambda }=-\frac{3}{2\lambda} <w_1^1>, \\
    \mathcal{D}_1&=\frac{1}{2} [<w_1^3>+<w_1^1>\mathcal{M}_2], \\
    \mathcal{D}_2&=\frac{1}{2}\rho [3<w_1^2>+<w_1^0>\mathcal{M}_2].
  \end{aligned}
\end{equation*}
The transformation between primitive and conservative flow variables gives
\begin{equation*}
  \frac{\partial \mathbf{q}}{\partial \mathbf{W}} = \left[\begin{matrix}
    1 & 0 & 0 & 0 & 0\\
 -W_1/\rho & 1/\rho & 0&0&0 \\
 -W_2/\rho & 0 &1/\rho & 0&0 \\
 -W_3/\rho &0 &0&1/\rho & 0\\
    \lambda/\rho-1/2\mathcal{C}(W_1^2+W_2^2+W_3^2) &\mathcal{C}W_1 &\mathcal{C}W_2 &\mathcal{C}W_3 &-\mathcal{C}
  \end{matrix}\right],
\end{equation*}
where
\begin{equation*}
  \mathcal{C}=\frac{4\lambda^2}{(K+3)\rho}.
\end{equation*}

\subsection{Jacobian matrix for viscous flux}
 The thin shear layer (TSL) \cite{steger1977implicit,pulliam1980implicit} approximation is used for viscous flux with the assumption that the viscous effect is only significant in the normal direction of the cell surface. The viscous flux is approximated as
  \begin{equation*}
    \mathbf{F}_v =\left[\begin{matrix}
      0\\
 n_x\tau_{xx}+n_y\tau_{xy}+n_z\tau_{xz}\\
 n_x\tau_{xy}+n_y\tau_{yy}+n_z\tau_{yz}\\
 n_x\tau_{xz}+n_y\tau_{yz}+n_z\tau_{zz}\\
 n_x\Theta_x + n_y\Theta_y + n_z\Theta_z
    \end{matrix}\right],
  \end{equation*}
 where $\mathbf{n}=(n_x,n_y,n_z)$ is the normal vector of the cell surface. To obtain the viscous stress tensor, the derivatives are approximated by the normal derivatives via
  \begin{equation*}
    \begin{aligned}
      \frac{\partial q}{\partial x} &= \frac{q_R-q_L}{d}n_x,\\
      \frac{\partial q}{\partial y} &= \frac{q_R-q_L}{d}n_y,\\
      \frac{\partial q}{\partial z} &= \frac{q_R-q_L}{d}n_z,
    \end{aligned}
  \end{equation*}
 where $q$ is the variable such as $V_1, V_2, V_3, T$, $q_R$ and $q_L$ are the cell-averaged values of $q$ at the right and left of the cell surface, respectively, and $d$ is the distance between the cell centers of left and right cells.
 The viscous stress tensor and heat flux are given by
  \allowdisplaybreaks
  \begin{alignat*}{2}
    \tau_{xx}&=2\mu\frac{\partial V_1}{\partial n}+\frac{2}{3}\mu(n_x\frac{\partial V_1}{\partial n}+n_y\frac{\partial V_2}{\partial n}+n_z\frac{\partial V_3}{\partial n}), \\
    \tau_{yy}&=2\mu\frac{\partial V_2}{\partial n}+\frac{2}{3}\mu(n_x\frac{\partial V_1}{\partial n}+n_y\frac{\partial V_2}{\partial n}+n_z\frac{\partial V_3}{\partial n}), \\
    \tau_{zz}&=2\mu\frac{\partial V_3}{\partial n}+\frac{2}{3}\mu(n_x\frac{\partial V_1}{\partial n}+n_y\frac{\partial V_2}{\partial n}+n_z\frac{\partial V_3}{\partial n}), \\
    \tau_{xy}&=\tau_{yx}=\mu(n_x\frac{\partial V_2}{\partial n}+n_y\frac{\partial V_1}{\partial n}), \\
    \tau_{xz}&=\tau_{zx}=\mu(n_x\frac{\partial V_3}{\partial n}+n_z\frac{\partial V_1}{\partial n}), \\
    \tau_{yz}&=\tau_{zy}=\mu(n_y\frac{\partial V_3}{\partial n}+n_z\frac{\partial V_2}{\partial n}), \\
    \Theta_x&=V_1\tau_{xx}+V_2\tau_{xy}+V_3\tau_{xz} + \kappa n_x\frac{\partial T}{\partial n}, \\
    \Theta_y&=V_1\tau_{xy}+V_2\tau_{yy}+V_3\tau_{yz} + \kappa n_y\frac{\partial T}{\partial n}, \\
    \Theta_z&=V_1\tau_{xz}+V_2\tau_{yz}+V_3\tau_{zz} + \kappa n_z\frac{\partial T}{\partial n},
  \end{alignat*}
 and $\kappa$ is the thermal conductivity. Thus, the Jacobian matrix of the viscous flux is given by
  \begin{equation*}
    \frac{\partial \mathbf{F}_v}{\partial \mathbf{W}_R} = \frac{\mu}{d}\left[\begin{matrix}
      0 & 0 & 0 & 0 &0 \\
 b_{21} &a_1/\rho &a_2/\rho &a_3/\rho &0 \\
 b_{31} &a_2/\rho &a_4/\rho &a_5/\rho &0 \\
 b_{41} &a_3/\rho &a_5/\rho &a_6/\rho &0 \\
 b_{51} &b_{52} &b_{53} &b_{54} &a_7/\rho
    \end{matrix}\right],
  \end{equation*}
 and $\partial \mathbf{F}_v/\partial \mathbf{W}_L = - \partial \mathbf{F}_v/\partial \mathbf{W}_R$,
 where
  \begin{equation*}
    \begin{aligned}
 b_{21} &= -(a_1V_1 +a_2V_2 +a_3V_3)/\rho,\\ 
 b_{31} &=-(a_2V_1 +a_4V_2 +a_5V_3)/\rho, \\
 b_{41} &= -(a_3V_1 +a_5V_2 +a_6V_3)/\rho, \\
 b_{51} &= -\left[a_7(V_1^2+V_2^2+V_3^2-E)+a_1V_1^2 +2a_2V_1V_2 +2a_3V_1V_3 +a_4V_2^2 +2a_5V_2V_3 +a_6V_3^2\right]/\rho, \\
 b_{52} &= -a_7V_1/\rho -b_{21},\quad b_{53} = -a_7V_2/\rho -b_{31},\quad b_{54} = -a_7V_3/\rho -b_{41},
    \end{aligned}
  \end{equation*}
 and
  \begin{equation*}
    \begin{aligned}
 a_1 &= 4/3n_x^2+n_y^2+n_z^2, \\
 a_2 &= 1/3 n_x n_y, \\
 a_3 &= 1/3 n_x n_z, \\
 a_4 &= n_x^2+4/3 n_y^2+n_z^2, \\
 a_5 &= 1/3 n_y n_z, \\
 a_6 &= n_x^2+n_y^2+4/3n_z^2, \\
 a_7 &= \gamma/\text{Pr}(n_x^2+n_y^2+n_z^2).
    \end{aligned}
  \end{equation*} 
\end{document}